\documentclass[12pt]{article}
\usepackage{amsmath, amssymb, amsfonts,euscript,mathrsfs}
\usepackage{bm,bbm,latexsym,amsthm}
\usepackage{psfig, epsfig, verbatim}
\usepackage{natbib}
\usepackage{times}
\usepackage{setspace}
\usepackage{multirow}
\usepackage{color}
\usepackage{wrapfig}
\usepackage[caption = false]{subfig}
\usepackage{graphicx}
\usepackage{array}
\usepackage{float}
\usepackage{xfrac}
\usepackage{hyperref}

\newcommand\independent{\protect\mathpalette{\protect\independenT}{\perp}}
\def\independenT#1#2{\mathrel{\rlap{$#1#2$}\mkern2mu{#1#2}}}
\newcolumntype{C}[1]{>{\centering}m{#1}}

\newcommand{\PMTwo}{$\text{PM}_{2.5}$}
\newcommand{\SOTwo}{$\text{SO}_{2}$}
\newcommand{\NOx}{$\text{NO}_{x}$}
\newcommand{\COTwo}{$\text{CO}_{2}$}

\newcommand{\noteCMK}[1]{{\footnote{\color{blue} \bf{from Chanmin:} #1} }}

\title{\Huge\bf Bayesian Longitudinal Causal Inference in the Analysis of the Public Health Impact of Pollutant Emissions}
\date{\today}
\author{\small Chanmin Kim\\
   \small Department of Biostatistics, Boston University School of Public Health,\\
       \small Corwin M. Zigler\\
    \small Department of Statistics and Data Sciences and Department of Women's Health, \\
    \small The University of Texas at Austin,\\
    \small Michael J. Daniels\\
    \small  Department of Statistics, University of Florida,\\
    \small Christine Choirat\\
    \small Department of Biostatistics, Harvard T. H. Chan School of Public Health,\\
    \small Jason A. Roy\\
    \small Department of Biostatistics and Epidemiology, Rutgers School of Public Health}

\begin{document}
\maketitle
\doublespacing

\abstract{Pollutant emissions from coal-burning power plants have been deemed to adversely impact ambient air quality and public health conditions. Despite the noticeable reduction in emissions and the improvement of air quality since the Clean Air Act (CAA) became the law, the public-health benefits from changes in emissions have not been widely evaluated yet. In terms of the chain of accountability (HEI Accountability Working Group, 2003), the link between pollutant emissions from the power plants (\SOTwo) and public health conditions (respiratory diseases) accounting for changes in ambient air quality (\PMTwo) is unknown. We provide the first assessment of the longitudinal effect of specific pollutant emission (\SOTwo) on public health outcomes that is mediated through changes in the ambient air quality. It is of particular interest to examine the extent to which the effect that is mediated through changes in local ambient air quality differs from year to year. In this paper, we propose a Bayesian approach to estimate novel causal estimands: time-varying mediation effects in the presence of mediators and responses measured every year. We replace the commonly invoked sequential ignorability assumption with a new set of assumptions which are sufficient to identify the distributions of the natural indirect and direct effects in this setting. }

\section{Air Pollution Study}
\SOTwo\, is not only an important air pollutant linked with a number of adverse effects to human health, but also is a highly reactive gas that contributes to the formation of fine particle pollution (\PMTwo) which is a complex mixture of solid particles and liquid droplets with diameters less than 2.5 micrometers that impacts human health. Emissions from coal combustion - in particular \SOTwo\, - are potentially among the most harmful sources of \PMTwo\, and subject to many regulations, but the extent to which reductions initiate reduction in total mass \PMTwo\, and health is unknown. 

A series of regulations have been enforced by the US Environmental Protection Agency (EPA) to reduce \SOTwo\, emissions from the US coal-fired power plants. While there has been many papers \citep{ware1986effects,bernard2001potential,pope2006health,bell2007spatial} supporting the necessity of those regulations based on findings about positive associations between \SOTwo\, emission and adverse health outcome or between \PMTwo\, concentration and adverse health outcome, there has been none for the causal relationship between \SOTwo\, emission and public health outcomes accounting for any potential changes in ambient \PMTwo\, concentration as one integral chain, which links pollutant emissions from the power plants (\SOTwo) to public health outcomes accounting for changes in ambient air quality (\PMTwo). This type of research is crucial for implementing a new policy targeting reduction in air pollution by providing insight into understanding the exact mechanism of the air pollution epidemiology.

In particular, title IV of the 1990 Amendments to the Clean Air Act (CAA)  set a clear goal of reducing annual \SOTwo\, emissions by implementing a two-phase cap and trade program for coal-fired power plants. The second phase began in 2000 and required most of power plants including almost all coal-fired facilities to reduce their emissions across the nation. The program has been evaluated extensively and is lauded as a success especially due to the fact the states with the highest emitting sources have seen the greatest \SOTwo\, emission reduction annually under the program \citep{EPA:2015}. 
However, depending on changes in energy demands or different regulations imposed on power plants over the years, the power plants generate more (less) electricity and therefore emit more (less)  air pollutants. Thus, it is of great interest as well that we learn about exactly how \SOTwo\, emission reduction affects improvement of ambient air quality (\PMTwo\, concentrations) and in turn enhances public health conditions in a `longitudinal way'.While, in general, a significant change in \SOTwo\, emissions exposure level at each location subsequently changes ambient \PMTwo\, concentrations, many different factors intervene on this process from exposure level to ambient air quality such that the potential value of \PMTwo\, remains at the value that would have been observed under the opposite level of exposure to \SOTwo\, emissions. For example, in addition to the national programs, local governments have implemented their own policies targeting local pollutant sources which can reduce or sometimes even increase local ambient \PMTwo\, concentrations. Also, many different human activities can cause changes in ambient \PMTwo\, concentrations (e.g., domestic heating). Therefore, the health effects through changes in \PMTwo\, concentrations may not be constant over periods. Thus, a longitudinal mediation analysis would a natural choice to formally examine these questions.

\subsection{Literature}
In mediation analysis, there have been many approaches proposed to extend the simple/conventional approach \citep{Baro:Kenn:1986} to more complex settings, but most of the approaches focus on a setting with a single exposure, a single mediator, and a single outcomes \citep{Robi:Gree:1992, Pear:2001,vand:pete:2008,Goet:Vans:Goet:2008,Vand:Vans:2009,Imai:2010,Tche:2011,Vans:2012,vanderweele2015explanation,kim2018bayesian}. Some further allow a time-dependent confounder (or multiple mediators) that is impacted by the exposure \citep{albert2011generalized,Imai:2013,daniel:2014,vanderweele2015explanation} under the non time-varying exposure, mediator and outcome setting. Currently, there is very little work in the mediation analysis literature with time-varying exposures and mediators under a formal causal framework.  

\cite{bind2016causal} propose an approach for estimating the mediated effects when there is no time-varying confounding; that is, they assume no confounding relationship affected by prior treatments, mediators or outcomes. \cite{vanderweele2016mediation} relax this assumption by introducing a randomized interventional analogue of natural direct and indirect effects and \cite{lin2017mediation} extend this approach to time-to-event data; however, interventional effects are not directly comparable to the conventional natural direct and indirect effects. Also, none of these accommodate time varying outcomes.

These restrictive settings are mainly due to upholding a series of conditional independence assumptions, the so-called {\it sequential ignorability assumptions} \citep{robi:hern:brum:00,imai2013identification}. The second part of these assumptions states that two variables, the potential outcome and mediator, are unconfounded conditional on the past observations and all confounders. If any of these confounders are affected by the past exposures, then the sequential ignorability assumptions do not hold in general. See \cite{robi:hern:brum:00} and \cite{vanderweele2016mediation} for more details. In our application, we need to allow the time-varying confounder(s) to be affected by the past exposures: for example, pollutant emissions can affect temperature (time-varying confounder) by climate change.

In this paper, we propose a new set of assumptions tailored to our application which are sufficient to estimate the causal effects of our interest defined in Section 3 and reasonable to conceive in our setting. With these assumptions, we can incorporate time-varying confounders which are affected by the past exposures.

\section{Data Science}
Our aim is to assess the effect of exposure to \SOTwo\, emissions from coal-fired power plants on public health outcomes (measured from the Medicare claims data) which is presumed to be mediated through ambient \PMTwo\, concentrations. To achieve this goal, we gather all relevant information: power plants data from EPA's Air Markets Program Data (AMPD) and the US Energy Information Administration (EIA), data-fused estimates of \PMTwo\, concentrations\citep{di2016assessing} and ground level Ozone concentrations\citep{di2017hybrid}, hospitalizations data from Centers for Medicare \& Medicaid Services (CMS) and other supplementary data for multiple years. A salient challenge is how to create one integrated dataset as the raw datasets are spatially and temporally misaligned without any single common identifier as described in Table \ref{data}
\begin{table}[h]
\scriptsize
\centering
\begin{tabular}{c|p{4.5cm}|p{4.5cm}|p{1.7cm}|p{1.7cm}}
Data & Description & Variables & Spatial Unit & Temporal Unit\\
\hline
Power Plants Data & 406 Coal-filed power plants in the US. & Pollutant emissions (\SOTwo, \NOx, \COTwo), Other power plants characteristics (e.g., Heat Inputs, Operating Times) & Longitude \& Latitude & Month\\ \hline
Ambient Air Quality Data & Data-fused estimates of \PMTwo\, and Ozone concentrations & Daily averages for 43,014 locations & ZIP code & Day\\ \hline
Health Data & Respiratory diseases related hospitalizations among all Medicare beneficiaries & Hospitalizations for each condition, Basic Demographics & ZIP code & Month\\ \hline
Census data & Baseline demographics and socio-economic information & Population by many factors (age, gender, race), Income, Urban area, etc. & Zip code & Year (2000)\\ \hline
Smoking Data & Small area estimation based on the data from CDC (ref) & Smoking rates & County & Year (2000)\\ \hline
Meteorologic data & Data-fused estimates of temperature and relative humidity (RH) & Daily averages of Temperature and RH & ZIP code & Annual \\ \hline
\end{tabular}\label{data}
\caption{Description of Data Sources}
\end{table}

\subsection{Linking Power Plants to Exposed Populations: Characterization of Atmospheric Pollution Transport}
The linkage between the power plant data and the other data measured at ZIP code level can be linked via identifying local areas under pathways of emissions from each power plant. However, attention must be paid because of complexity of long range emission transportation. Here, we leverage an atmospheric transport and dispersion model, the HYSPLIT from the National Oceanic Atmospheric Administration (NOAA), to simulate forward air mass trajectories from each power plant over multiple years.

\begin{figure}[h]
    \centering
    \subfloat[]{{\includegraphics[width=7cm]{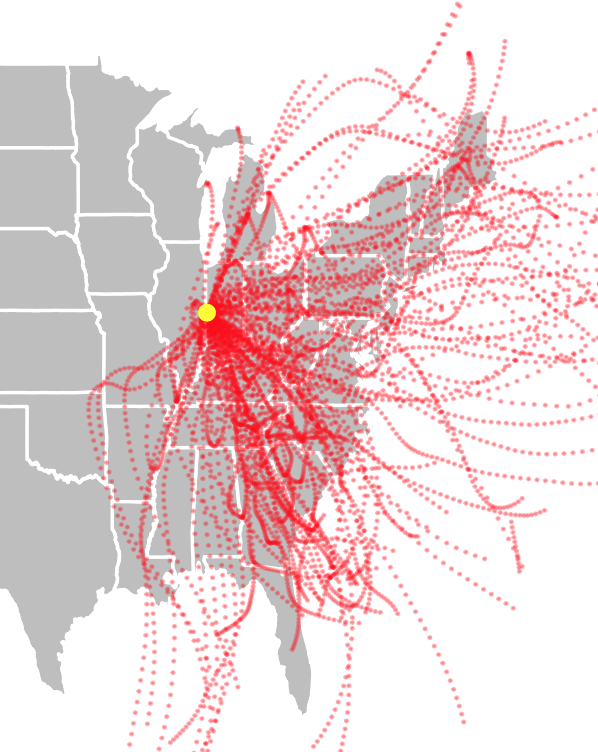} }}%
    \quad \quad \quad \quad \quad
    \subfloat[]{{\includegraphics[width=4.5cm]{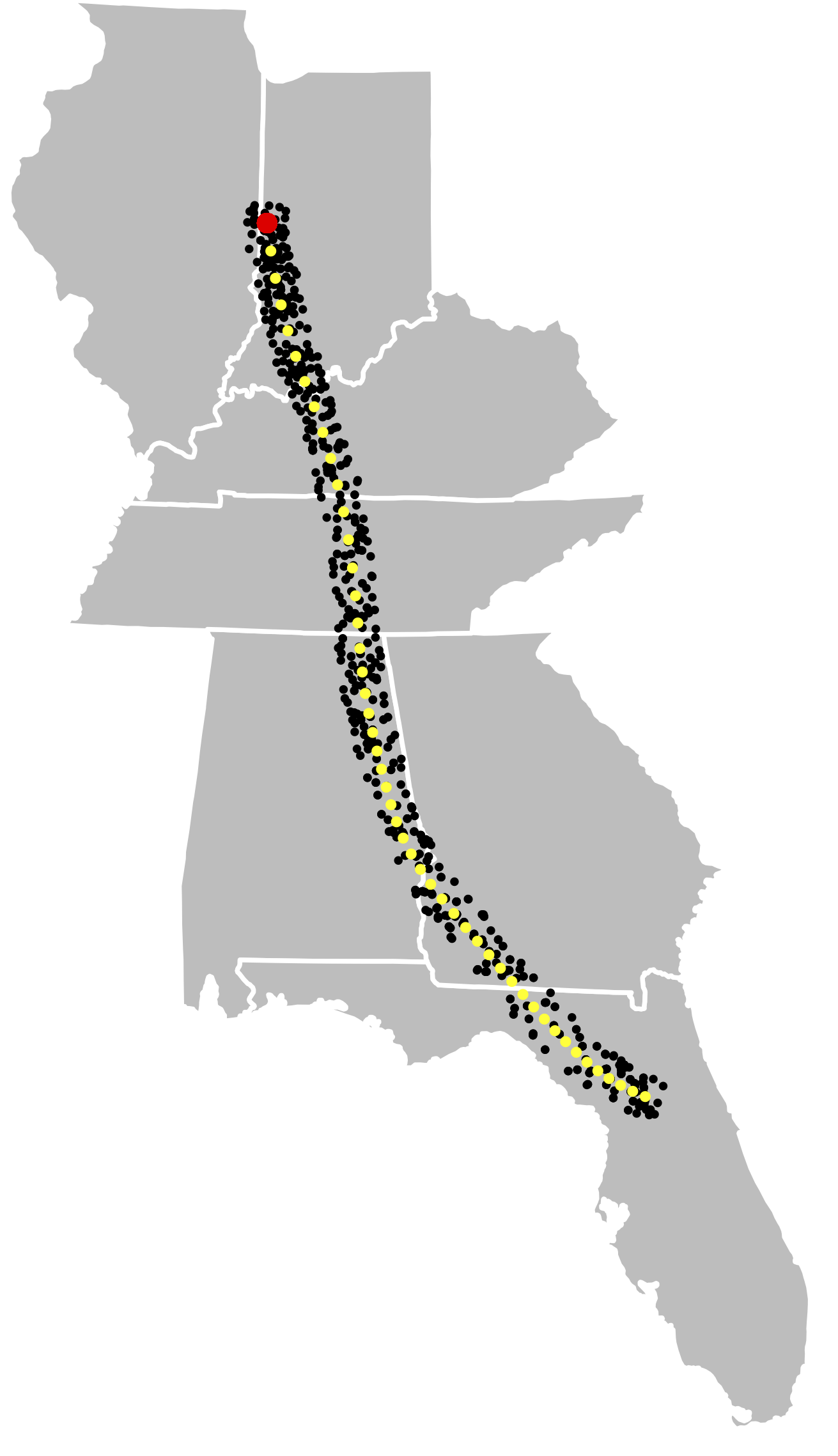} }}%
    \label{hysplit}%
    \caption{(a) All simulated trajectories from the power plant located in Indiana during Jan, 2003; (b) The centroids of all zip codes (black dots) within a 30km radius from the trajectory starting at 18:00 (EST) on Jan. 1st, 2003}
\end{figure}

Using the HYSPLIT model, we simulate forward trajectory paths of \SOTwo\, emissions from all coal-fired power plants four times every day over multiple years (2003-2007). In total, we simulate $4 \times 5\times 365$ trajectories for each power plant per year where we believe uncertainty around emission transportation is considered (Figure \ref{hysplit}a) . Then, we identify all ZIP codes on or near each trajectory path (Figure \ref{hysplit}b) and connect those ZIP codes to the power plant of the corresponding trajectory; that is, those ZIP codes are the locations affected by the power plant. For more details about the datasets and linkage, see the appendix.

\subsection{Exposure Levels}
Once the datasets are integrated, we define a level of exposure to \SOTwo\, emissions for each ZIP code location (note that \SOTwo\, emissions were originally recorded at the power plant locations). For each zip code location $C_i$, we extract monthly average \SOTwo\, emission for month $h$ ($E_{j,h}$) of power plant $j$ for $j \in \mathcal{J}_i$ where $\mathcal{J}_i$ is a set of power plants that are connected to the zip code location $C_i$ based on our linkage scheme (see Figure \ref{hysplit}). We count the number of times that air mass trajectories of power plant $j$ are on or near the zip code location $C_i$ and divide by the total number of the HYSPLIT trajectories of given month $h$  (4 $\times$ \# of days in month $h$). $W_{j,i,h}$ denotes this quantity which implies the weight of the linkage between $C_i$ and $j$ in month $h$. 
Then, the \SOTwo\, emission level of zip code location $C_i$ over multiple months $\mathcal{H}$ is defined as
\begin{equation}
\text{Level of Exposure to \SOTwo\, emission at } C_i = \sum_{j \in \mathcal{J}_i}\sum_{h \in \mathcal{H}} E_{j,h} \times W_{j,i,h}. \label{Level}
\end{equation}
Throughout the paper, we only take into account zip code locations within three EPA regions (Northeast, Southeast, Industrial Midwest) since there are few coal-fired power plants in the remaining regions and simulated trajectories from the power plants within the selected four regions barely cross the Rocky mountains. Also, among all power plants data available for years 1997-2016, we only consider the data from year 2003 since some of the power plants did not come into the air monitoring system during the earlier years, and the analysis continues for the next four years (2003-2007).

For the main analysis, we only consider the data during the warm season (June - October) each year since \SOTwo\, emissions are highly interactive to form secondary \PMTwo\, under high temperature conditions. Specifically, \SOTwo\, exposure levels are derived based on the June data, \PMTwo\, concentrations (mediator) are defined during June and July, and health outcomes are measured from June to October. 
A continuous \SOTwo\, exposure level is dichotomized at the median (12437) over all five years. Each continuous exposure level is not interpretable in an absolute sense (e.g., unit-less concentrations making hard to interpret in relation to other observed ambient air quality measurements), but to provide a relative quantity of how a given location is impacted by pollution emissions (\SOTwo) from the power plants. In fact, our choice of the median over all five years divide two modes of each year exposure distribution nicely. Additionally, we conduct an analysis based on a different dichotomization (the mean) to assess sensitivity .
\begin{figure}[h]
    \centering
    \subfloat[June 2003]{{\includegraphics[width=4cm]{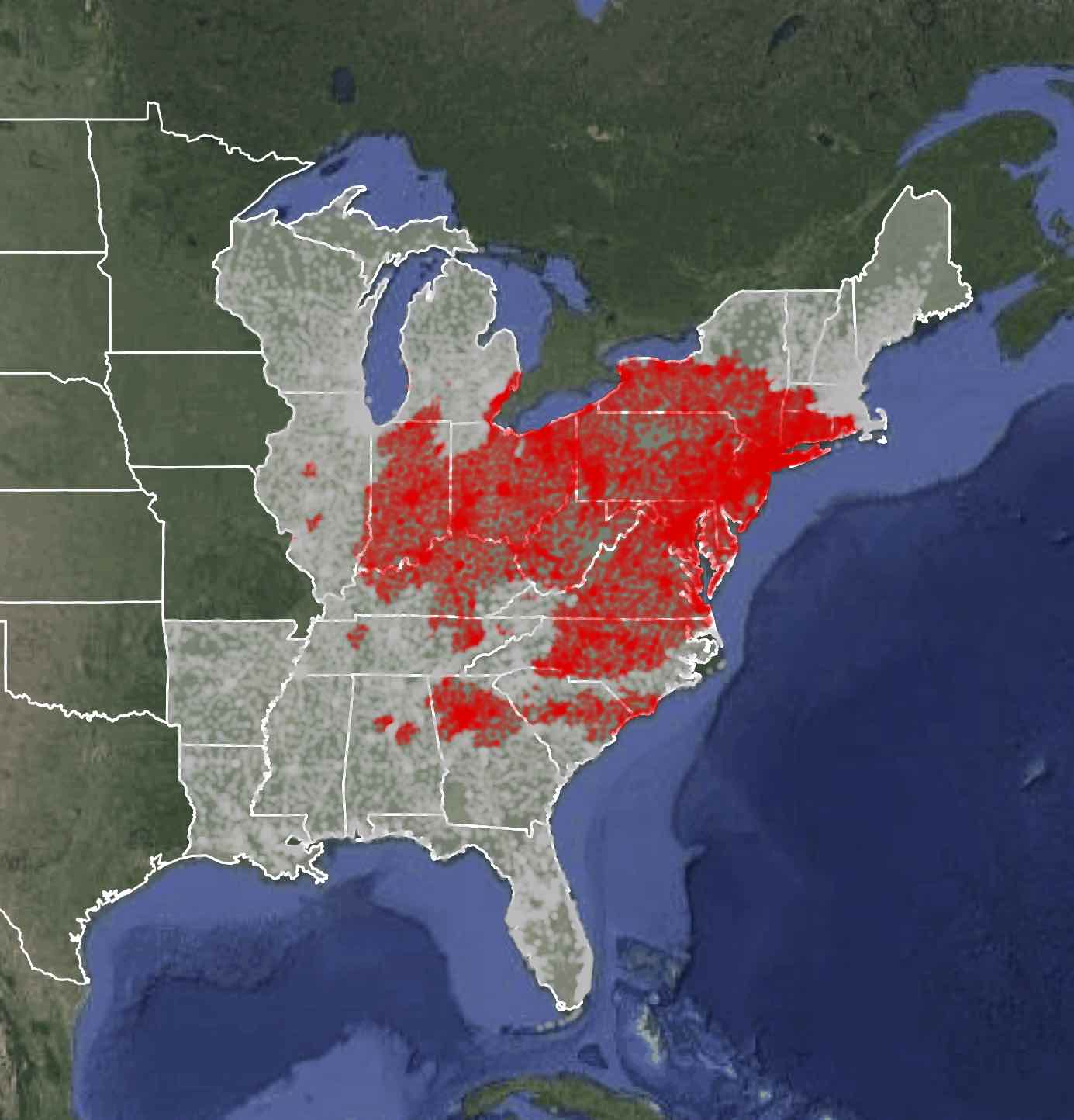} }}%
     \subfloat[June 2004]{{\includegraphics[width=4cm]{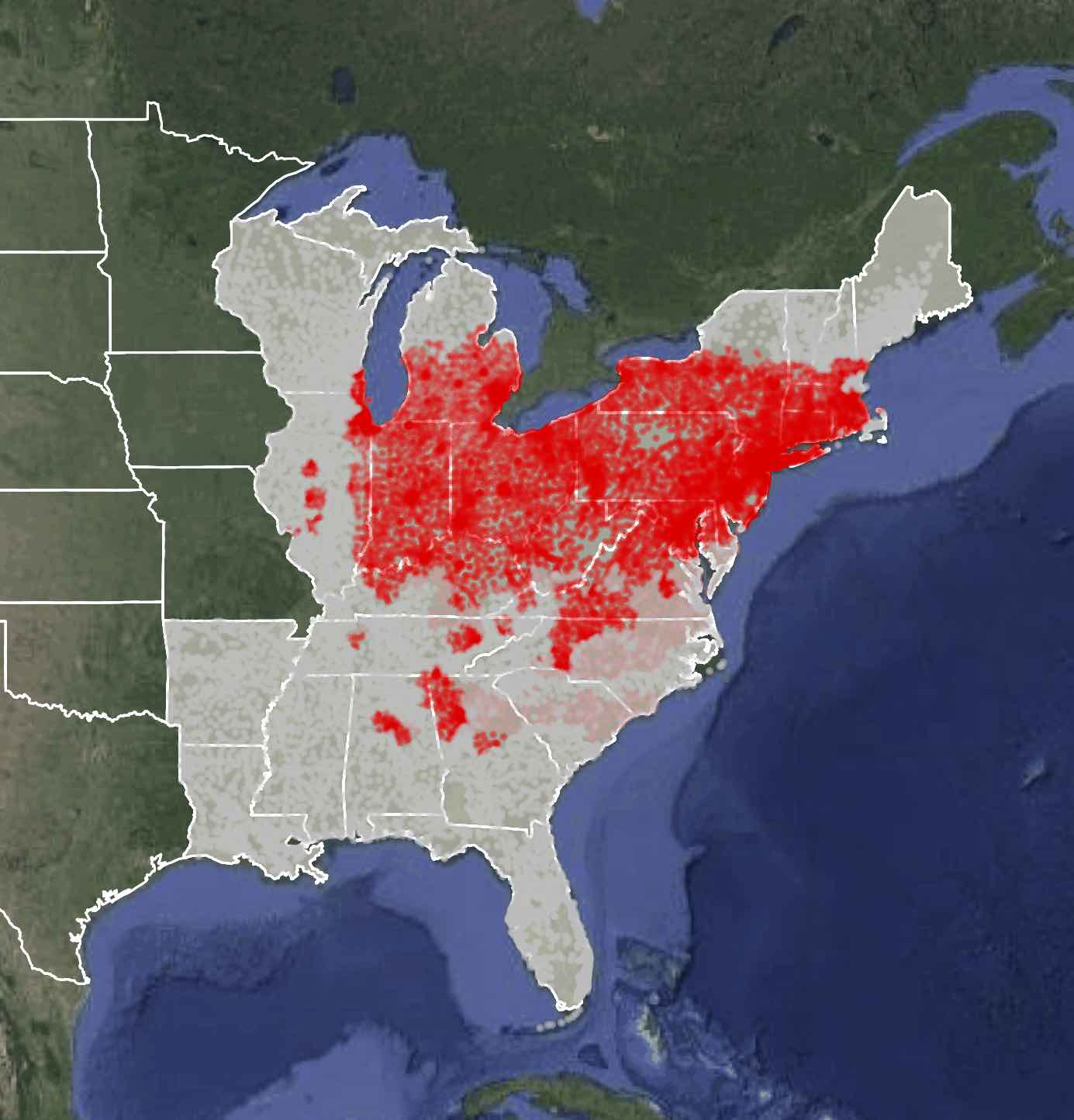} }}%
     \subfloat[June 2005]{{\includegraphics[width=4cm]{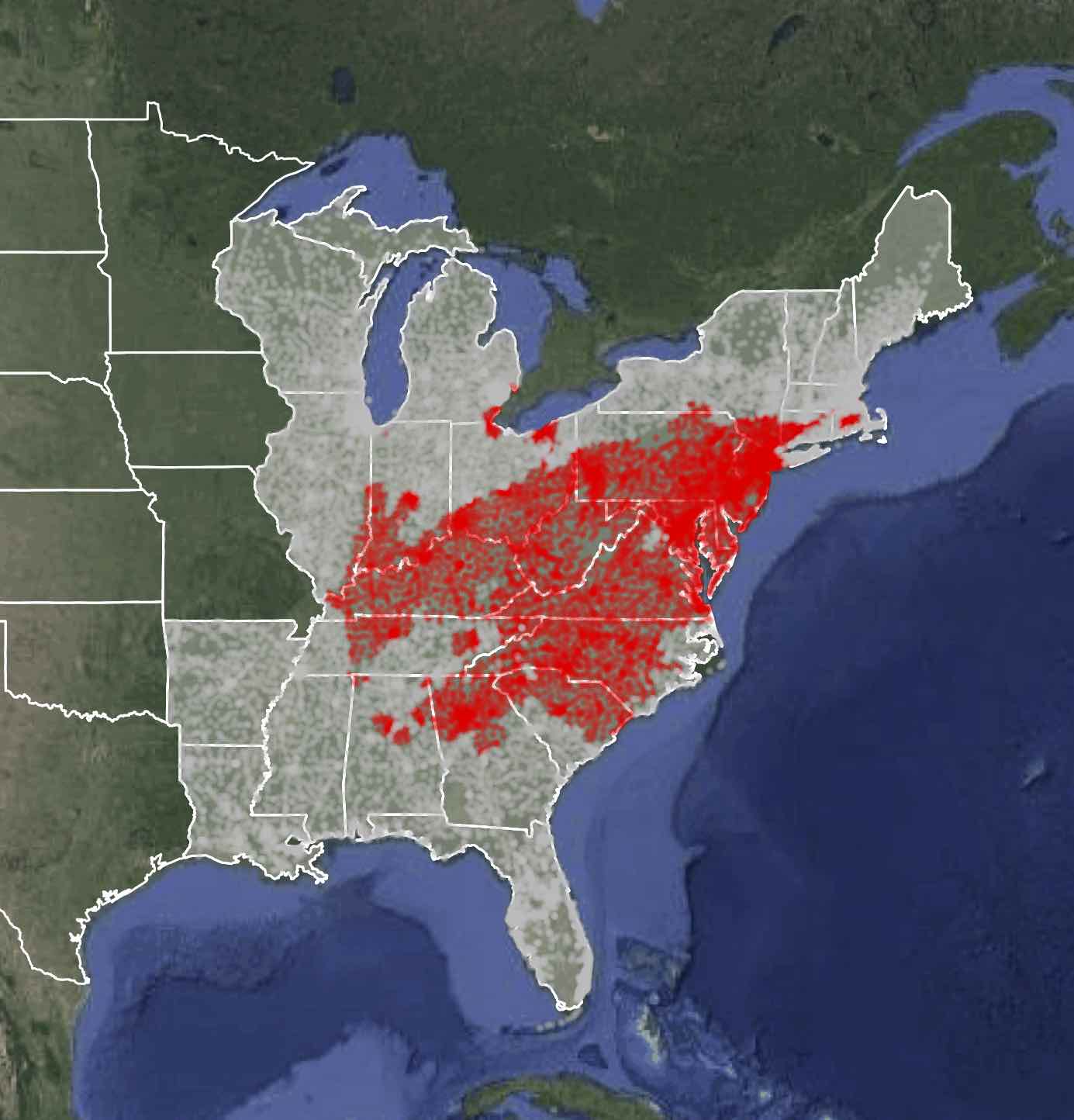} }}%
     
      \subfloat[June 2006]{{\includegraphics[width=4cm]{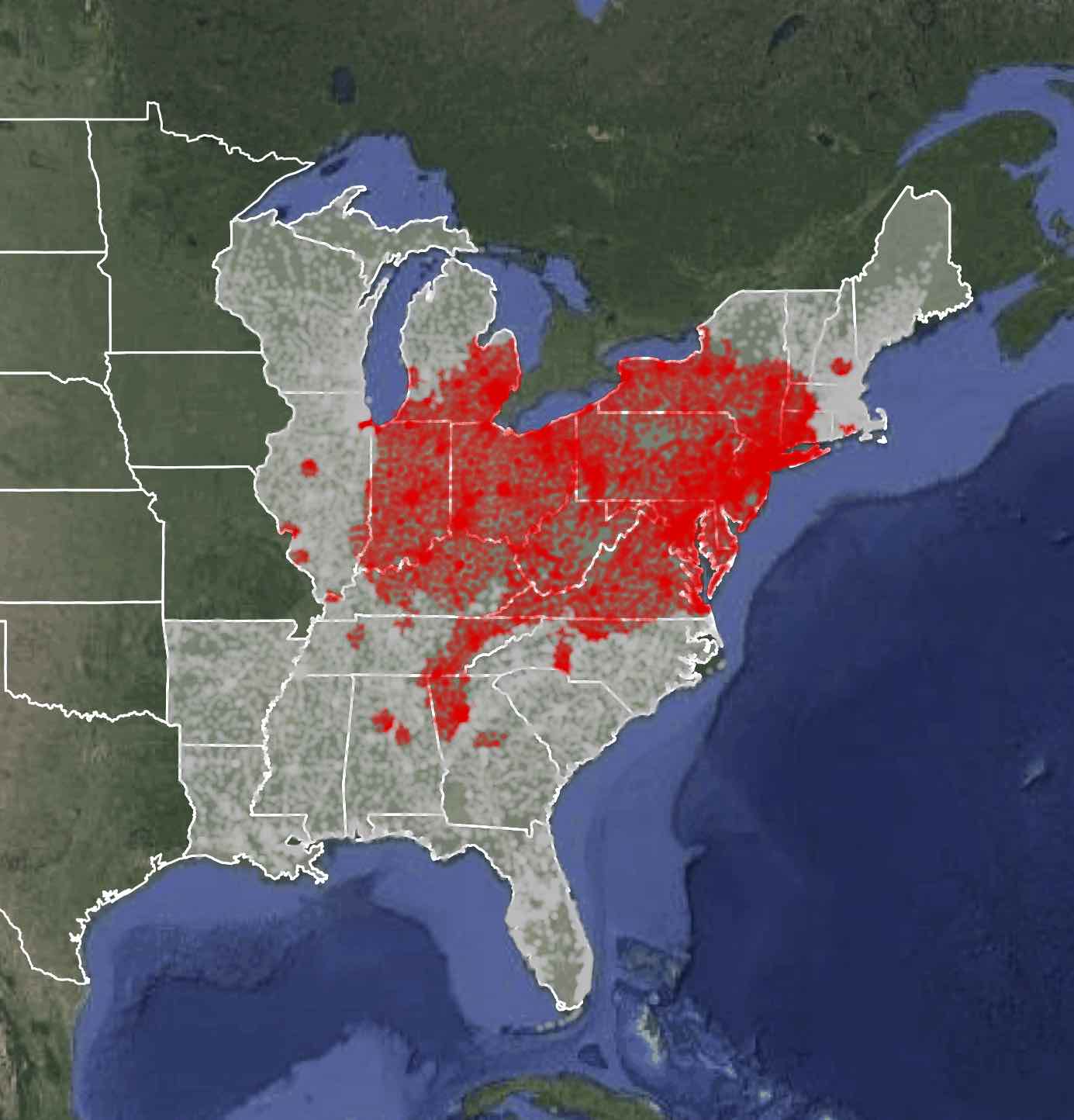} }}%
            \subfloat[June 2007]{{\includegraphics[width=4cm]{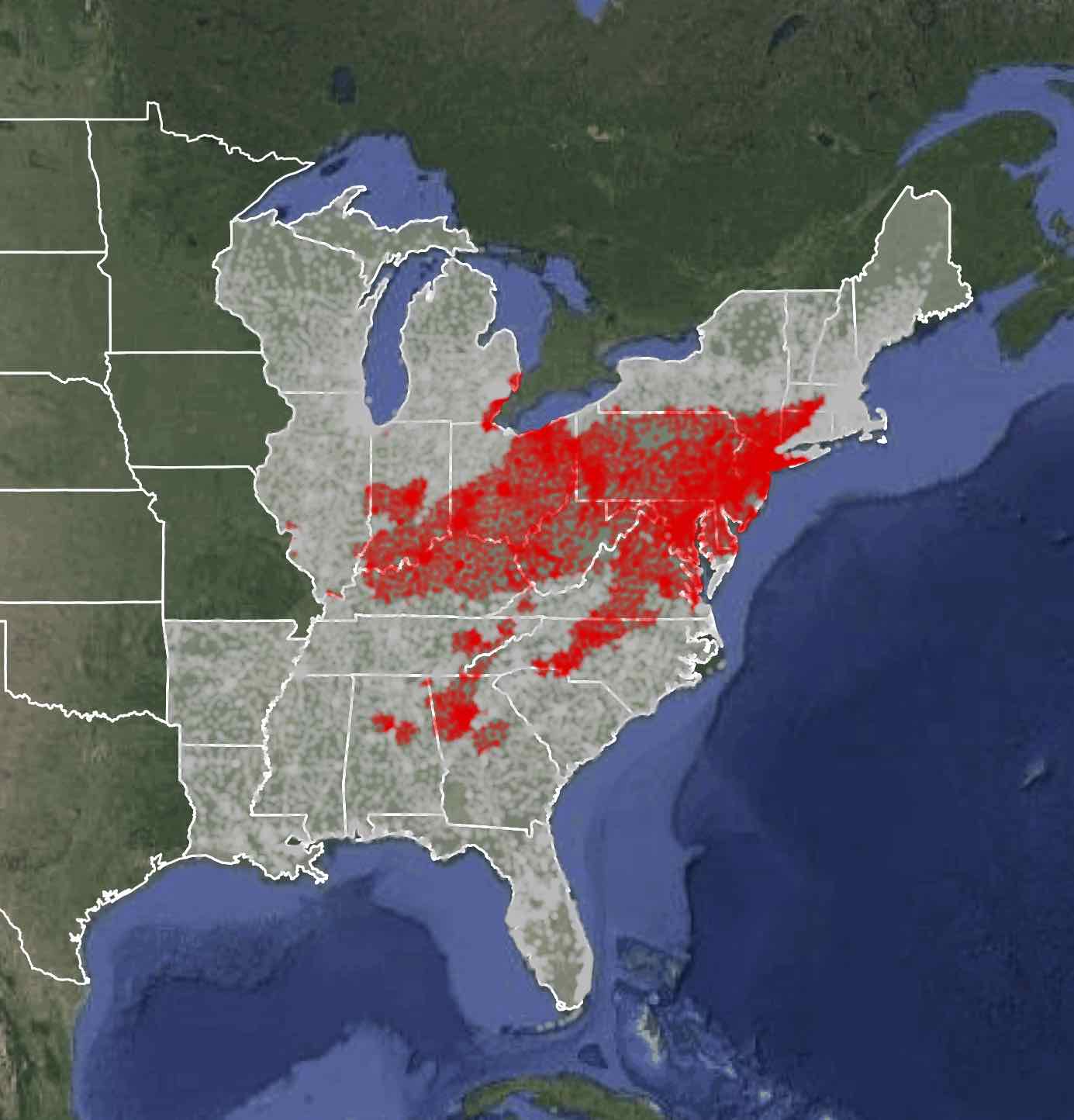} }}%
    \caption{Illustration of exposures to \SOTwo\,emissions at zip code locations changed by seasons. High emissions exposure (red) vs. Low emissions exposure (gray).}\label{fig2}
\end{figure}

Figure \ref{fig2} depicts locations of high (red) vs. low (gray) levels of exposure to \SOTwo\, emissions for every June from 2003 to 2007. As can be seen, exposure levels vary from year to year in many locations as do ambient \PMTwo\, concentrations. This motivates an analysis of the air pollution study  that considers time-varying exposures and \PMTwo\, concentrations.

\section{Notation}
\begin{figure}[h]
    \centering
\includegraphics[width=8cm]{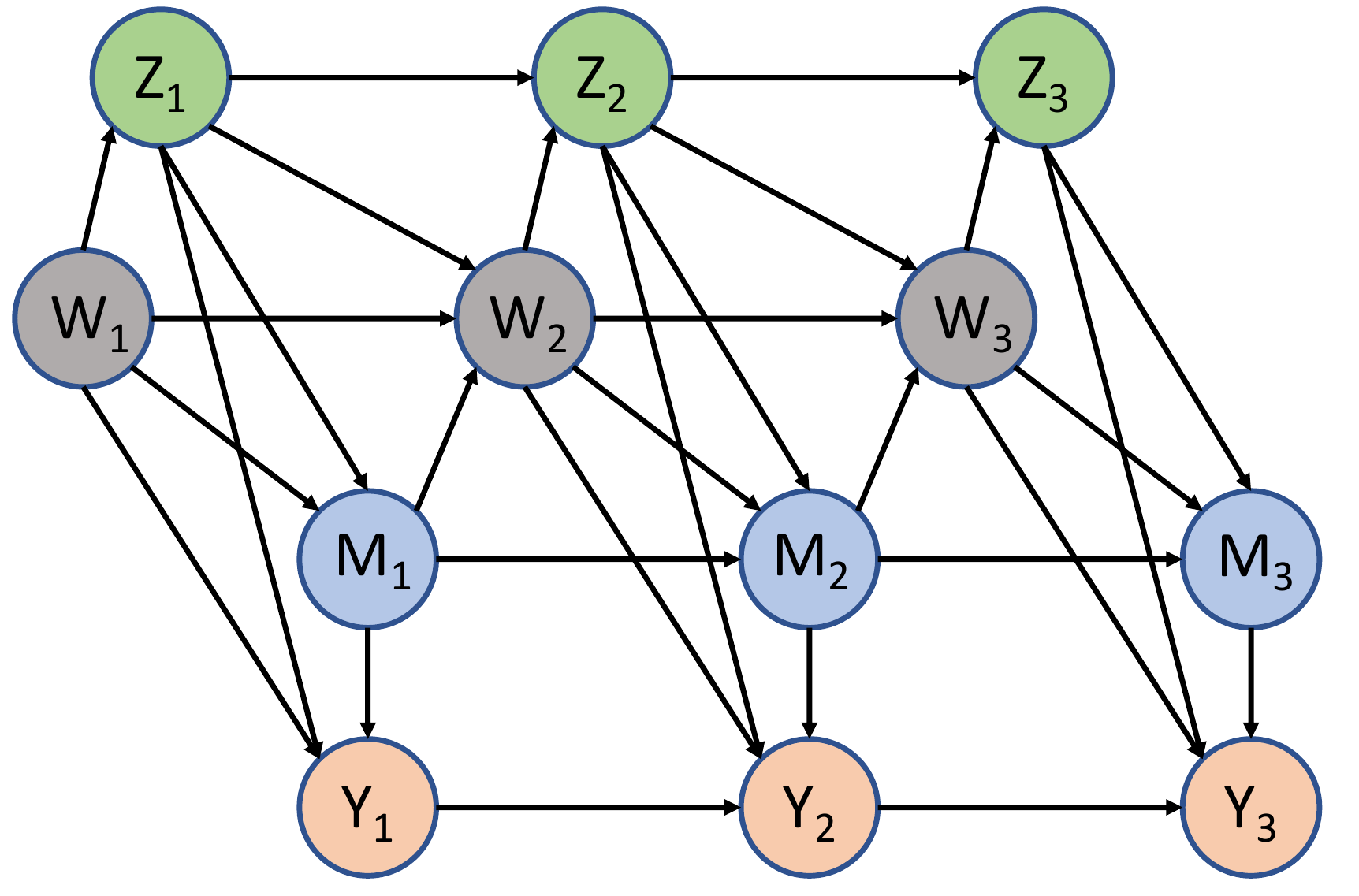}
    \caption{Directed acyclic graph (DAG) for $T=3$}    \label{fig}

\end{figure}

We start by defining notation. Let $Z(t), W(t), M(t)$ and $Y(t)$ denote the binary exposure, confounders, mediator and outcome at time $t$ for $t=1,\cdots, T$ with temporal ordering as given in Figure \ref{fig}. Initial baseline covariates, $W(1)$, include non-time varying baseline covariates. Let $\overline{Z}(T) = (Z(1), Z(2), \cdots, Z(T))$. Other variables follow the similar definition for histories up to time $T$. In our analysis, we consider one time-varying confounder, temperature, but, it is straightforward to include more than one by specifying a multivariate distribution. In the appendix, we introduce a strategy to include multiple spatially-correlated time-varying confounders.

In Figure \ref{fig}, we illustrate temporal ordering of the variables for time points 1,2, 3:
(1) $W(t)$ affects $Z(t), W(t+1), M(t), Y(t)$; (2) $M(t)$ affects $W(t+1), M(t+1), Y(t)$;
(3) $Y(t)$ affects $Y(t+1)$; and (4) $Z(t)$ affects $Z(t+1), W(t+1),M(t), Y(t)$. This ordering is tailored to our application since (1) weather conditions, especially temperature, affect power plants' operation strategies which in turn affect emissions exposure, promote/impede chemical reactions to form secondary \PMTwo\,  and directly impact health conditions; (2) ambient air pollution potentially affects weather conditions by climate change, but not emissions exposure; (3) individual hospitalizations do not directly affect \SOTwo\, emissions exposure, ambient \PMTwo levels, or weather conditions. 

\subsection{Potential Outcomes Framework}
We utilize the potential outcomes framework \citep{rubi:1974} which has been widely used to provide easy interpretation to causal inference. Let $M_{\overline{z}}(t)$ denote the potential value of $M(t)$ that would have been observed if $\overline{Z}(t)$ were set to $\overline{z}$ and $Y_{\overline{z},\overline{m}}(t)$ denote the potential value of $Y(t)$ that would have been observed if $\overline{Z}(t)$ were set to $\overline{z}$ and $\overline{M}(t)$ were set to $\overline{m}$. If we assume the consistency assumption \citep{vanderweele2016mediation} holds, then $M(t) = M_{\overline{z}}(t)$ given the observed $\overline{Z}(t) = \overline{z}$,  and $Y(t) = Y_{\overline{z},\overline{m}}(t)$ given the observed $\overline{Z}(t) = \overline{z}$ and $\overline{M}(t) = \overline{m}$. The history of the potential mediators up to time $t$ under exposures $\overline{Z}(t)=\overline{z}$ is defined as $\overline{M}_{\overline{z}}(t)$ and the history of the potential outcomes up to time $t$ under exposures $\overline{Z}(t)=\overline{z}$ and mediators $\overline{m}$ is defined as $\overline{Y}_{\overline{z},\overline{m}}(t)$. Potential outcomes of the form $Y_{\overline{z},\overline{M}_{\overline{z}^\prime}(t)}(t)$ for $\overline{z} \neq \overline{z}^\prime$ are \emph{a priori} counterfactuals (the potential value of the outcome at time $t$ that would have been observed under exposures $\overline{z}$ but the potential mediator values up to time $t$ set to what they would have been under exposures $\overline{z}^\prime$). However, this hypothetical situation is conceivable in our air pollution study when ambient \PMTwo\, concentrations are impacted by factors other than \SOTwo\, emissions from the coal-fired power plants. For example, ambient \PMTwo\, concentrations can remain at what it would have been observed under the low exposure level even if an actual \SOTwo\, emissions exposure level increases at a certain location, if the local authority starts imposing rules to reduce local \PMTwo\, sources (e.g., traffic related pollutant), which effectively controls the total ambient \PMTwo\, concentrations in that location. Thus, we postulate a situation that individual health outcomes are affected by high 
\SOTwo\, emissions exposure but the \PMTwo\, level is still set to what it would have been under low \SOTwo\, emissions exposure level, or vice versa.

A further assumption is made throughout the paper, which is referred to as the stable unit treatment value assumption (SUTVA; Rubin, 1980)\nocite{rubin1980randomization}; in the presence of a mediator,
this needs a more careful treatment \citep{mattei2011augmented}. Here, we assume that 1) there is no ``no interference'', in the sense that potential mediator and outcome values of one subject do not depend on the histories of the exposures and the mediators of the other subjects for every $t$ 
and 2) there are not ``multiple versions'' of the history of the exposures up to  $t$; whenever $\overline{z} = \overline{z}^\prime$, $\overline{M}_{\overline{z}}(t) = \overline{M}_{\overline{z}^\prime}(t)$ and $Y_{\overline{z}, \overline{M}_{\overline{z}}(t)}(t) = Y_{\overline{z}^\prime, \overline{M}_{\overline{z}^\prime}(t)}(t)$. Additionally, we assume ``no multiple versions'' of the mediators so that if $\overline{z} = \overline{z}^\prime$ and $\overline{m} = \overline{m}^\prime$, then $Y_{\overline{z}, \overline{m}}(t) = Y_{\overline{z}^\prime, \overline{m}^\prime}(t)$.

\section{Causal Estimands}
To evaluate the overall health impact of the level of exposure to  \SOTwo\, emissions through changes in \PMTwo\, concentrations, we need to acknowledge that the health effect resulting from changes in \PMTwo\, concentrations can vary over time.
In particular, we are interested in evaluating the causal effect of the air pollution exposure where a location switches from high to low level of exposure to \SOTwo\, emissions at time $t$; that is, the \SOTwo\, emissions exposure level from high ($z=0)$ to low ($z=1$) at time $t$, from which we can assess the health impact of the high air pollution exposure recursively (i.e., comparison of the health outcomes under $\overline{z} = \{0,0,0,\cdots, 0,1\}$ and $\overline{z}^\prime =\{0,0,0,\cdots, 0,0\}$) and examine how the impact changes over the time periods, which never been discussed in the literature. We primarily focus on the effect that is through any reduction in ambient \PMTwo\, concentrations: the natural indirect effect or the NIE \citep{Robi:Gree:1992,Pear:2001}. However, since \SOTwo\, also directly causes adverse health outcomes such as asthma \citep{guarnieri2014outdoor}, we expect a certain amount of the effect is directly related to \SOTwo\, emissions exposure: the natural direct effect or the NDE: 
\begin{eqnarray*}
NDE_t (\overline{z}, \overline{z}^\prime) &=& E[Y_{\overline{z} \overline{M}_{\overline{z}^\prime}(t)}(t) - Y_{\overline{z}^\prime \overline{M}_{\overline{z}^\prime}(t)}(t) ],\\
NIE_t (\overline{z}, \overline{z}^\prime) &=& E[Y_{\overline{z} \overline{M}_{\overline{z}}(t)}(t) - Y_{\overline{z} \overline{M}_{\overline{z}^\prime}(t)}(t) ],
\end{eqnarray*}
where $\overline{z} = \{0,0,0,\cdots, 0,1\}$ vs. $\overline{z}^\prime =\{0,0,0,\cdots, 0,0\}$. The sum of the natural direct and indirect effects is equal to the total effect at time $t$
\[
TE_t (\overline{z}, \overline{z}^\prime) = E[Y_{\overline{z} \overline{M}_{\overline{z}}(t)}(t) - Y_{\overline{z}^\prime \overline{M}_{\overline{z}^\prime}(t)}(t) ].\]
In addition to the contrasts of primary interest above, we can consider contrasts between any histories of exposures that alternate between high and low over the years:
\[\overline{Z}(t) = \{Z(1),Z(2),Z(3),\cdots, Z(t-1),Z(t)\} \text{ vs.  } \overline{Z}^\prime(t) = \{Z^\prime(1),Z^\prime(2),Z^\prime(3),\cdots, Z^\prime(t-1),Z^\prime(t)\}, \]
where $Z(j), Z^\prime(j) \in \{0,1\}^{\otimes 2}$ for $j=1, \cdots, t$. However, we only consider \emph{a priori} counterfactuals that have the form 
$Y_{\overline{z} \overline{M}_{\overline{z}^\dagger}(t)}(t)$
where $\overline{z} = \{z_1,_2,z_3,\cdots, z_{t-1},z_t\}$ and $\overline{z}^\dagger(t) = \{z_1,_2,z_3,\cdots, z_{t-1},1-z_t\}$; that is, the mediators are exposed to the same levels of \SOTwo\, emissions as the outcomes up to time $t-1$.

In what follows, we describe the flexible observed data models for the longitudinal outcomes, mediators and confounders, and provide a set of assumptions sufficient to identify the causal effects of interest.

\section{Models}
To specify a flexible and relatively parsimonious observed data model for multiple time points, we use a Bayesian dynamic model \citep{harrison1999bayesian} which does not require the model at time $t$ to contain the entire history (up to time $t$), but to depend only on the data just prior to the current observation given the vector of state parameters (through which the past information is carried forward). 
\subsection{Bayesian dynamic model}
\begin{figure}[h]
    \centering
    \subfloat[]{{\includegraphics[width=6cm]{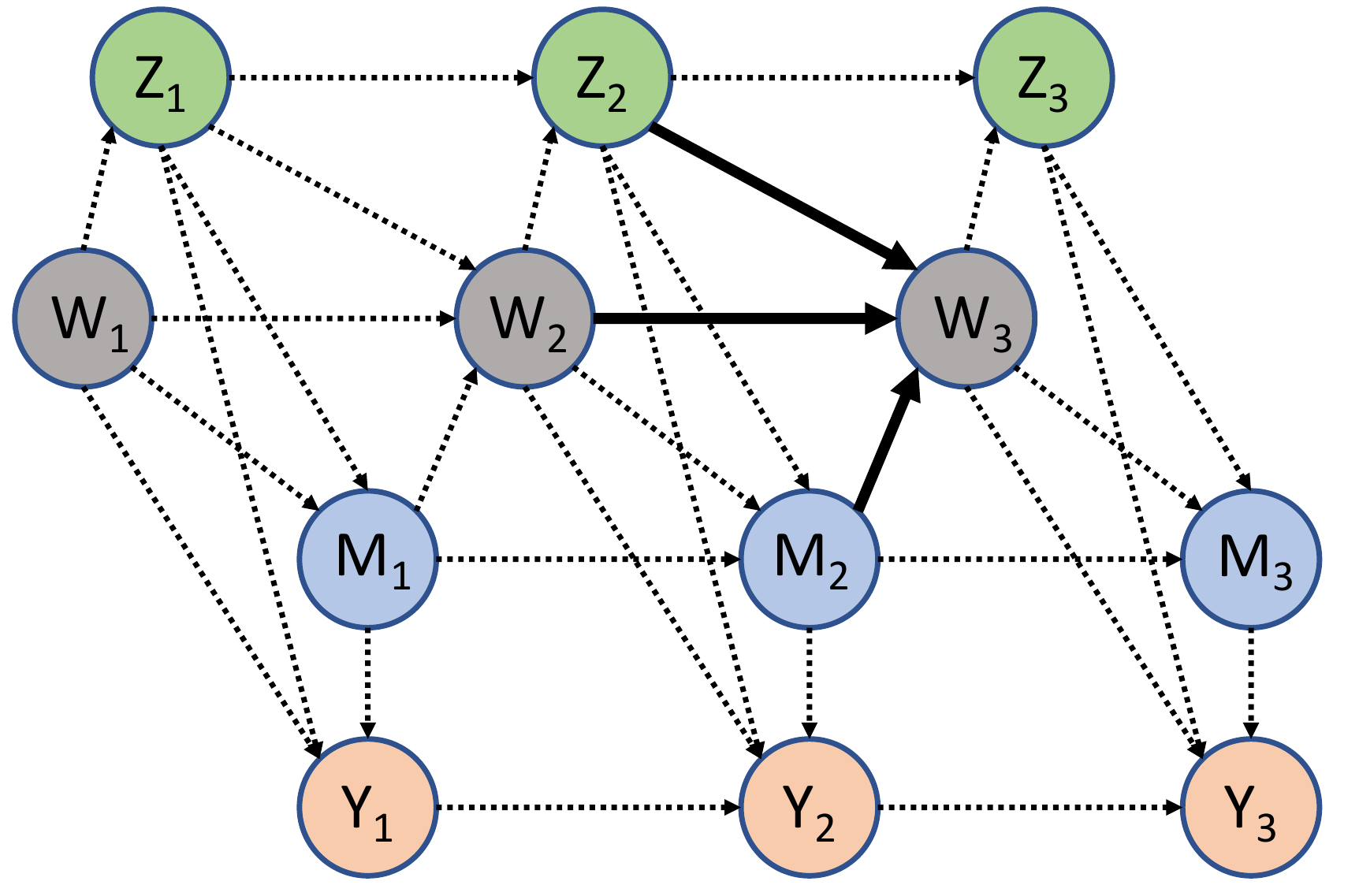} }}%
    \subfloat[]{{\includegraphics[width=6cm]{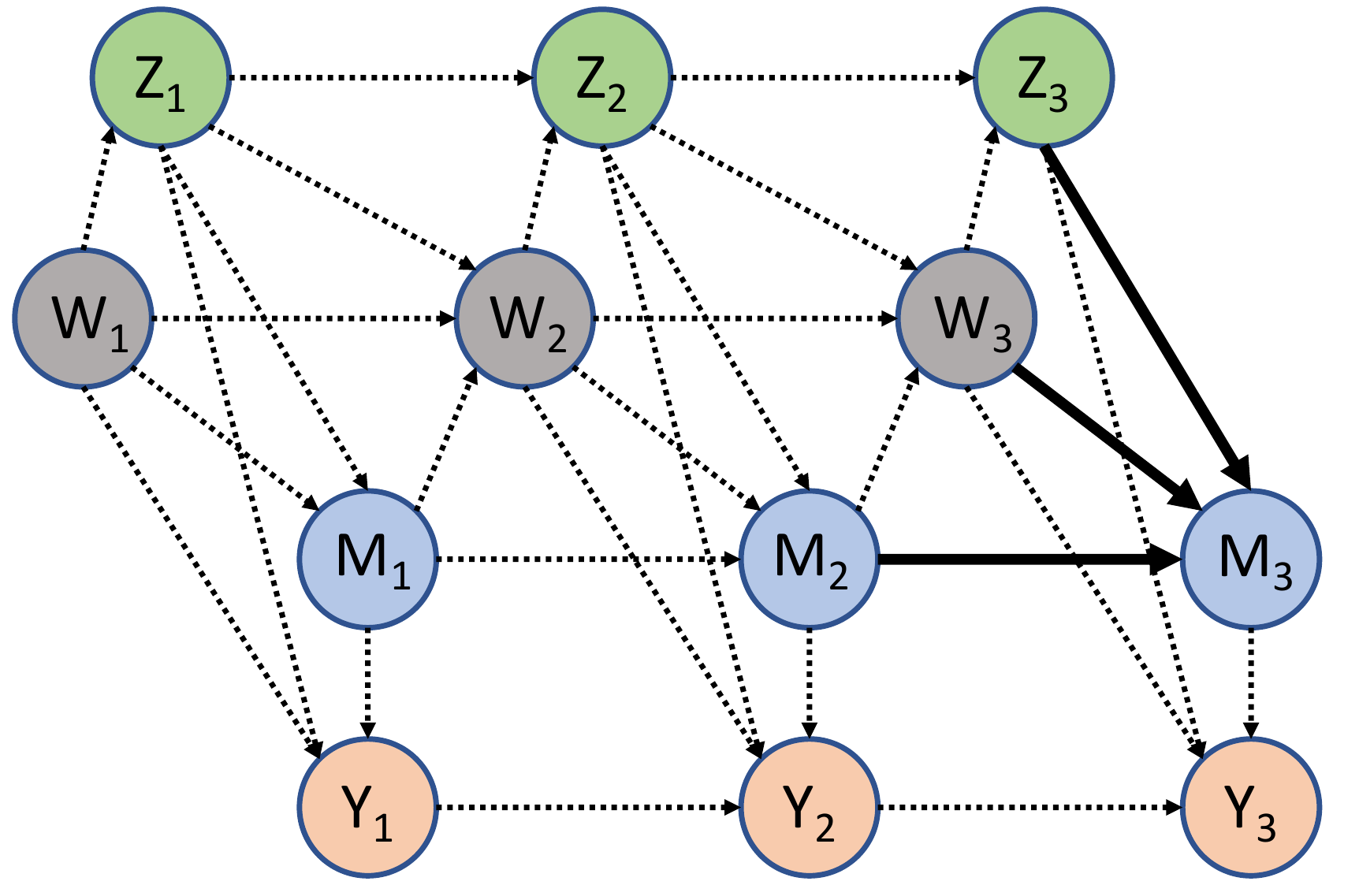} }}%
    \subfloat[]{{\includegraphics[width=6cm]{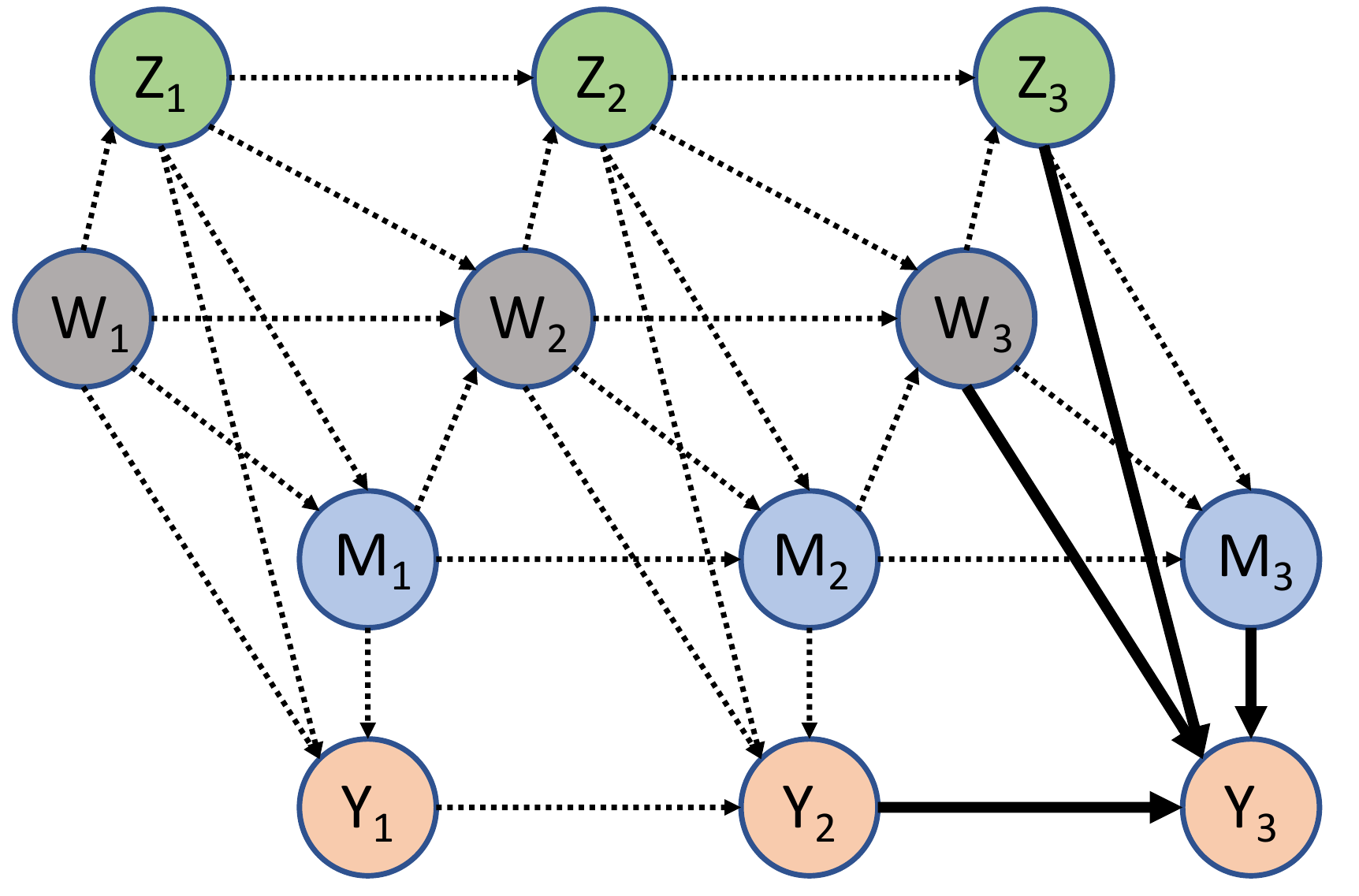} }}%
    \caption{Bayesian dynamic model for the `observed data' at $t=3$. Solid lines indicate effects from direct preceding observations for $W(3), M(3), Y(3)$ that are represented as state parameters $\boldsymbol{\theta}(3)$. Dashed lines are all previous and future relationships among observations that are independent on $W(3), M(3), Y(3)$ based on the Bayesian dynamic model assumption. All previous dashed lines (i.e., all previous relationships among observations; i.e., $\boldsymbol{\theta}(1)$ and $\boldsymbol{\theta}(2)$ ) are implicitly embedded in the current state parameters $\boldsymbol{\theta}(3)$ through the evolution model for $t=3$.}    \label{bdm}
\end{figure}

For each exposure level at time $t$, ${Z}(t)$, we assume that conditional on direct preceding observations (see Figure \ref{bdm}) and the vector of state parameters $\boldsymbol{\theta}(t)$, $Y(t)  (\text{or  } M(t) \text{  or  } W(t))$ is  independent of all future and past observations at time $s$ and $\boldsymbol{\theta}(s)$ for all $s\neq t$:
\begin{eqnarray}
\text{Observation Model} & : & Y(t) \sim p_o(Y(t)|{Z}(t), M(t), W(t), Y(t-1), \boldsymbol{\theta}(t)) \\  
&  & M(t) \sim p_o(M(t)|{Z}(t), M(t-1), W(t),  \boldsymbol{\theta}(t))\nonumber\\  
&  & W(t) \sim p_o(W(t)|{Z}(t), M(t-1), W(t-1),  \boldsymbol{\theta}(t))\nonumber\\  
\text{Evolution Model} & : & (\boldsymbol{\theta}(t) | \boldsymbol{\theta}(t-1)) \sim p_e(\boldsymbol{\theta}(t)|\boldsymbol{\theta}(t-1)),  
\end{eqnarray}
where the vector of state parameters $\boldsymbol{\theta}(t)$ evolves via the evolution model in (3). For example at $t=3$,  in Figure \ref{bdm} (c), the relationships between observed $Y(3)$ and its direct preceding observations $\{Z(3),M(3),W(3), Y(2)\}$ are represented as $\boldsymbol{\theta}(3)$ (solid lines), and given those direct preceding observations and $\boldsymbol{\theta}(3)$, the observed $Y(3)$ is independent of all other previous/future observations and relationships ($\boldsymbol{\theta}(1)$ and $\boldsymbol{\theta}(2)$). However, the current state parameters $\boldsymbol{\theta}(3)$ contain the past information of $\boldsymbol{\theta}(s)$ for $s=1,2$ via the evolution model. That is, we are updating information of the current relationship among observations through the state parameters $\boldsymbol{\theta}(t)$.

The Bayesian dynamic model \citep{harrison1999bayesian} provides several advantages over an observation model that depends on the full history, including allowing us to simplify the specification of assumptions described in Section 5. Also, it provides additional information from the past data for estimation of the current time point causal effects. Thus, even if the number in the treated (or control) group is decreasing over the course of the study such that there are not enough data to conduct inferences at the end, the estimation with the Bayesian dynamic model provides stability; for more details, see Section \ref{Sec:Simulation} The exact specification of observation and evolution models and details of the updating step are provided in Section 5.2 and Section 7, respectively.

\subsection{Observed Data Model}\label{sec:obsmodel}
In this sub-section, we specify the modeling strategy aligned with the Bayesian dynamic model assumption (see the DAG in Figure \ref{bdm}). We try to minimize parametric assumption in the observation models by specifying Dirichlet process priors \citep{maceachern1994estimating,escobar1994estimating,maceachern1998estimating} on the distributions of some parameters. 

\subsubsection{Observation Model}
For each exposure level at time $t$, ${Z}_i(t)=z$,  the observation model for the mediator is specified as
\[M_i(t) \,|\, {Z}_i(t)=z, M_i(t-1), W_i(t),\boldsymbol{\theta}(t) \sim N(\alpha_{t0,i}^{z} + \alpha_{t1,i}^{z} M(t) + \alpha_{t2,i}^{z} W(t), \sigma_{M,i}^z),\]
for $t=1, \cdots, T$, where we place Dirichlet process priors on the distributions of all coefficients and variance components, $(\alpha_{t0,i}^{z},\alpha_{t1,i}^{z},\alpha_{t2,i}^{z},\sigma_{M,i}^z)$ for $i=1, \cdots, n^z$. Here, the coefficient of $W_i(t-1)$ can be either a scalar or a vector depending on the number of covariates in the model. For $Y_i(0), M_i(0)$ and $W_i(0)$, we use their baseline values. 

For the observation model for the outcome (respiratory disease related hospitalizations), 
 we specify a log linear model with Dirichlet process priors on the distributions of all coefficient parameters, ($\beta_{t0,i}^{z},\beta_{t1,i}^{z},\beta_{t2,i}^{z},\beta_{t3,i}^{z}$) for $i=1, \cdots, n^z$, which induces a Dirichlet process mixture of generalized linear models \citep{hannah2011dirichlet}
\begin{eqnarray*}
\lefteqn{\log(E[Y_i(t) \,|\, {Z}_i(t)=z, M_i(t), W_i(t), Y_i(t-1),\boldsymbol{\theta}(t)])}\\
&=& \log(\text{person-years})+ \beta_{t0,i}^{z} + \beta_{t1,i}^{z}M(t) +  \beta_{t2,i}^{z} W(t)+  \beta_{t3,i}^{z} Y(t),
\end{eqnarray*}
where `person-years' is the offset. 

For the time-varying covariate $W_i(t)$, we specify the same DPM model that is used in the mediator model, but conditioning on $Z_i(t)$ and $M_i(t)$:
\[W_i(t) \,|\, Z_i(t)=z, M_i(t-1), W_i(t-1), \boldsymbol{\theta}(t) \sim N(\gamma_{t0,i}^{z}  + \gamma_{t1,i}^{z} M_i(t-1) + \gamma_{t2,i}^{z} W_i(t-1), \sigma_{W_i}^{z}).\] If multiple time-varying covariates are considered, we can specify a multivariate regression model with Dirichlet process priors on all coefficients. 


\subsubsection{Dirichlet Processes Priors}
We place Dirichlet process priors on the distributions of coefficient parameters and variance parameters of the models from the previous section. For instance, at each time $t$ and exposure level $z$,
\begin{eqnarray*}
(\boldsymbol{\alpha}_{t,i}^{z},\sigma_{M,i}^z)  & \sim & F_t^{z} \qquad \text{ for } \,\, i=1,\cdots,n\\
F_t^{z} & \sim & DP (\lambda_t^{z}, \mathcal{F}_t^{z})\\
\mathcal{F}_t^{z} & = & \prod^{\mathbf{card}(\boldsymbol{\alpha}_{t,i}^{z})}_{h=1} N\left(\mathcal{A}_{t,h}^{z}, \frac{1}{\tau_{t,h}^{z}}\right)  \times Inv.Gamma(a,b)
\end{eqnarray*}
where $\boldsymbol{\alpha}_{t,i}^{z} = \{\alpha_{t0,i}^{z},\alpha_{t1,i}^{z},\alpha_{t3,i}^{z} \}$ and $DP$ denotes the Dirichlet process with two parameters, a mass parameter ($\lambda_t^{z}$) and a base measure ($\mathcal{F}_t^{z}$). 
Here, $\mathcal{A}_{t,h}^{z}$ and $\tau_{t,h}^{z}$ denote the mean and the precision parameters of the $h$-th base distribution (i.e., the base distribution for the $h$-th element of $\boldsymbol{\alpha}_{t,i}^{z}$). We update the mean parameters of the base measure for all $h$ (that is, $\boldsymbol{\theta}(t)=\boldsymbol{\mathcal{A}}_{t}^{z} = \{\mathcal{A}_{t,1}^{z}, \mathcal{A}_{t,2}^{z}, \cdots, \mathcal{A}^{z}_{t, \mathbf{card}(\boldsymbol{\alpha}_{t,i}^{z})}\}$ for the mediator model) via the evolution model in (3). Similarly, we update the mean parameters of the base measure  for the covariate and the outcome models. Note that the outcome model does not have any variance components.

\subsubsection{Evolution Model}\label{Evolution}
The mean parameters of the base measure for the mediator model, ($\mathcal{A}_{t,h}^{z}$), are updated via the following evolution model
\[\boldsymbol{\mathcal{A}}^z_{t}|\boldsymbol{\mathcal{A}}^z_{t-1} \sim MVN(\boldsymbol{\mathcal{A}}^z_{t-1}, \Sigma^z_{\mathcal{A}}),\]
where $MVN()$ denotes a multivariate normal distribution with dimension $\mathbf{card}(\boldsymbol{\alpha}_{t,i}^{z})$. Note that we only update the mean parameters of the base measure.

Based on these specifications, the mean of each parameter at time $t$ is the corresponding parameter at time $t-1$, so that the past information are directly carried forward to the present  parameters. Further details on the priors is given in Section \ref{Sec:Post} 

We specify similar evolution models for the mean parameters of the base measures for the outcome and covariates models.

\section{Identification}

\subsection{Identifying Assumptions}
We start by assuming that, conditional on the past observed data, the treatments, and $\overline{\boldsymbol{\theta}}(t)$ (all previous and current pathways information in the observed data; see Figure \ref{fig}), assignment to high (low) \SOTwo\, exposure status at time $t$ is independent of the potential outcome and mediator at time $t$:

\noindent{\bf Assumption 1} ({\bf Ignorability of Treatments}) For every $t = 1, \cdots, T$,
\[\{Y_{\overline{Z}, \overline{M}_{\overline{Z}}}(t), M_{\overline{Z}}(t)\} \perp Z(t) \,|\, \overline{Y}(t-1), \overline{M}(t-1),\overline{W}(t-1), \overline{Z}(t-1),\overline{\boldsymbol{\theta}}(t). \] 
This assumption states that the exposure level at time $t$, $Z(t)$, is randomized given all past observations and allows us to estimate the distribution of potential outcomes (and mediators) given treatment history, $\overline{Z}(t-1)$, from the observed data. In our application, this assumption is plausible if we adjust for  all confounding relationships between the exposure level and health outcomes, and between the exposure level and ambient air quality at each time $t$.

\bigskip

\noindent{\bf Assumption 2}  For each time $t$, define $\overline{\mathcal{Z}} = \{\overline{Z}(t-1), Z(t)\}$ and $\overline{\mathcal{Z}}^\prime = \{\overline{Z}(t-1), 1-Z(t)\}$. Then, we assume 
\begin{eqnarray*}
Y_{\overline{\mathcal{Z}}, \overline{M}_{\overline{\mathcal{Z}}}(t)}(t) &\independent& M_{\overline{\mathcal{Z}}^\prime}(t) \,|\, M_{\overline{Z}}(t), \overline{M}(t-1), \overline{Y}(t-1), \overline{W}(t-1), \overline{Z}(t-1),\overline{\boldsymbol{\theta}}(t),\\
Y_{\overline{\mathcal{Z}}, \overline{M}_{\overline{\mathcal{Z}}^\prime}(t)}(t) &\independent& M_{\overline{\mathcal{Z}}}(t) \,|\, M_{\overline{Z}^\prime}(t), \overline{M}(t-1), \overline{Y}(t-1), \overline{W}(t-1), \overline{Z}(t-1),\overline{\boldsymbol{\theta}}(t).
\end{eqnarray*}
This is an extended version of the homogeneity and generalized weak principal ignorability assumptions introduced in \cite{forastiere2015identification,forastiere2017}. The first part of the assumption states that  the conditional distribution of the health outcome exposed to a certain levels of \SOTwo\, emissions up to time $t$, $\overline{\mathcal{Z}}$, is the same regardless of the potential value of ambient \PMTwo\, concentration that would have been observed under the opposite exposure status at time $t$, $\overline{\mathcal{Z}}^\prime$, given the potential value of \PMTwo\, under $Z(t)$ and the other past observations. The second part assumes the conditional distribution of the health outcomes exposed to a certain levels of \SOTwo\, emissions up to time $t$, $\overline{\mathcal{Z}}$, but with ambient \PMTwo\, concentration at time $t$ set to what it would have been under the opposite exposure level at time $t$, $\overline{\mathcal{Z}}^\prime$, is the same regardless of the potential value of ambient \PMTwo\, concentration under the exposures $\overline{\mathcal{Z}}$. Alternatively, we can express this as follows
\begin{eqnarray}
\lefteqn{f(Y_{\overline{Z}, \overline{M}_{\overline{Z}^\prime}(t)}(t)\,|\, M_{\overline{Z}^\prime}(t)=m, M_{\overline{Z}}(t), \text{Remainder}, \overline{\boldsymbol{\theta}}(t)) } \nonumber\\
 & = &  f(Y_{\overline{Z}, \overline{M}_{\overline{Z}}(t)}(t) \,|\,M_{\overline{Z}^\prime}(t), M_{\overline{Z}}(t)=m, \text{Remainder}, \overline{\boldsymbol{\theta}}(t)) \label{A2}
\end{eqnarray}
where $\text{Remainder} = (\overline{M}(t-1)=\overline{m}(t-1), \overline{Y}(t-1)=\overline{y}(t-1), \overline{W}(t-1)=\overline{w}(t-1), \overline{Z}(t-1)=\overline{z}(t-1))$. 
The key point is that the health outcome with the potential value of \PMTwo\, concentration level set to a value, $m$, at time $t$ is assumed to have the same distribution as that of the health outcome that had, in reality, \PMTwo\, concentration level $m$ observed, regardless of an observed level of exposure to \SOTwo\, emission. 
As mentioned above, this assumption has close ties to the assumption of homogeneity across principal strata in Forastiere et al. [2016, 2017]; for example, it implies $Y_{\overline{Z}, \overline{M}_{\overline{Z}}(t)}$ is homogeneous across all principal strata with $M_{\overline{Z}}=m$ regardless of the value of $M_{\overline{Z}^\prime}$. This connection aids interpretation and justification of Assumption 2 in the context of the air pollution study since the potential health outcome value of a certain location is related to the ambient air quality value tied to that location only, and not to whether the value is induced under $\overline{\mathcal{Z}}^\prime$ or $\overline{\mathcal{Z}}$.
Note that Eq. (\ref{A2}) is an extension of one of main identifying assumptions that has also appeared in the literature \citep{Dani:Roy:Kim:Hoga:Perr:2012,Kim:2016,kim2016framework}. In Section \ref{sec:sen}, we assess sensitivity to violation of Assumption 2 based on a similar sensitivity analysis technique introduced in \cite{kim2016framework,Kim:2016}.

\subsection{Identification of the Effects}
With the Bayesian dynamic model  and the identifying assumptions in Section 5.1, we can identify the posterior distributions of the potential outcomes and the corresponding posterior means. The conditional posterior mean of the \emph{a priori} potential outcome can be identified using the following:
\begin{eqnarray}
\lefteqn{E[Y_{\overline{z} \overline{M}_{\overline{z}^\prime}(t)}(t) | \overline{\boldsymbol{\theta}}(t)] }\nonumber\\
& = & \int E[Y(t)|M(t)=m_t, Y(t-1)=y_{t-1}, W(t)=w_t, V=v, Z(t)=z_t, \theta(t)]\nonumber\\
&  & \,\times\, p(M(t)=m_t | M(t-1)=m_{t-1}, W(t)=w_t, V=v, Z(t)=z^\prime_t, \theta(t))\label{med}\\
& & \,\times\, \prod_{l=1}^{t-1} p(M(l)=m_l | M(l-1)=m_{l-1}, W(l)=w_l, V=v, Z(l)=z_l, \theta(l))\nonumber\\
& & \,\times\, \prod_{k=2}^{t} p(W(k)=w_k | M(k-1)=m_{k-1}, W(k-1)=w_{k-1}, V=v, Z(k-1)=z_{k-1}, \theta(l))\nonumber\\
& & \,\times\, \prod_{j=1}^{t-1} p(Y(j)=y_{j} |M(j)=m_j, Y(j)=y_{j}, W(j)=W_j, V=v, Z(j)=z_j, \theta(j))\nonumber\\
& & \,\times\, p(Y(0)=y_0, M(0)=m_0, W(1) = w_1, V=v, \theta(0)) \,d y_{t-1} \cdots \,d y_0 \,d m_t \cdots \,d m_0 \,dw_t \cdots \,dw_1 \,dv,\nonumber
\end{eqnarray}
where $\overline{z} = \{z_1, \cdots, z_{t-1}, z_t\}$ and $\overline{z}^\prime = \{z_1, \cdots, z_{t-1},z_t^\prime\}$.
The proof is in the appendix. The natural direct and indirect effects at each time $t$ are computed by integrating with respect to the posterior distribution of $\overline{\boldsymbol{\theta}}(t)$, 
$E_{\overline{\boldsymbol{\theta}}(t)}\left[E[Y_{\overline{z} \overline{M}_{\overline{z}}(t)}(t) | \overline{\boldsymbol{\theta}}(t)] - E[Y_{\overline{z} \overline{M}_{\overline{z}^\prime}(t)}(t) | \overline{\boldsymbol{\theta}}(t)]\right]$ and $E_{\overline{\boldsymbol{\theta}}(t)}\left[E[Y_{\overline{z} \overline{M}_{\overline{z}^\prime}(t)}(t) | \overline{\boldsymbol{\theta}}(t)] - E[Y_{\overline{z}^\prime \overline{M}_{\overline{z}^\prime}(t)}(t) | \overline{\boldsymbol{\theta}}(t)]\right]$, respectively. Estimation of $E[Y_{\overline{z} \overline{M}_{\overline{z}}(t)}(t) | \overline{\boldsymbol{\theta}}(t)]$ for any  $\overline{z}$ (i.e., observables counterfactuals) is straightforward by replacing $z_t^\prime$ in Eq. ($\ref{med}$) with $z_t$.

\section{Posterior Computations}\label{Sec:Post}
Define $\overline{D}(t-1) = \{\overline{Z}(t-1), \overline{Y}(t-1),\overline{M}(t-1), \overline{W}(t-1)\}$. Historical information is included in the posterior distribution of the state parameter $\boldsymbol{\theta}_{t-1}$, $p(\boldsymbol{\theta}(t-1)\,|\,\overline{D}(t-1))$, prior to the evolution via (2). In the posterior computation, the evolution step is
\begin{equation}
p(\boldsymbol{\theta}(t)\,|\,\overline{D}(t-1)) = \int p_e(\boldsymbol{\theta}(t)\,|\, \boldsymbol{\theta}(t-1))p(\boldsymbol{\theta}(t-1)\,|\,\overline{D}(t-1)) \, d\boldsymbol{\theta}(t-1), \label{evol}
\end{equation}
where $p_e()$ denotes the evolution model. After observing current information $D(t) = \{{Z}(t), {Y}(t),{M}(t),{W}(t))$, the posterior for time $t$ is obtained via the updating step
\begin{eqnarray}
p(\boldsymbol{\theta}(t) \,|\, \overline{D}(t)) &\propto& p(\boldsymbol{\theta}(t) \,|\, \overline{D}(t-1))p_o(D(t)\,|, \overline{D}(t-1), \boldsymbol{\theta}(t))\nonumber\\
& = & p(\boldsymbol{\theta}(t) \,|\, \overline{D}(t-1))p_o(D(t)\,|, {D}(t-1), \boldsymbol{\theta}(t)) \label{eq2},
\end{eqnarray}
where $p_o()$ denotes the observation model and the last equality follows from the Bayesian dynamic model assumption in (2).
Note that this updating proceeds sequentially over time and only $p(\boldsymbol{\theta}_{t-1}\,|\,\overline{D}_{t-1})$ contains the information from the past that is currently (at time $t-1$) available for further analysis. Since the process is repeated as time progresses, it is desirable for efficient computation that the form of the posterior after the updating step, $p(\boldsymbol{\theta}(t)\,|\,\overline{D}(t))$,  is the same as the form of the input, $p(\boldsymbol{\theta}(t-1)\,|\,\overline{D}(t-1))$, which is in general no the case; otherwise, we need to code a separate posterior computation step for each time point. Thus, instead of using the exact form of the posterior distribution at each time $t$, we only use posterior samples from time $t-1$ to approximate prior density at time $t$. Assume that $p(\boldsymbol{\theta}(t-1)|\overline{D}(t-1))$ has been saved through $\Lambda(t-1) = \{n_{t-1}, \Theta(t-1)\}$ where $n_{t-1}$ is the Monte Carlo sample size and $\Theta(t-1) = \{\boldsymbol{\theta}_i(t-1), i=1,\cdots,n_{t-1}\}$ is the posterior samples at time $t-1$.
The prior density $p(\boldsymbol{\theta}(t)\,|\,\overline{D}(t-1))$ is approximated by Monte Carlo integration and using the evolution model in (3),
\[p(\boldsymbol{\theta}(t)\,|\,\overline{D}(t-1)) \approx \sum_{i=1}^{n_{t-1}} p_e(\boldsymbol{\theta}(t)|\boldsymbol{\theta}_i(t-1)).\]
Then, we generate one value of $\boldsymbol{\theta}(t)$ from $ p_e(\boldsymbol{\theta}(t)|\boldsymbol{\theta}_i(t-1))$ for each $i=1, \cdots, n_{t-1}$ and define these resulting sample points as $\Theta^0(t) = \{\boldsymbol{\theta}^0_i(t), i=1,\cdots,n_{t}\}$ (prior points for time $t$) with $n_t = n_{t-1}$.  Thus,  $\Lambda^0(t) = \{n_{t}, \Theta^0(t)\}$ approximates the prior distribution $p(\boldsymbol{\theta}(t)\,|\,\overline{D}(t-1))$.
\begin{table}
\centering
\begin{tabular}{l}
\hline\hline
{\bf Algorithm 1} Posterior computation algorithm \citep{west1993mixture}\\
\hline
\\
1. Draw $n_{t-1}$ samples of $\boldsymbol{\theta}_{t-1}$ from $p(\boldsymbol{\theta}_{t-1}\,|\,\overline{D}_{t-1})$ at time $t-1$,\\
\quad\qquad  save $\Lambda(t-1) = \{n_{t-1}, \Theta(t-1)\}$ where $\Theta(t-1) = \{\boldsymbol{\theta}_i(t-1), i=1,\cdots,n_{t-1}\}$.\\
\\
2. For DP base measure at time $t$, $p(\boldsymbol{\theta}(t)\,|\,\overline{D}(t-1))$ in (\ref{evol}),\\

\quad\qquad set $p(\boldsymbol{\theta}(t)\,|\,\overline{D}(t-1)) \approx \sum_{i=1}^{n_{t-1}} p_e(\boldsymbol{\theta}(t)|\boldsymbol{\theta}_i(t-1));$\\
\quad\qquad draw one sample (or more) from $p_e(\boldsymbol{\theta}(t)|\boldsymbol{\theta}_i(t-1))$ for each $i=1, \cdots, n_{t-1}$;\\
\quad\qquad save $\Lambda^0(t) = \{n_{t}, \Theta^0(t)\}$ where $\Theta^0(t) = \{\boldsymbol{\theta}^0_i(t), i=1,\cdots,n_{t}\}$.\\
\\
3. We approximate the parameters of the base measure using a multivariate normal distribution and $\Lambda^0(t)$. 
\\
4. Sample $p(\boldsymbol{\theta}(t)\,|\,\overline{D}(t))$ using the algorithm in Hannah et al., 2011:\\
\quad\qquad save $\Lambda(t)= \{n_{t}, \Theta(t)\}$ for time $t+1$.\\ \\
\hline
\hline
\end{tabular}
\end{table}

To compute the posterior distribution of $\boldsymbol{\theta}(t)$ in (\ref{eq2}), we approximate the parameters of the base measure of DP at time $t$, $p(\boldsymbol{\theta}(t)\,|\,\overline{D}(t-1))$ using the summary $\Lambda_{t}^0$. Similarly, the posterior distribution $p(\boldsymbol{\theta}_{t}\,|\,\overline{D}(t))$ is summarized through $\Lambda(t) = \{n_{t}, \Theta(t)\}$ for inferring the prior distribution at time $t+1$.

\section{Simulation Study}\label{Sec:Simulation}
In this section, we assess the performance of the proposed model (Bayesian nonparametrics + Bayesian dynamic model; hereafter, BNP+BDM) via a simulation study based on the dataset from the air pollution study (restricted to the Northeastern US). In the simulated data (n=1573), we have 2 baseline covariates ($\mathbf{W}$; \% of urban, and terrain elevation) and one time-varying covariate (temperature). The outcome is generated based on \# of respiratory disease related hospitalizations and mediator is based on \PMTwo concentrations. Values for the outcome, mediator and covariate at time $t=2, 3, 4$ are generated from the following models
\begin{itemize}
\item Time-varying covariate (a mixture of normals) for $t=2,3,4$
\begin{eqnarray*}
X(t) &\sim& 0.5 N\left(\alpha_{t,0} + \sum_{h=1}^{t-1} \alpha_{h,1} M(h) + \sum_{h=1}^{t-1} \alpha_{h,2} Z(h) + \sum_{h=1}^{t-1} \alpha_{h,3} X(h) +  \mathbf{W}\boldsymbol{\alpha}_t, 0.5\right)\\
& & + \, 0.5 N\left(\alpha_{t,0}^\prime + \sum_{h=1}^{t-1} \alpha_{h,1}^\prime M(h) + \sum_{h=1}^{t-1} \alpha_{h,2}^\prime Z(h) + \sum_{h=1}^{t-1} \alpha_{h,3}^\prime X(h) +  \mathbf{W}\boldsymbol{\alpha}_t^\prime, 0.5\right).
\end{eqnarray*}
\item Mediator (a skew normal) for $t=1,2,3,4$
\begin{equation*}
M(t) \sim SN\left(\beta_{t,0} + \sum_{h=0}^{t-1} \beta_{h,1} M(h) + \sum_{h=1}^{t} \beta_{h,2} Z(h) + \sum_{h=1}^{t} \beta_{h,3} X(h) +  \mathbf{W}\boldsymbol{\beta}_t, \xi, \psi\right),
\end{equation*}
where $SN(\cdot, \xi, \psi)$ denotes a skew-normal distribution with scale parameter $\xi$, and shape parameter $\psi$. 
\item Outcome (a Poisson log-linear model with nonlinear terms) for $t=1,2,3,4$,
\begin{equation*}
Y(t) \sim Poi(\text{offset}_t \times \mu(t)),
\end{equation*}
and
\begin{eqnarray*}
\mu(t) &=&\exp \left(\gamma_{t,0} + \sum_{h=1}^{t} \gamma_{h,1} M(h) + \sum_{h=1}^{t} \gamma_{h,2} Z(h) +\sum_{h=0}^{t-1} \gamma_{h,3} \log (Y(h) + c)  \right.\\
 &  & \left.+ \sum_{h=1}^{t} \gamma_{h,4} X(h)+ \gamma_{t,5} M(t) Z(t) +\gamma_{t,6} M(t) X(t)+ \mathbf{W}\boldsymbol{\gamma}_t \right),
\end{eqnarray*}
where $c$ is a small constant.
\item Treatment assignment for $t=1,2,3,4$
\begin{equation*}
Z(t) \sim Bern\left\{\text{expit} \left(\delta_{t,0} + \sum_{h=1}^{t-1} \delta_{h,1} Z(h)+ \sum_{h=1}^{t} \delta_{h,2} X(h) +  \mathbf{W}\boldsymbol{\delta}_t \right)\right\}.
\end{equation*}
\end{itemize}
Note that the all models depend on the full histories of the predictors. Also, each model has some complexity: either a non-standard distribution (i.e., covariate and mediator) or including nonlinear terms (i.e., outcome).


The coefficients in the models for $t=1$ are obtained by fitting the observed data models to the air pollution study data. Then, the coefficients of the subsequent models are specified with attenuation of the effects. But, the coefficients of the one-time preceding variables (i.e., variables at time $t-1$) are set to the values that change by 15\% or by 30\% each time point. For example, in the mediator model at $t=2$, $(\beta_{1,1},\beta_{2,2},\beta_{3,3})$ coefficients in the model for $M(2)$ are equal to $(\beta_{0,1},\beta_{1,2},\beta_{2,3}) \times 0.85\, (\text{or } 0.7)$ coefficients in the model for $M(1)$. And $(\beta_{0,1},\beta_{1,2},\beta_{2,3})$ in the model for $M(2)$ are attenuation (1/10) of the same coefficients $(\beta_{0,1},\beta_{1,2},\beta_{2,3})$ in the model $M(1)$.
Thus, we consider two cases: 1) when the effects of variables at time $t-1$ on variable at time $t$ decrease by 15\% every time, and 2) when the effects of variables at time $t-1$ on variable at time $t$ decrease by 30\% every time. Also, the coefficients of the treatment assignment model at each time are specified in a way that the number of the treated samples is increasing over the time periods, which reflects the fact that the number of locations under the low \SOTwo\, exposure generally increases every year (in our application, the number increases during 2005-2007). 

We consider 5 different models: (I) standard regression models with all past predicting variables (i.e., containing all previous predictors); (II) standard regression models only with one-time preceding variables (at  $t-1$); (III) Generalized Additive Models with one-time preceding predictors; (IV) Our Bayesian nonparametric models w/o Bayesian dynamic model; (V) BNP+BDM (out model). 

\begin{figure}[p]
    \centering
\includegraphics[width=13cm]{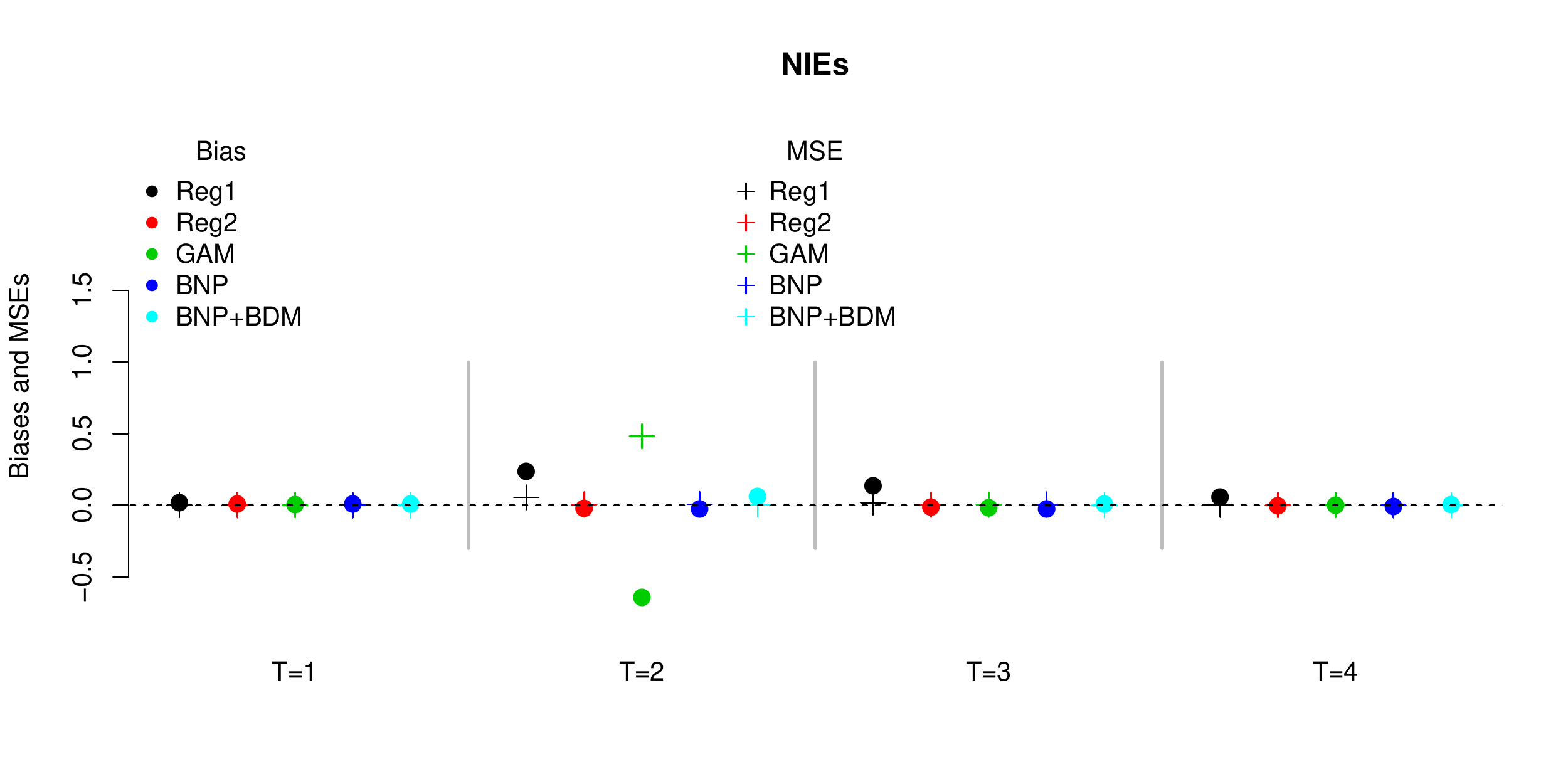}
\includegraphics[width=13cm]{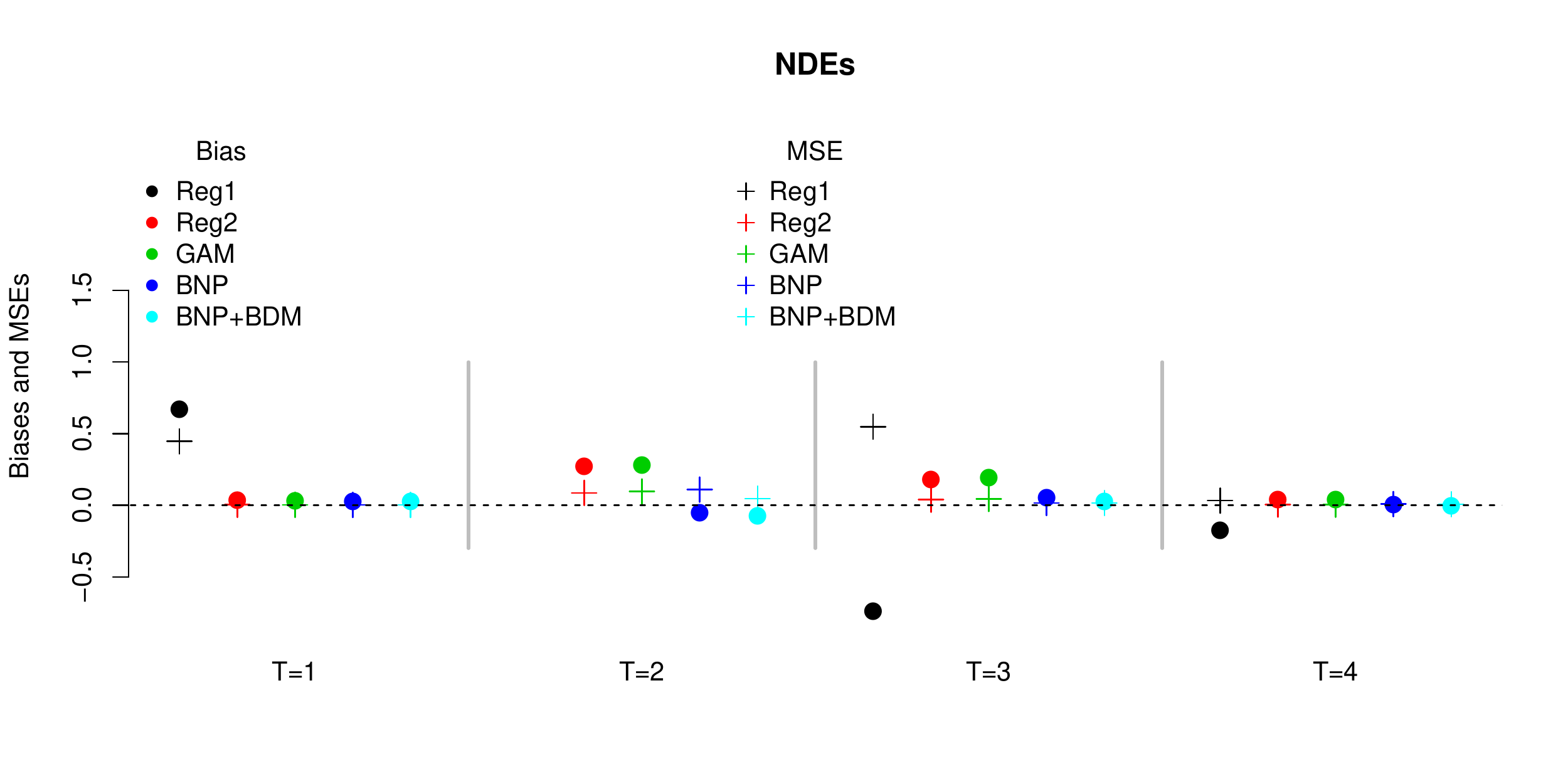}
\includegraphics[width=13cm]{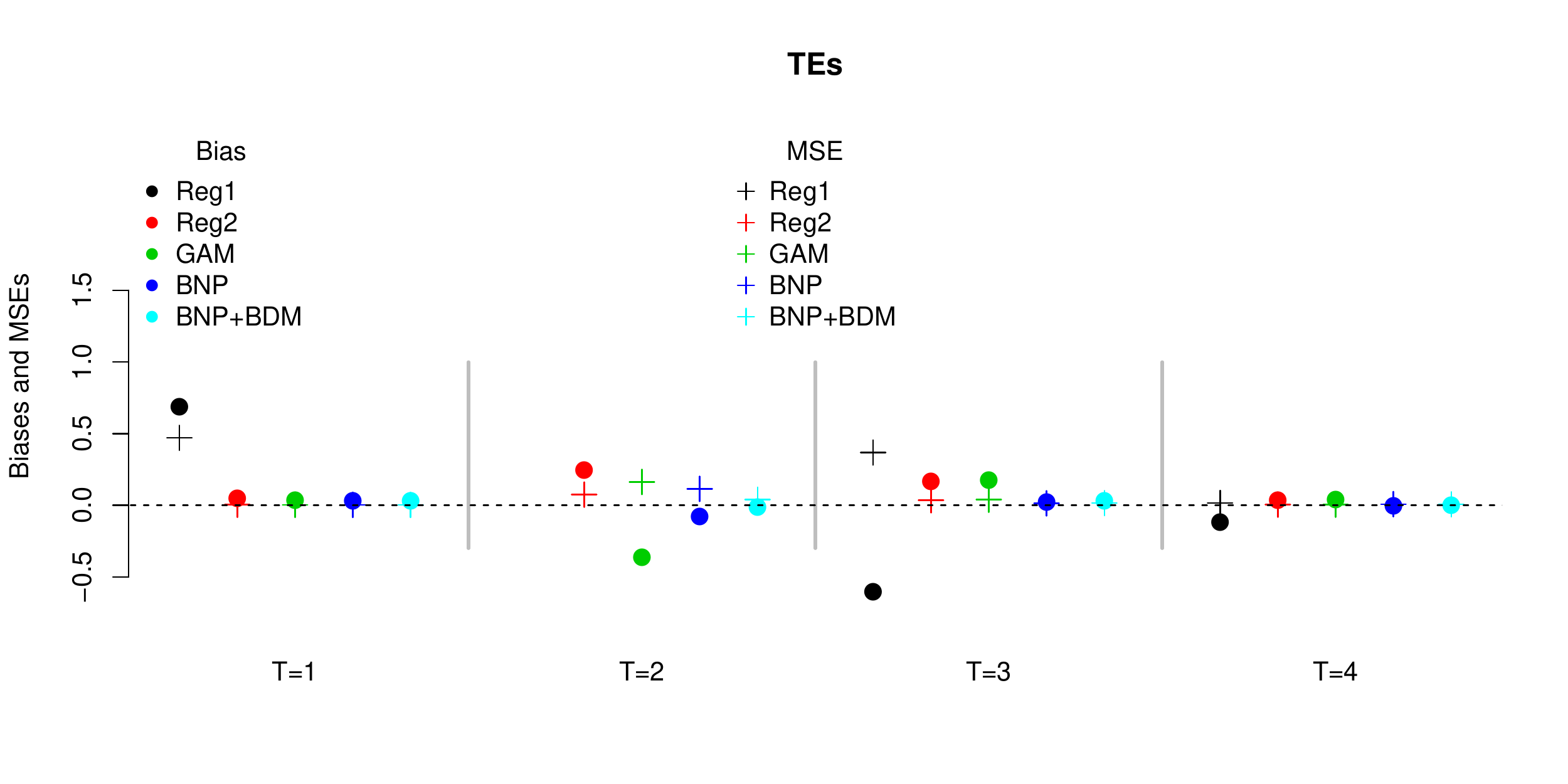}
\caption{Case 1 : Simulation results for natural indirect, direct and total effects over 400 replications. Solid circles are biases and crosses are mean squared errors (MSEs) from 5 different models: Reg 1 (regression models with all previous predictors); Reg 2 (regression models with one-time preceding predictors); GAM (Generalized Additive Models with one-time preceding predictors); BNP (Bayesian Nonparametric models w/o Bayesian dynamic model); BNP+BDM (Bayesian Nonparametric models w/ Bayesian dynamic model)}\label{simulation1}
\end{figure}

Figure \ref{simulation1} depicts the results from 5 different competing models for the case when the coefficients of the one-time preceding variables (variables at time $t-1$) change slowly (decrease by 15\%) across the time points. In terms of biases and MSEs, BNP+BDM (our model) performs best. In particular, from the comparison between BNP and BNP+BDM, it shows that BNP+BDM reduces biases (and MSEs) resulting from not using all previous histories but updating current parameters of the distributions from the past parameters information. Especially, for the total effects at time $t=2$, the bias (and the MSE) of BNP are 6 times (and 3 times) larger than that of BNP+BDM. From the comparison between GAM and BNP, since they use the same set of predictors in modeling, BNP's better performance in terms of biases and MSEs (especially at time $t=2$) suggests improved ability of BNP in fitting the observed data.

We also evaluate our model performance under the case that the coefficients of the one-time preceding variables (variables at time $t-1$) change more quickly (decrease by $30 \%$ each time) and the case that coefficients of the one-time predicting variables change irregularly (decrease by $30\%$ at time 2; increase by $15\%$ at time 3; decrease by $20\%$ at time 4). The results are in the appendix and they still suggest that BNP+BDM (and BNP) still performs best in terms of biases and MSEs. 

\section{Application}
We apply our method to the air pollution study described in Section 1. The study enrolled 15,472 US zipcodes with information for each zip code including estimated \SOTwo\, exposures, an ambient \PMTwo\, concentrations (mediator), respiratory diseases related hospitalizations among the Medicare beneficiaries (outcome), variables from the Census data : \% of white, \% of high school graduates, median household incomes, \% of poor, total population, \% of female, smoking rates from the CDC and terrain elevations from the USGS. Here, we are interested in evaluation of longitudinal changes in the health effects of \SOTwo\, exposures across every warm season from 2004 to 2007 (with 2003 as the baseline year)  since secondary \PMTwo's are synthesized from \SOTwo\, actively under high temperature. For each zip code location, a level of exposure to \SOTwo\, emissions is measured every June.  The mediator is the average \PMTwo\, concentration of June and July every year, and the outcome is the total number of hospitalizations between June and October every year. All covariates are measured in June 2003 except the time-varying covariate (temperature) which is measured in every June.  Table \ref{Data} in the Web Appendix include summary statistics of the variables used in the analysis. From Year 2005, more zip code locations were assigned to low \SOTwo\, emission exposure areas (see Section 1.1 for the dichotomization of continuous exposure levels). This was partially driven by multiple regulatory policies from local and federal governments, but also by EPA's 2005 Clean Air Interstate Rule (CAIR) which required 28 eastern states to make reductions in \SOTwo\, and \NOx\, emissions that contribute to unhealthy levels  of \PMTwo\, (and ozone) pollution in downwind states. This is also evidenced by the fact that the number of coal-fired power plants equipped with \SOTwo\, scrubbers (a technology to reduce \SOTwo\, emissions) increased from years 2005-2006.

Here, we are interested in how switching an exposure level to low at time $t$ from a chronic high \SOTwo\, emission exposure ($1,\cdots, t-1$) affects health conditions:  $\bar{z} = \{0\}$ vs. $\bar{z}^\prime = \{1\}$ at time 1,$\bar{z} = \{0,0\}$ vs. $\bar{z}^\prime = \{0,1\}$ at time 2, $\bar{z} = \{0,0,0\}$ vs. $\bar{z}^\prime = \{0,0,1\}$ at time 3, $\bar{z} = \{0,0,0,0\}$ vs. $\bar{z}^\prime = \{0,0,0,1\}$ at time 4.  Through this, we try to understand whether a policy to reduce levels of exposure to \SOTwo\, emissions is effective for immediate individual health improvements even if they are under high  \SOTwo\, emission exposures for years. If that is effective, we also expect to see how much of the effect is attributable to changes in ambient \PMTwo\, concentrations. Under these contrasts of interest, the number of observed data under each treatment regime is decreasing: 7606 for $\bar{z}=\{0\}$ vs. 7866 for $\bar{z}^\prime=\{1\}$ at time 1, 6576 for $\bar{z} = \{0,0\}$ vs. 1030 for $\bar{z}^\prime = \{0,1\}$ at time 2, 4702 for $\bar{z} = \{0,0,0\}$ vs. 1874 for $\bar{z}^\prime = \{0,0,1\}$ at time 3, 4431 for $\bar{z} = \{0,0,0,0\}$ vs. 271 for $\bar{z}^\prime = \{0,0,0,1\}$ at time 4. Thus, the causal effect estimates might not be sufficiently informed by the data especially at the last time point (Year 2007). Our proposed dynamic model can help address this issue.

\subsection{Prior Specifications}

\subsubsection{Priors for the parameters in the observation model}

For scale and shape parameters of the inverse gamma distributions, $Inv.Gamma(a,b)$, in base measures $\mathcal{F}_t^z$ in Section 5.2.2, we set $a=5$ (shape) and $b=1$ (scale) to avoid large clusters with heterogeneous memberships. For the precision parameters in the base measure, $\tau_{t,h}^z$, we specify $Gamma(2,1)$ priors for $h=1, \cdots, \mathbf{card}(\boldsymbol{\alpha}_{t,i}^{z}); \,\, t=1, \cdots, T$. For the mass parameters of the Dirichlet processes, $\lambda_t^z$, we specify a $Gamma(1,1)$ prior . Alternatively we can specify uniform priors, however it is known that the results are typically insensitive to reasonable changes in the prior specification on the DP mass parameter \citep{Tadd:2008}.

\subsubsection{Parameters in evolution model}
We specify a multivariate normal distribution $MVN(\boldsymbol{\theta}(t-1), \Sigma_{\boldsymbol{\theta}})$ for the evolution model in Section \ref{Evolution} where the state parameters at time $t-1$, $\boldsymbol{\theta}(t-1)$, are the mean parameters ($\boldsymbol{\mathcal{A}}_{t-1}$) of the base measures at time $t-1$. We set $\Sigma_{\boldsymbol{\theta}}$ to the posterior covariance matrix of $\boldsymbol{\theta}(t-1)$. Note again that we only update the mean parameter of the base measure.

\subsection{Model Fits and Comparison to GAM}
\begin{figure}[p]
    \centering
\includegraphics[width=17cm]{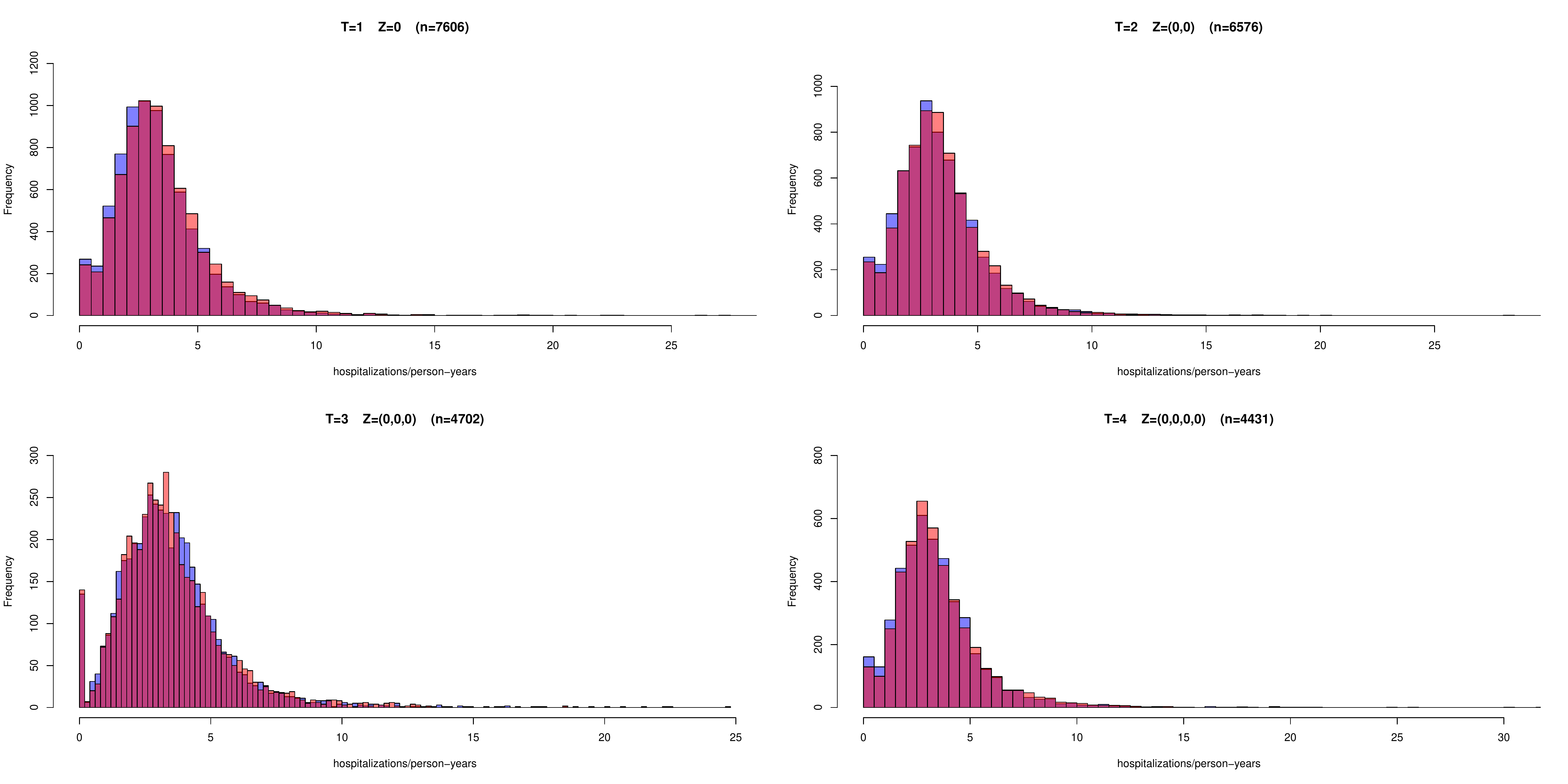}
\includegraphics[width=17cm]{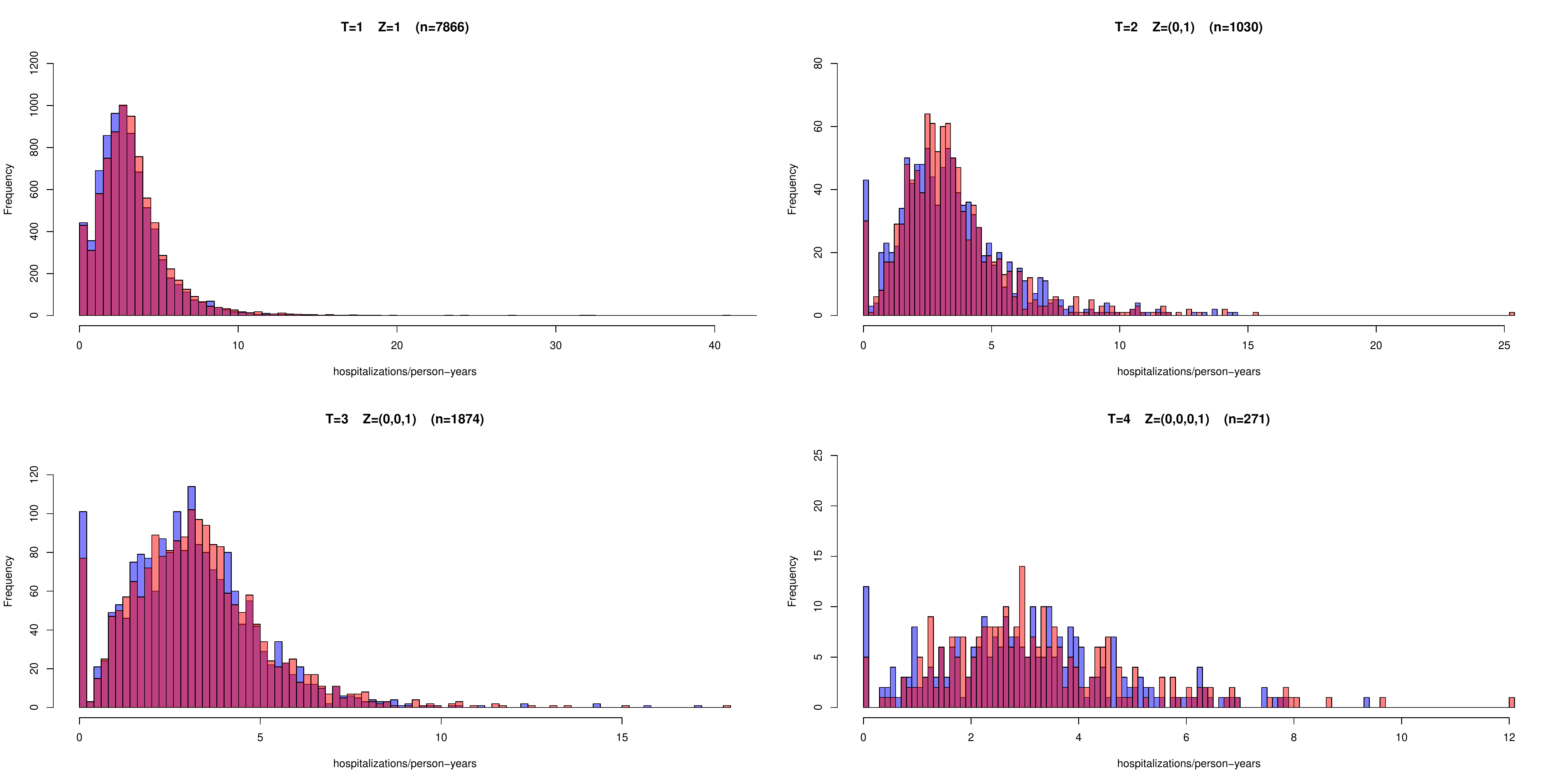}
    \caption{Blue (predictive values) vs. Red (data). The first row: 1 replication of the outcome under $Z=0$, $Z=(0,0)$, $Z=(0,0,0)$ and $Z=(0,0,0,0)$ from the posterior predictive distribution of the proposed model. The second row: 1 replication of the outcome under $Z=1$, $Z=(0,1)$, $Z=(0,0,1)$, and $Z(0,0,0,1)$ from the posterior predictive distribution of the BNP+BDM.}
 \label{ModelFit1}
 \end{figure}

\begin{figure}[p]
    \centering
\includegraphics[width=17cm]{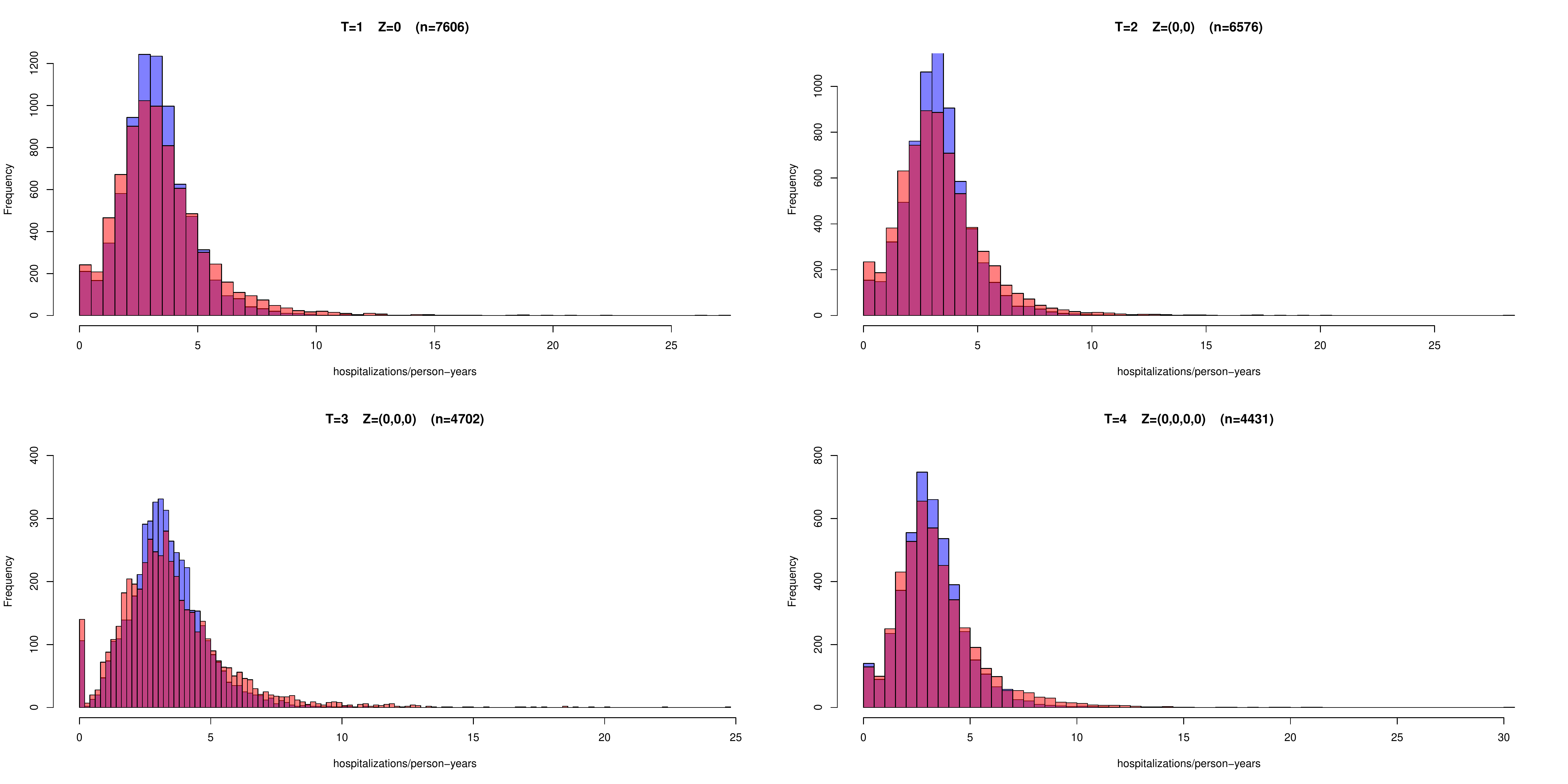}
\includegraphics[width=17cm]{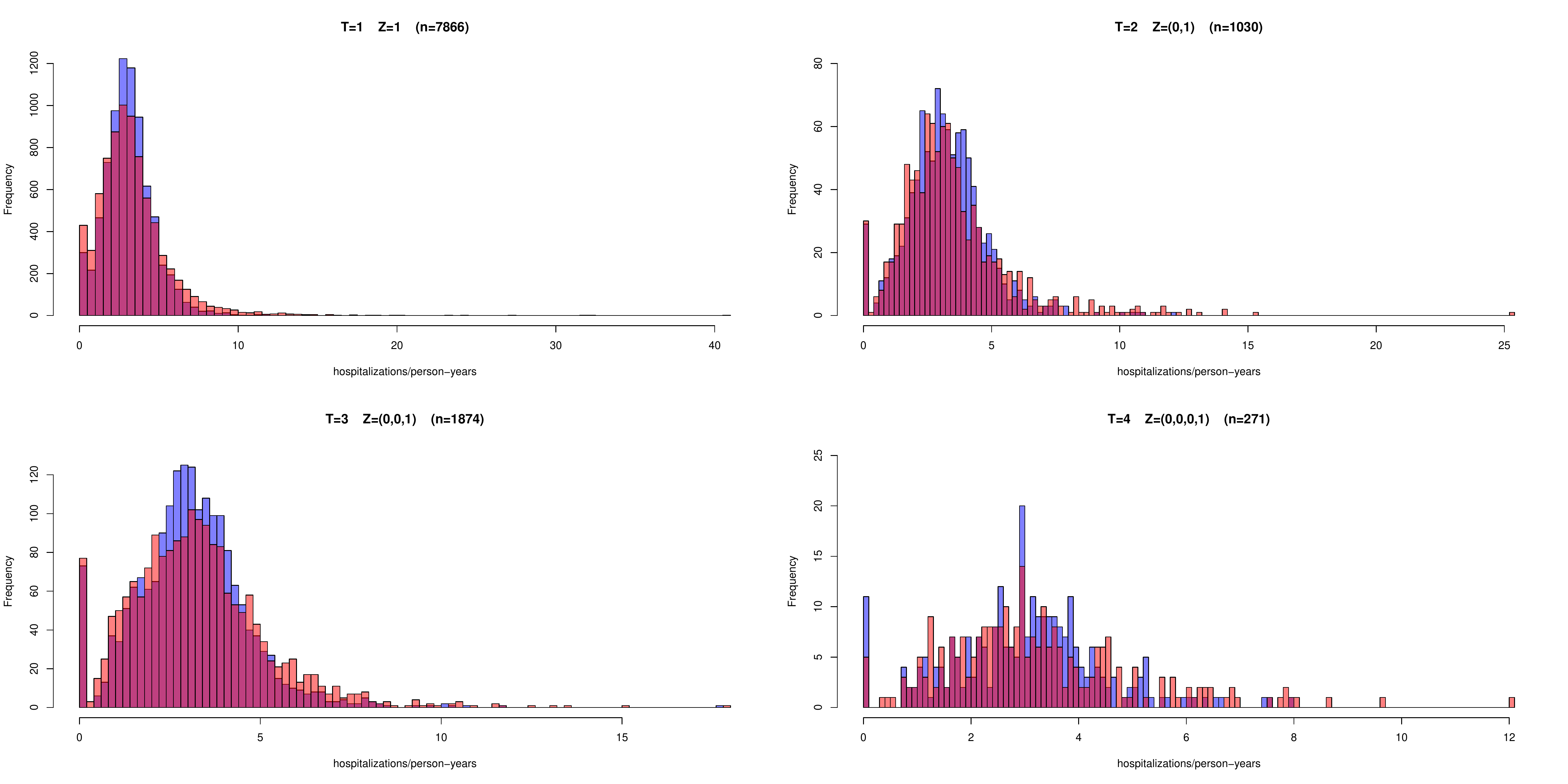}
    \caption{Blue (predictive values) vs. Red (data). The first row: 1 prediction of the outcome under $Z=0$, $Z=(0,0)$, $Z=(0,0,0)$ and $Z=(0,0,0,0)$ from the generalized additive model (GAM). The second row: 1 prediction of the outcome under $Z=1$, $Z=(0,1)$, $Z=(0,0,1)$, and $Z(0,0,0,1)$ from the generalized additive model (GAM).}
    \label{ModelFit2} 
\end{figure}

To illustrate the need for our flexible modeling strategy (BNP+BDM) for the observed data in the air pollution study, we compare the fit of our proposed model against a generalized additive model (GAM) with the same set of predictors used in our model (i.e., one time preceding observations as the predictors) using posterior predictive checks (Figure \ref{ModelFit1}). We display observed data $y$ along with $y^{\text{pred}}$ from the posterior predictive distributions under high \SOTwo\, emission exposures and low \SOTwo\, emission exposures. Figure \ref{ModelFit2} illustrates the observed data $y$ along with predictions of $y$ from the GAM. Under $z=0$ for all time points, differences between the means of predictive values and the means of observed data for time $t=1,2,3,4$ are 0.17 (0.20), 0.07 (0.13), 0.05 (0.21), 0.03 (0.27) for the BNP+BDM (for the GAM), respectively. Under $z=1$ for the last time points, differences between the means of predictive values and the means of observed data for time $t=1,2,3,4$ are 0.16 (0.20), 0.05 (0.21), 0.12 (0.14), 0.15 (0.17) for the BNP+BDM (for the GAM), respectively. Based on these mean differences and the visual inspection of histograms, our model fits better for all times $t$, particularly at time $t=2,3,4$.

\subsection{Results}
\begin{figure}[p]
    \centering
\includegraphics[width=8.5cm]{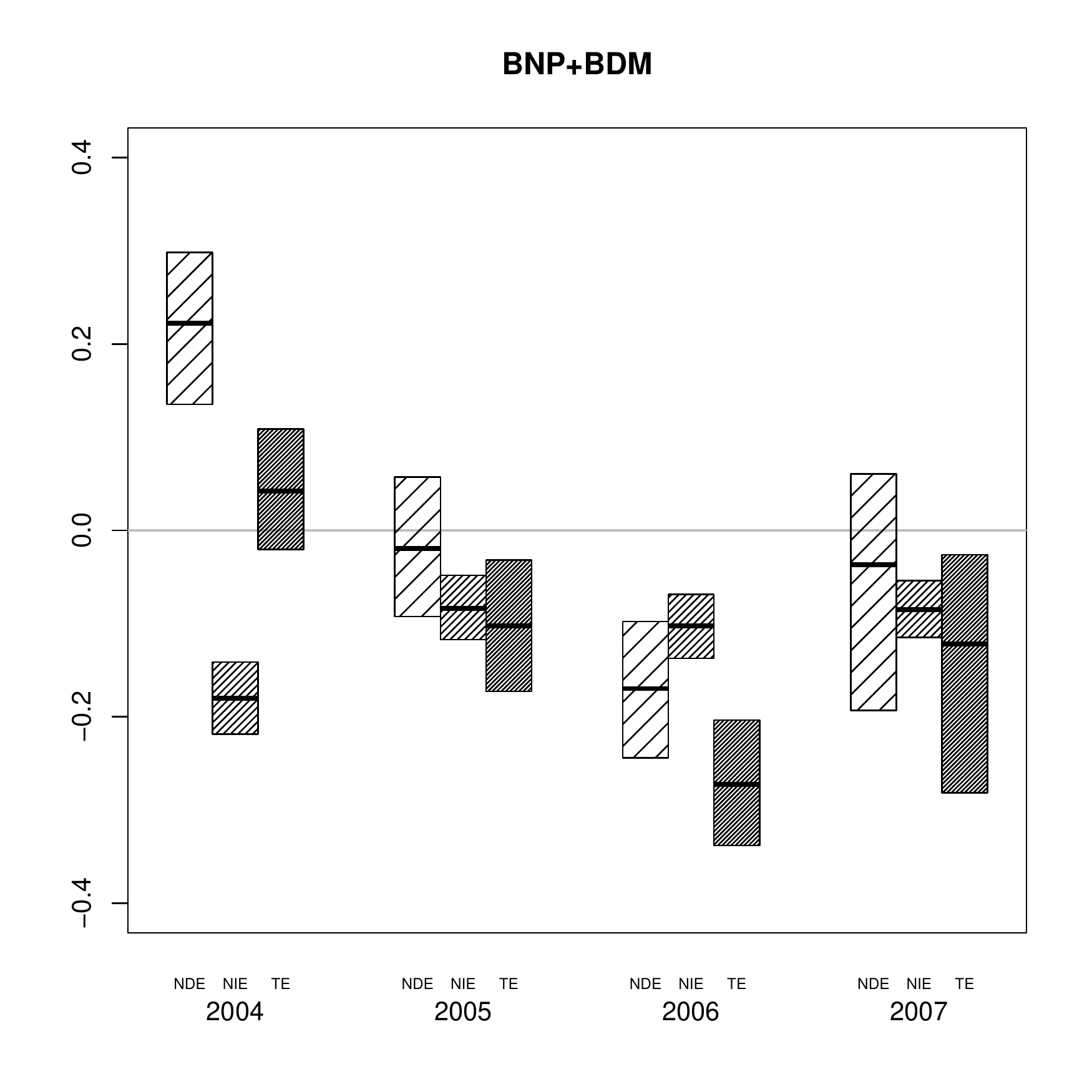}
\includegraphics[width=8.5cm]{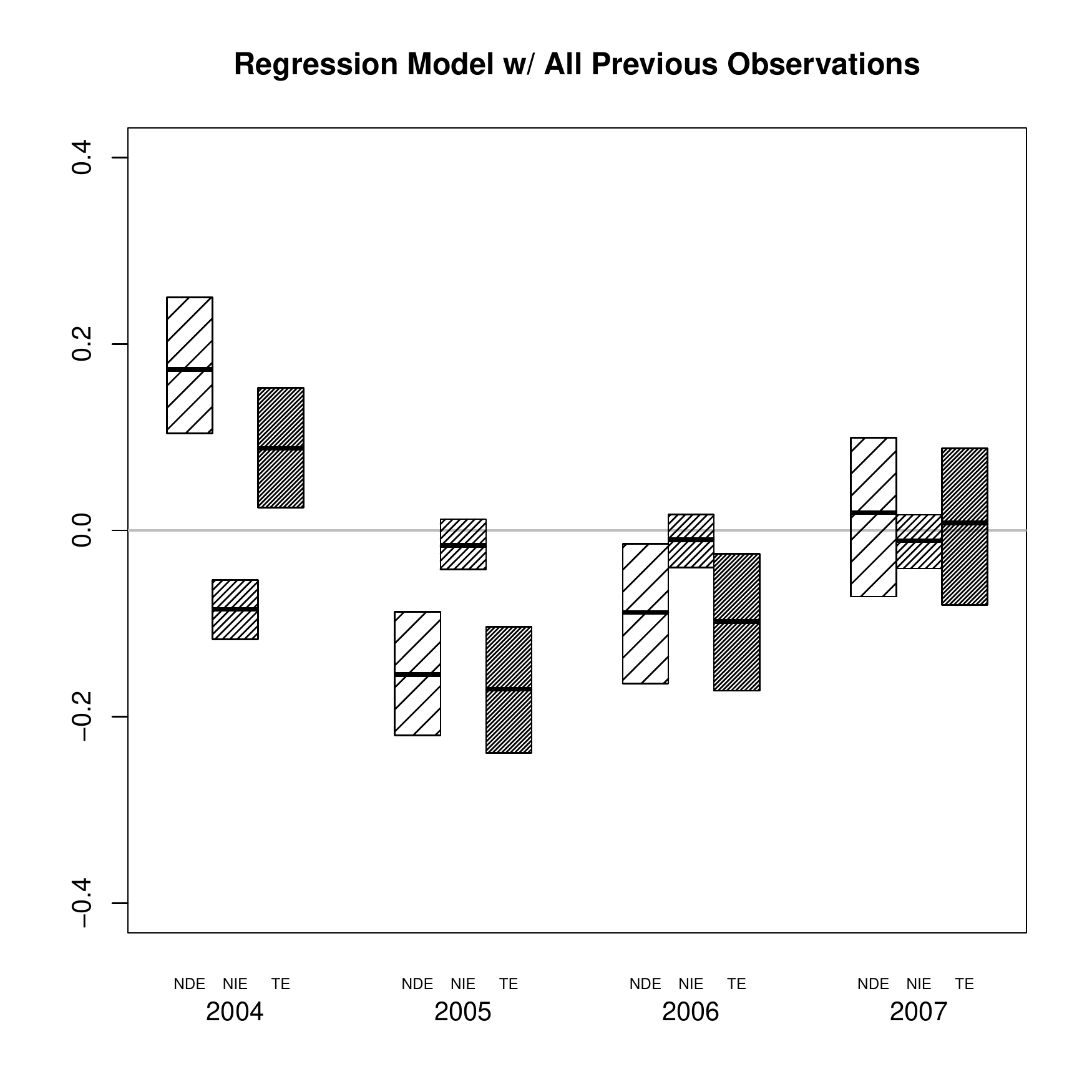}
\includegraphics[width=8.5cm]{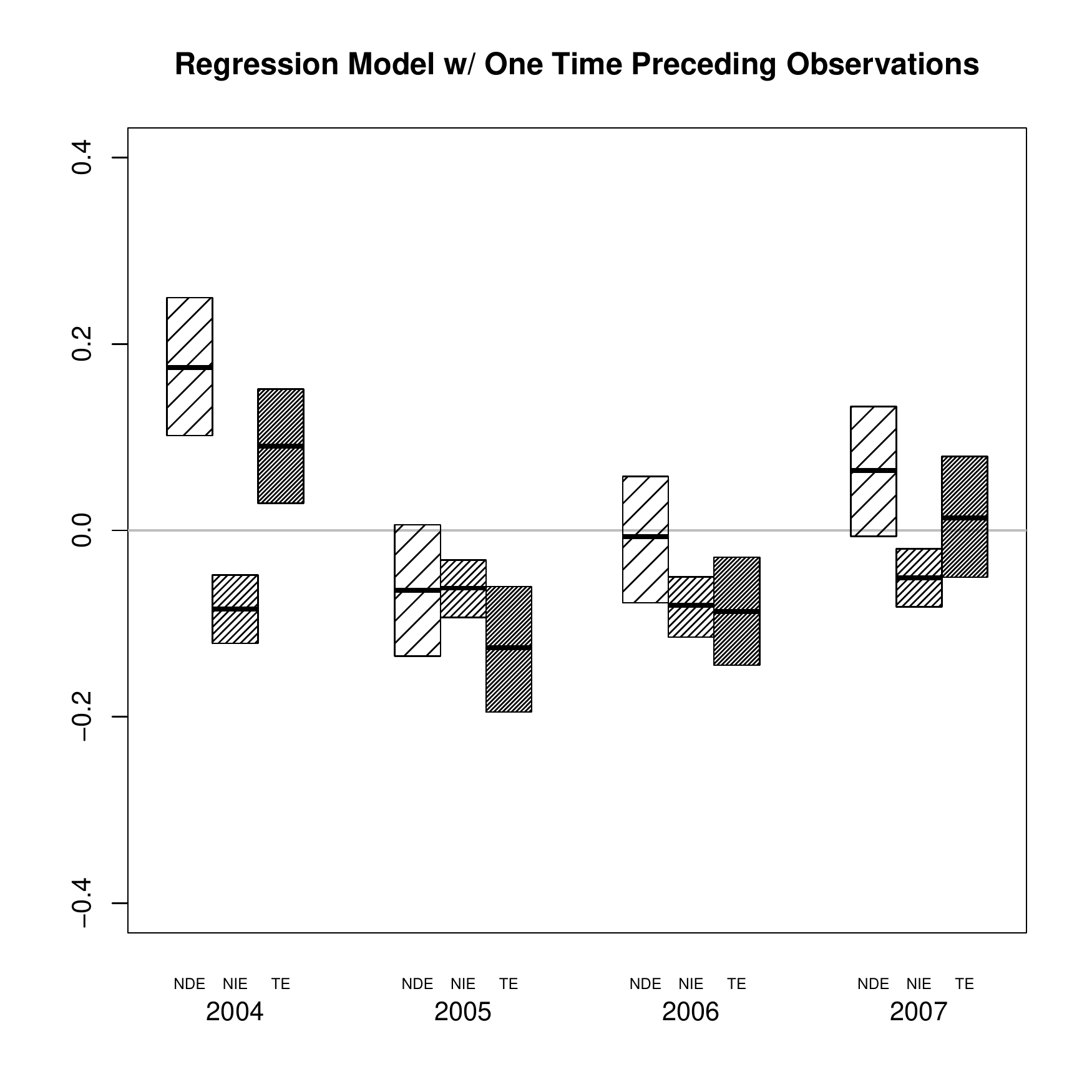}
\includegraphics[width=8.5cm]{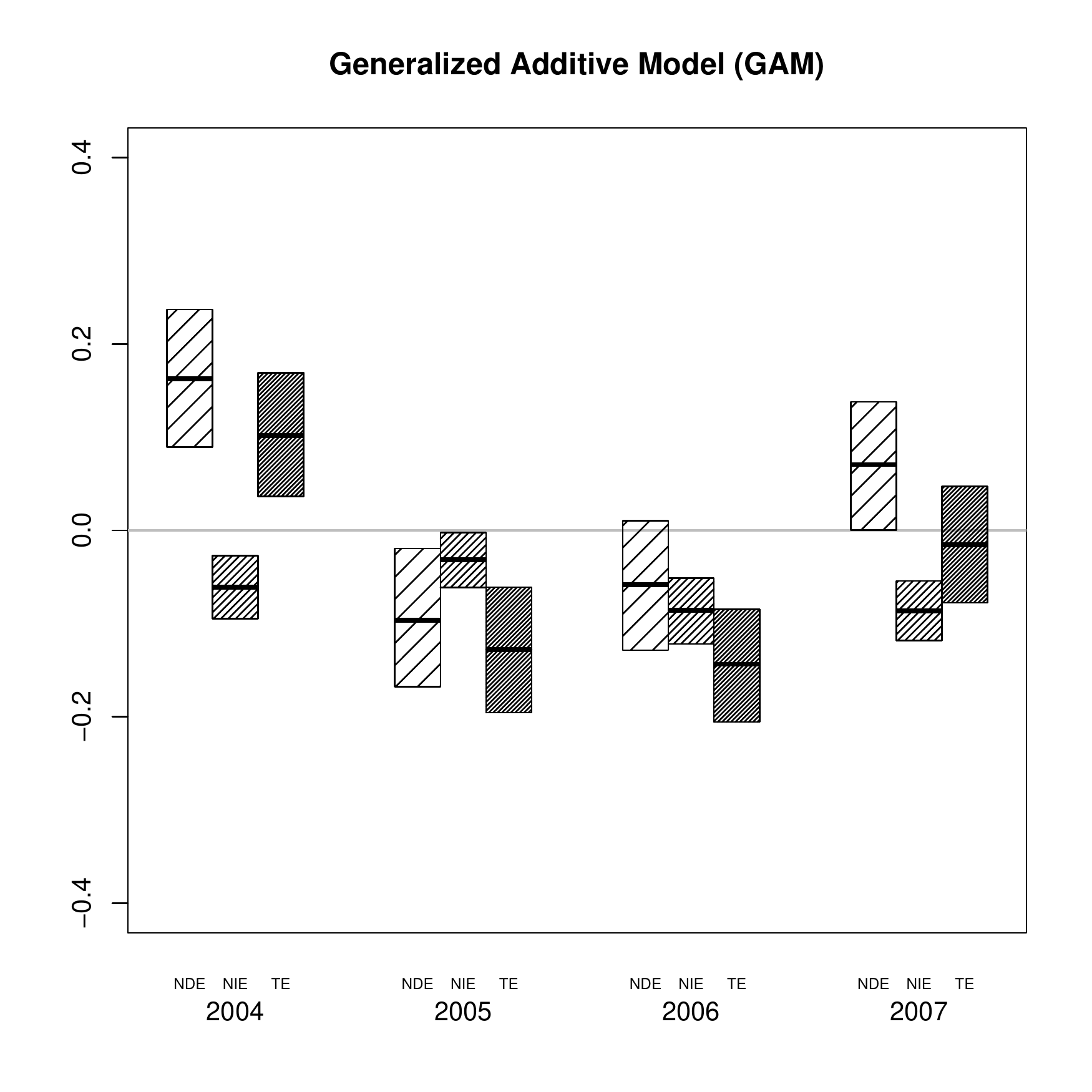}
    \caption{The natural direct, indirect and total effects of exposure to \SOTwo\, emissions on respiratory hospitalizations in warm seasons during 2003-2007; The solid black lines indicate the posterior means of the effects in each season and the boxes around the solid lines indicate the 95\% C.I.s. The upper left plot is for the BNP+BDM (our proposed model); the upper right plot is for the regression model with all previous predictors; the lower left plot is for the regression model with one-time preceding predictors; the lower right plot is for the GAM with one-time preceding predictors.}   \label{fig3}
\end{figure}

To obtain the posterior for each $\boldsymbol{\theta}(t)$, we ran the MCMC algorithm for 15000 iterations and discarded the first 5000 as burn-in, which requires 2209.5 seconds to run for each time $t$ on Mac OS X with 3.7 GHz Intel Xeon E5 processor and 32 GB RAM. For the posterior computation in Section 6, we set $N=1000$ (by taking every 10th posterior sample in order to minimize autocorrelation).
Figure \ref{fig3} shows the posterior means and the 95\% credible intervals of the causal effects from 4 different models: (a) Bayesian nonparametric model with Bayesian dynamic modelling (upper left); (b) the regression model with all previous histories as the predictors (upper right); (c) the regression model with one-time preceding observations as the predictors (lower left); (d) the generalized additive model with one-time preceding observations as the predictors (lower right). 

The total effect of low \SOTwo\, emission exposure versus high \SOTwo\, emission exposure corresponds to reduction in an average hospitalization rate (hospitalization / 200 person-years) among the Medicare beneficiaries in 2005 and in 2006 (ranging from -0.27 to -0.09) for all four approaches. Our approach shows no initial total effect (in 2004) and the significant total effect in 2007, while other approaches estimate increased hospitalization rates in the initial year and no total effects at the end. Under the low \SOTwo\, exposure status, it is hard to envisioning scenario where the overall hospitalizations increase. Also, since there exists a short-term impact of air pollution on respiratory outcomes\citep{schwartz1995short}, we expect reduction in the average hospitalization rate in 2007 even after chronic high \SOTwo\, exposures until 2006. Thus, our estimates of the total effects are more reasonable. 

The indirect effects measure the causal effects of low \SOTwo\, emission exposure that are through the effects on local \PMTwo\, concentrations. As the chemical/medical pathways from \SOTwo\, emissions to local \PMTwo\, and from \PMTwo\, to respiratory diseases are well known, the negative indirect effects were expected. We estimate the effects of low \SOTwo\, emission exposure on \PMTwo\, concentrations as -1.26 (-1.31, -1.20), -1.17 (-1.24, -1.09), -1.57 (-1.63, -1.51), -0.86 (-0.89, -0.83) for years 2004, 2005, 2006, 2007, respectively.
All four approaches estimate the most pronounced indirect effects in 2004. The locations exposed to high \SOTwo\, emissions in 2004 cover the rust belt of the United States and Massachusetts (see Figure \ref{fig2}) where \PMTwo\, concentrations are relatively high. Thus, the effect of high \SOTwo\, emissions on the mediator (\PMTwo) is relatively larger: the estimates are -1.26 (-1.31, -1.20), -1.17 (-1.24, -1.09), -1.57 (-1.63, -1.51), -0.86 (-0.89, -0.83) for years 2004, 2005, 2006, 2007, respectively. Also, in those regions under high \SOTwo\, emission exposure in 2004, \PMTwo\, concentrations and hospitalization rates are highly correlated (0.14 vs -0.01, 0.05, 0.01 in other years). These result in the larger indirect effect in 2004. 
Our estimates of the indirect effects are ranging from -0.18 to -0.09 which are twice as large as the estimates from other models. Since we expect a significant amount of the causal effects though this indirect pathway, our model again provides more realistic results. \SOTwo\, emissions are also known to cause asthma. Thus, we can expect some significant direct effects (e.g., 2006). However, in most approaches, the direct effects are not pronounced and the indirect effects represent most of the total effect. 
\begin{figure}[h]
    \centering
\includegraphics[width=10cm]{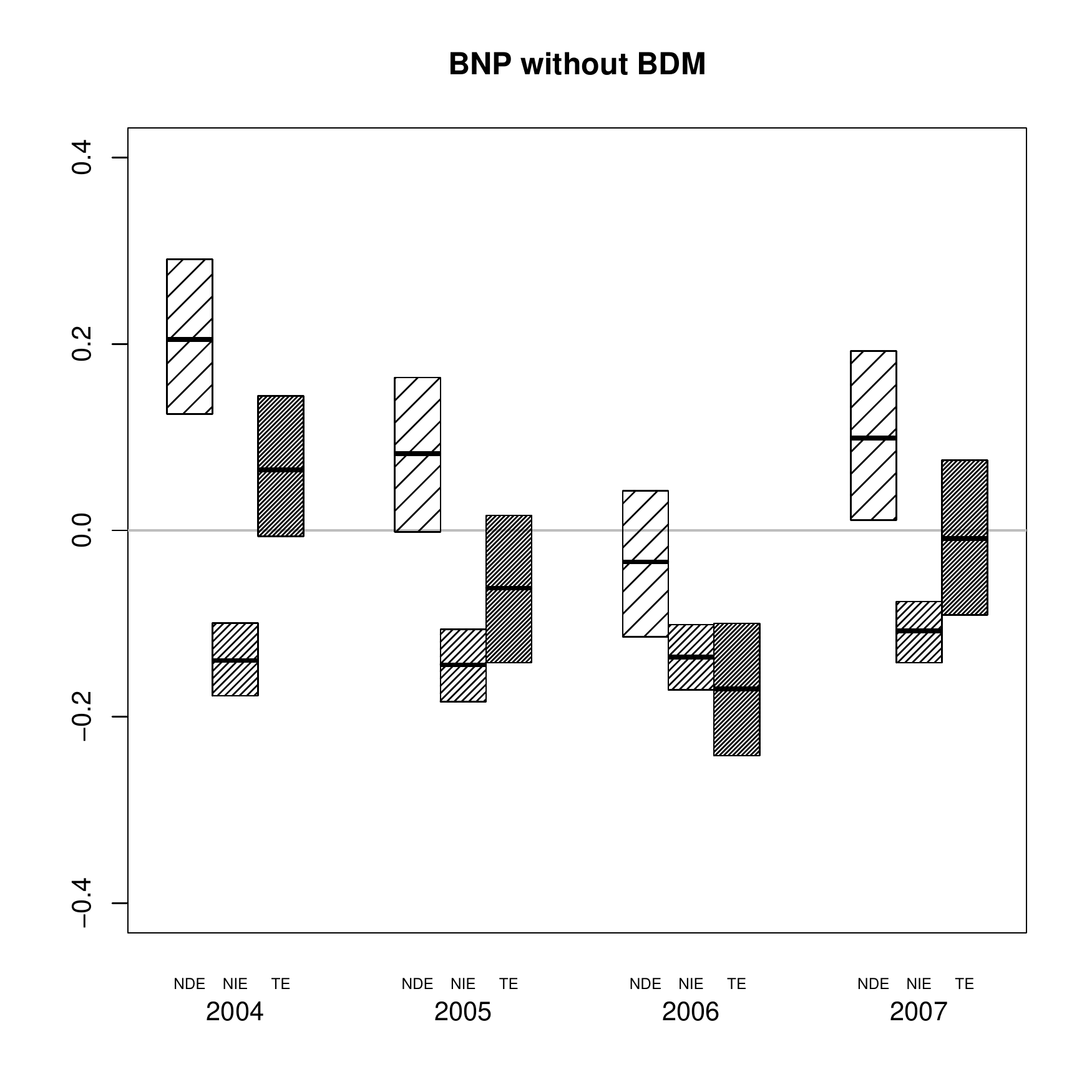}
    \caption{The natural direct, indirect and total effects from the BNP model without Bayesian dynamic modelling.}   \label{Sens1}
\end{figure}

\subsubsection{Sensitivity Analyses}\label{sec:sen}
\begin{figure}[h]
    \centering
\includegraphics[width=10cm]{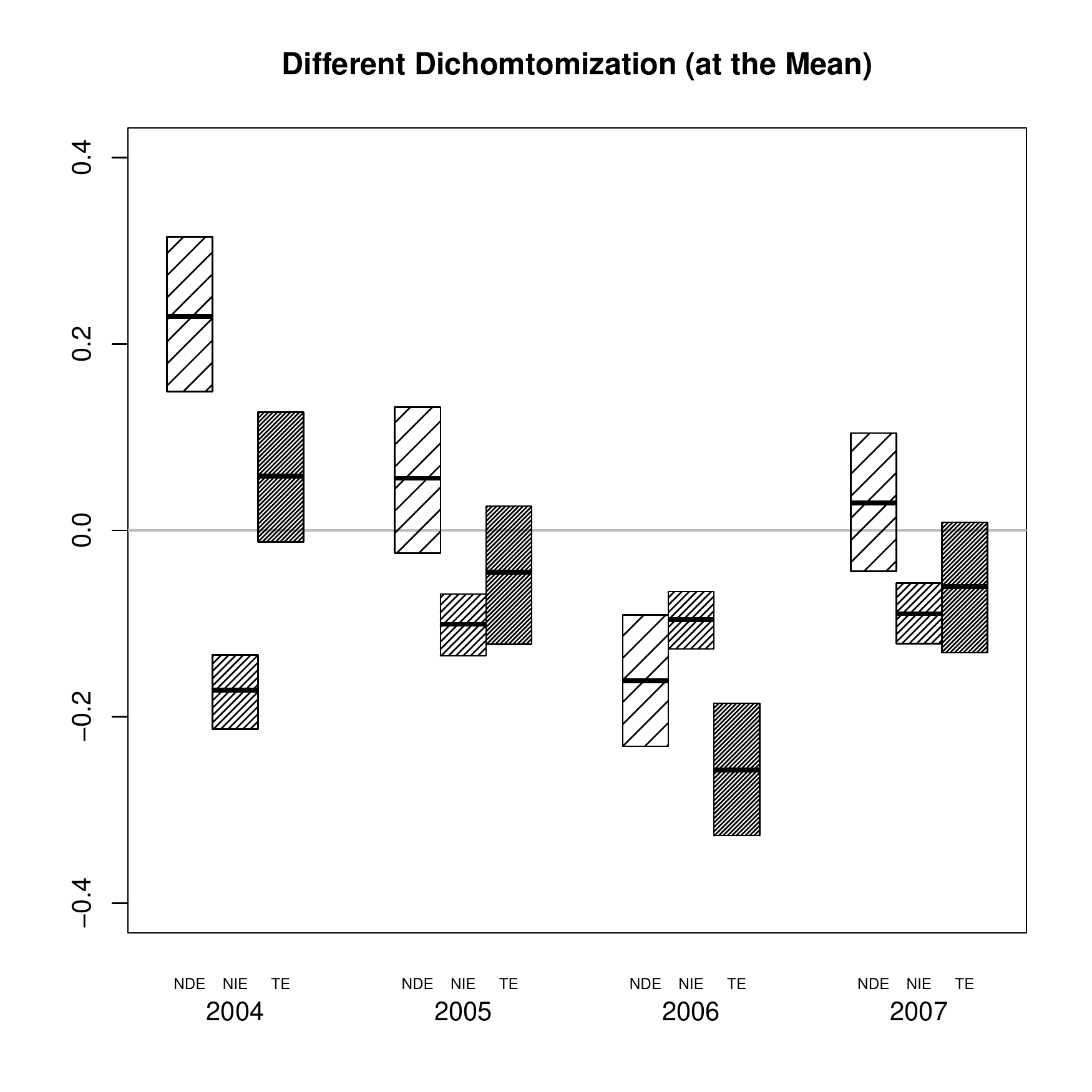}
    \caption{The natural direct, indirect and total effects with cutoff based on the mean \SOTwo\, exposure.}   \label{Sens2}
\end{figure}

To assess the sensitivity to the current dichotomization of the exposure level, we conducted the analysis under a different cutoff that distinguishes zip codes under low level of \SOTwo\, emission exposures from zip codes under high level of \SOTwo\, emission exposures.  In the Web Appendix, Figure \ref{Sens2} shows the causal effects for a cutoff based on the mean (12,953; as opposed to the median) of the exposures over the years. This new cutoff resulted in similar estimates.

We also consider deviations from Assumption 2 in terms of a sensitivity analysis technique similar to the one introduced in \cite{Dani:Roy:Kim:Hoga:Perr:2012} and \cite{kim2016framework}. We weaken the Assumption 2 by assuming the conditional distributions in (4) are not equal, but are proportional up to a multiplicative function (an exponential tilt) with some sensitivity parameter $\chi$ if the difference between the mediators at time $t$ is larger than a threshold (which we  here set to $0.5 \times SD(M_{\overline{Z}^\prime}(t)-M_{\overline{Z}}(t)))$
\begin{eqnarray*}
\lefteqn{f(Y_{\overline{Z}, \overline{M}_{\overline{Z}^\prime}(t)}(t)=y\,|\, M_{\overline{Z}^\prime}(t)=m, M_{\overline{Z}}(t), \text{Remainder}, \overline{\boldsymbol{\theta}}(t)) } \nonumber\\
 & \propto &  \exp\{\log (\chi^{\text{sgn}(y/\text{offset}-\text{med}(Y_{\overline{Z}}(t)/\text{offset})) \times \text{sgn}(d)}) \}f(Y_{\overline{Z}, \overline{M}_{\overline{Z}}(t)}(t)=y \,|\,M_{\overline{Z}^\prime}(t), M_{\overline{Z}}(t)=m, \text{Remainder}, \overline{\boldsymbol{\theta}}(t)) 
\end{eqnarray*}
where $\text{Remainder} = (\overline{M}(t-1)=\overline{m}(t-1), \overline{Y}(t-1)=\overline{y}(t-1), \overline{W}(t-1)=\overline{w}(t-1), \overline{Z}(t-1)=\overline{z}(t-1))$ and sgn$()$ is the sign function and med$(Y_{\overline{Z}}(t)/\text{offset})$ indicates the median of the outcome (rate) under treatments $\overline{Z}$. 

\begin{figure}[p]
    \centering
\includegraphics[width=8cm]{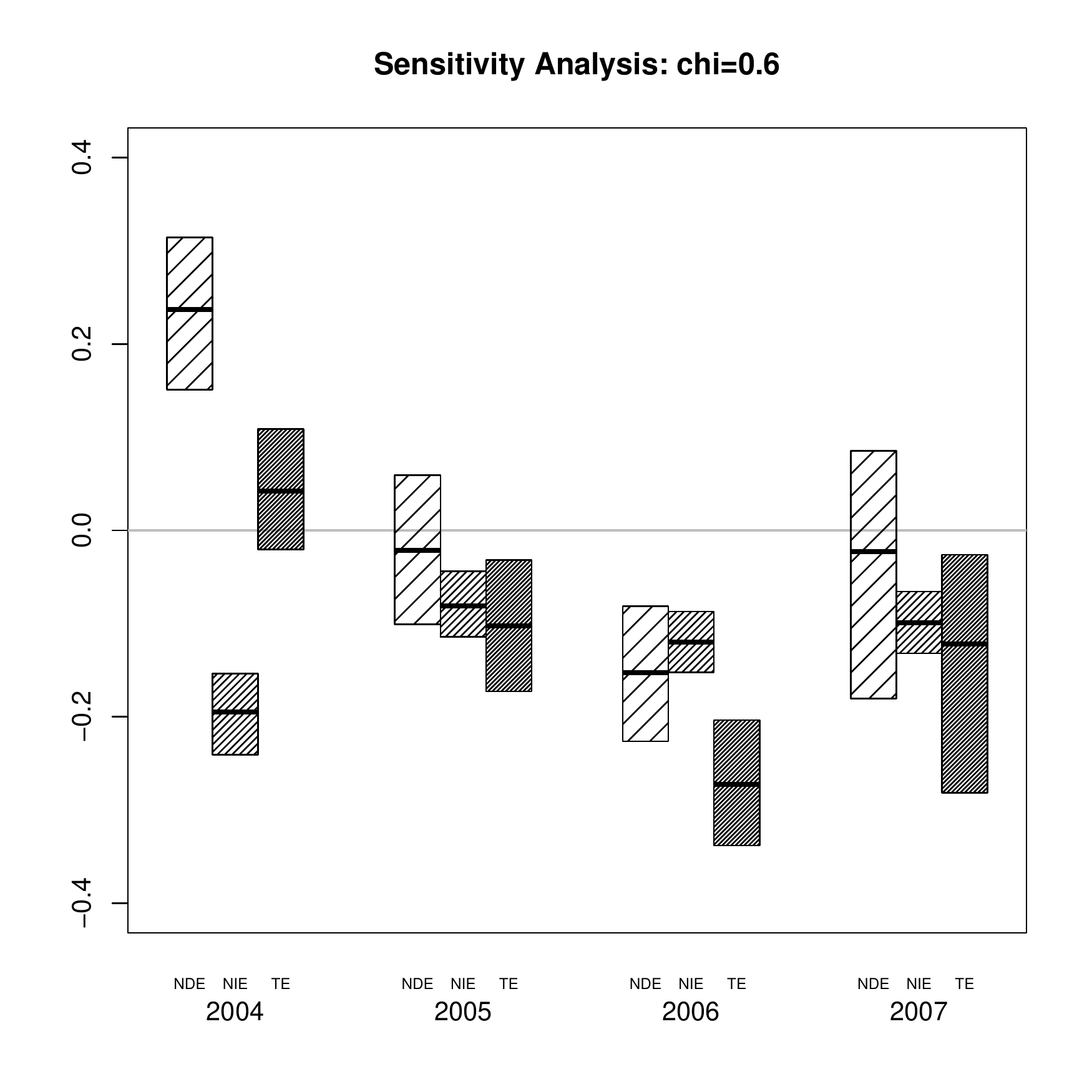}
\includegraphics[width=8cm]{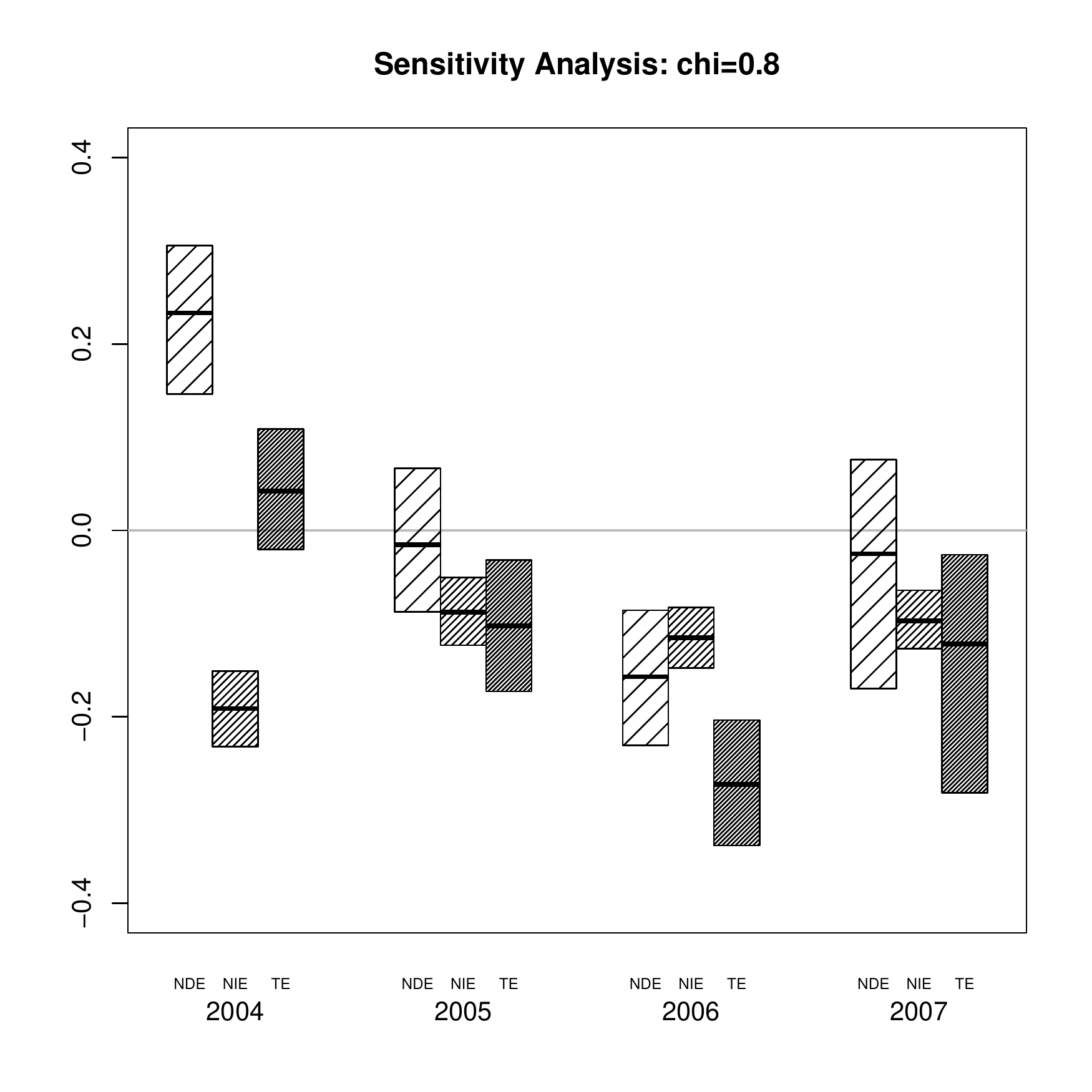}
\includegraphics[width=8cm]{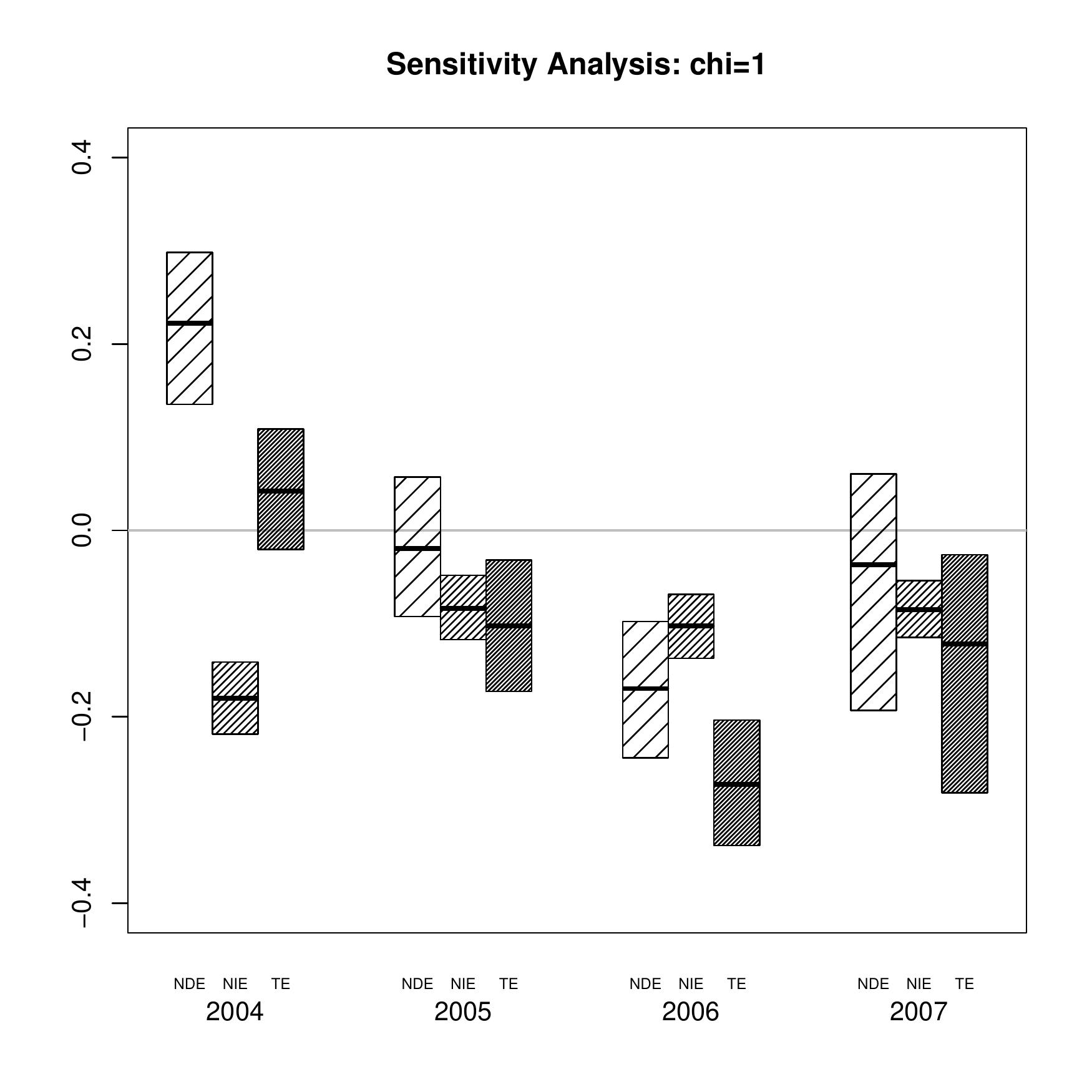}
\includegraphics[width=8cm]{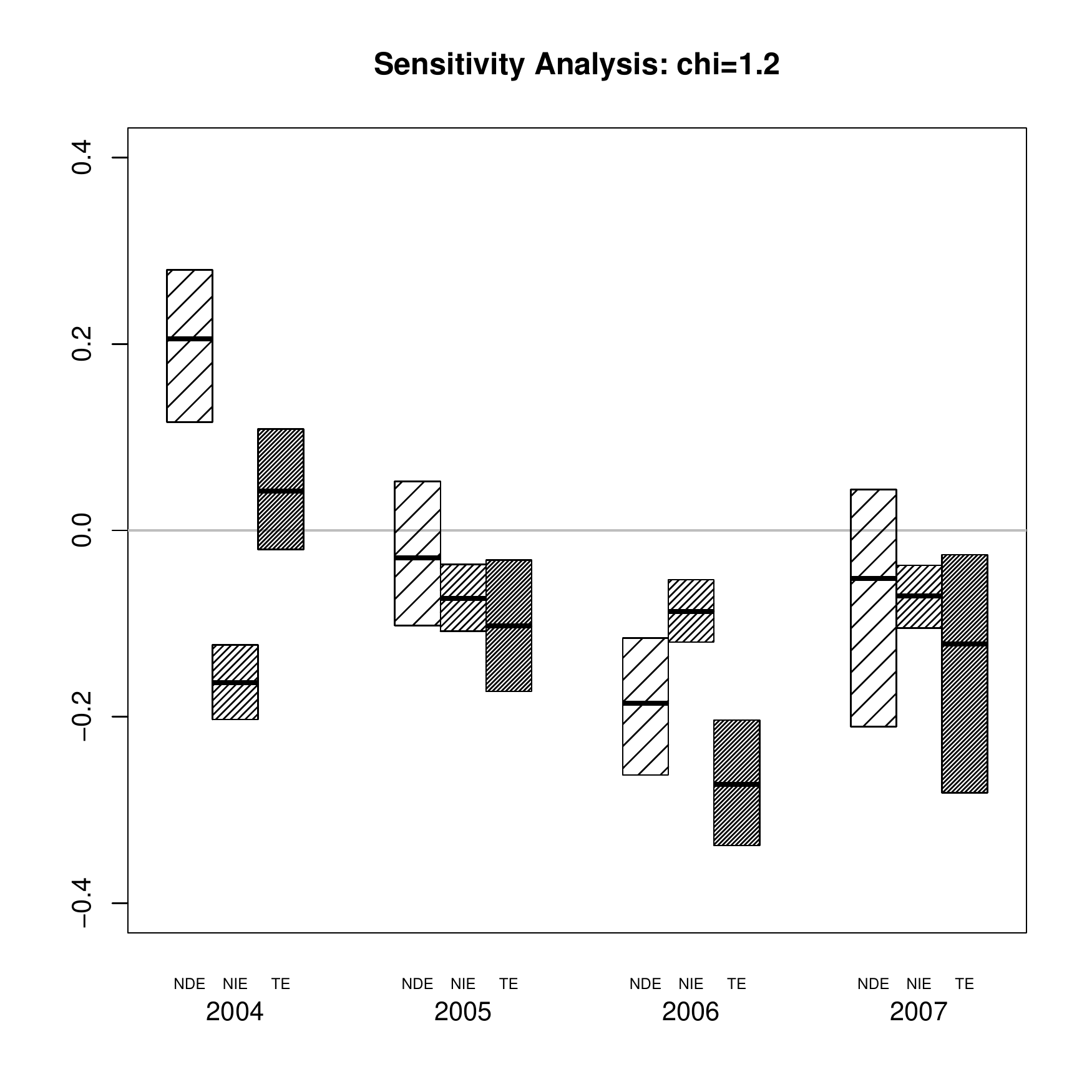}
    \caption{The natural direct, indirect and total effects when the sensitivity parameter $\chi$ takes values $\{0.6,0.8,1.0,1.2\}$ } \label{Sens3}
\end{figure}

This implies that, for the subgroup of subjects who have a large effect of the treatment on the mediator at time $t$, the conditional distributions are unequal. See \cite{kim2016framework} for further rationale behind this specification. 
The posterior means and 95\% credible intervals for the total, indirect and direct effects are displayed in Figure \ref{Sens3} in the Web Appendix for different values of the sensitivity parameter, $\chi=0.5, 0.8, 1.0, 1.2$. 
Note that the total effect is invariant to the values of the sensitivity parameter $\chi$. In addition, setting $\chi=1$ implies Assumption 1. Overall, the NIE (NDE) decreased (increased) as the value of $\chi$ decreased; however, the differences are minimal.

\section{Conclusion}
We have proposed a causal framework for longitudinal mediation which assesses time-varying direct and indirect effects of mediators and have attempted to minimize parametric assumptions for the observed data via the Bayesian nonparametric models within a Bayesian dynamic model. Several assumptions necessary to identify the effects of interest are specified (potentially up to sensitivity parameter). A limitation with our approach is that eliciting a plausible range for the sensitivity parameter $\chi$ is challenging. However, in causal inference settings, there is always subjectivity regarding the assumptions and sensitivity parameters. A major advantage of the proposed approach over cross-sectional mediation is the ability to estimate how direct and indirect effects change over time. Figure \ref{Sens1} shows estimates of the effects with the same Bayesian nonparametric model without the Bayesian dynamic model (i.e., cross-sectional analysis for each time point). In comparison with the indirect effect estimates estimates from the BNP+Bayesian dynamic model, the indirect effects estimates from the BNP without Bayesian dynamic model show no or less change over the time periods, which seems counterintuitive since the effects are expected to decrease as subjects are under the high exposure level longer. This is the main advantage in air pollution studies, where there is a great deal of interest in whether a regulatory policy effects persist and in what ways (through local air quality).

For the air pollution study, there was evidence that the total effect of low \SOTwo\, emission exposure decreased respiratory diseases hospitalizations over the time periods considered here. In our analysis, for every time $t$, we only compare the longitudinal outcomes under two treatment regimes which are different only at time $t$. In future work, we will evaluate the effects of all plausible and interesting combinations of treatment regimes. 
The effect of low \SOTwo\, emission exposure through local ambient air qualities measured by \PMTwo concentrations, on respiratory health outcome was significantly negative for all years. This is expected based on the proven physical/chemical pathway from \SOTwo\, emissions to \PMTwo\, concentrations and from \PMTwo\, concentrations to respiratory diseases. However, this is the first statistical analysis to prove this pathway as a single chain (the chain of accountability \citep{samet2003health}) within a longitudinal setting. 

The results of this air pollution study should be interpreted in light of an important limitation. We assume that the factors listed in Section 8 and Table \ref{Data} in the Web Appendix are sufficient to adjust for confounding relationships among longitudinal exposures, mediators and outcomes. However, as is the case with the for the positive direct effect in 2004, there might be some unmeasured confounder(s).

Future work will incorporate a new covariate model to accommodate spatial correlations among multiple time-varying covariates. These covariates could potentially be used to weaken spatial/regional effects in air pollution studies. A multivariate Gaussian Process (MCGP) model with cross-covariance function proposed in \citep{banerjee2008gaussian} and \cite{Zigl:2012} could be considered;  however, they may be too computationally expensive for a large sample size. Extending the model to allow multiple time-varying mediators is also of interest, as is allowing second order observation models. Finally, we will also explore the impact of the homogeneity in Assumption 2.

\section*{Acknowledgements}
This work was supported by NIH R01ES026217, NIH R01GM112327, NIH R01CA183854 and EPA ACE Center grant: EPA RD-835872. Its contents are solely the responsibility of the grantee and do not necessarily represent the official views of the USEPA.  Further, USEPA does not endorse the purchase of any commercial products or services mentioned in the publication.

\bibliographystyle{plainnat}
\bibliography{ms}

\section*{Appendix A}
We have constructed a database encompassing information on power plant emissions, ambient air quality, meteorology, population demographics, and health outcomes covering the Eastern U.S. across the years 1995 - 2013.  In addition to the unprecedented level of granularity and spatial coverage, construction of this database entailed two major innovations. First, for each data category (e.g., power plants data, Medicare data, Census data etc.), we download the data from almost all available sources, process (cleaned / rearranged / merged) them, and will make them open to the public (except estimated ambient air quality data and Medicare data). Therefore, each of data can be easily used for any other studies as well as researches trying to reproduce our results. Second, we use state-of-the-art techniques and tools to connect spatially and temporally misaligned data sources together: e.g., power plants data are recorded monthly at each electric generating unit (EGU) while ambient air quality data are estimated daily at all US zip code locations. Many geospatial techniques and tools are used to overcome this issue, which are described in Section \ref{sec:linkage}.

\subsection*{Data Sources}
For the power plant data, from the US Energy Information Administration (EIA) and EPA's Air Markets Program Data (AMPD), we are able to access monthly data of the EGU or facility level from 1995 to 2015 (AMPD) or from 1985 to 2012 (EIA). We merge two raw data  by the unique IDs for the EGUs and facilities, which leaves 533 coal-fired facilities or 1359 EGUs. The integrated data encompasses coal-fired facilities not only under the Acid Rain Program (ARP) but under other programs as well: e.g, the \NOx\, Budget Program (NBP), Clean Air Interstate Rule (CAIRSO2, CAIRNOX, CAIROS),  Mercury and Air Toxics Standards (MATS), Transport Rule (TRSO2G1, TRSO2G2, TRNOX, TRNOXOS), Regional Greenhouse Gas Initiative (RGGI), and SIP \NOx\, Program (SIPNOX).  Especially, we collect the power plants level `\SOTwo\, emissions' recorded monthly during 2003-2007 (where there were 406 active coal-fired power plants during the years). The exact coordinates of each facility location are provided.

For the ambient \PMTwo\, exposure data, daily estimates at a 1 km $\times$ 1km grid resolution are constructed over the years 2000-2014 \citep{di2016assessing,di2017hybrid}. A neural network method is used to predict ambient \PMTwo\, exposures with multiple input variables: GEOS-Chem outputs of \PMTwo, Scaling factor from GEOS-Chem (vertical distribution), AOD, OMI absorbing aerosol index, Land-use variables (including NDVI), Meteorological fields, Surface reflectance, Monthly/regional dummy. Through complex hidden layers, the neural network is able to capture nonlinearity and interaction among the input variables for predicting \PMTwo. To train and validate this prediction model, 10-fold cross validation is used on \PMTwo\, monitoring stations data. Cross-validated R$^2$ for each sub-region is between 0.8 and 0.95 (except Mountain region : 0.7-0.8), which suggests the outstanding performance of the neural network estimation. For the zip code level \PMTwo\, data, the four nearest grid values are linked to each zip code by the inverse distance weighted average scheme.

For the health outcomes data (e.g., hospitalization and mortality), we use annual health outcomes among all Medicare beneficiaries obtained from Centers for Medicare \& Medicaid Services (CMS). Restrictively accessible data spans from 1999 to 2015 and is recorded at the zip code level. This data contains information on all Medicare enrollees and their hospitalization and death records. Among many variables, we collect respiratory- and cardiovascular disease- (CVD) related hospitalization and all cause mortality records as our primary health outcomes of interest. In addition to the primary health outcomes, we collect Medicare enrollees' age, gender, race and zip codes of their residence. 

To strengthen our analysis, we collect additional information. We obtain the demographic information of the year 2000 from the US Census Bureau (\href{http://factfinder.census.gov/faces/nav/jsf/pages/index.xhtml}{http://factfinder.census.gov/faces/nav/jsf/pages/index.xhtml}). This annual data includes total population by age, sex, race, economical status, and educational status at the zip
code level. We also obtain the annual average smoking rate data in 2000 from the CDC Behavioral Risk Factor Surveillance System, which contains estimated smoking rates of 3109 counties by using small area estimation methods \citep{dwyer2014cigarette}.

All the dataset can be easily aligned to a zip-code level dataset except the power plant dataset. A smoking rate data for each county is duplicated for all the zip codes within the county. To facilitate this procedure with massive datasets, we use \verb|data.table|, the R package for fast aggregation/manipulation of large data (e.g., 100GB in RAM), and \verb|arepa|, a newly developed R package for EPA data retrieving and processing. The biggest challenge remains that we need to link the power plant dataset recorded with the exact coordinates to this zip-code level dataset in the reality of long-range chemical transport of pollutant emissions. This will be discussed in detail in the subsequent section.

\subsection*{Mapping Emissions to Exposures to Accommodate Long-Range Pollution Transport}\label{sec:linkage}

To convert power-plant level emissions to zip-code level exposures, we use HYSPLIT to simulate forward air mass trajectory paths originating from each of the  coal-fired power plants over the years 2003-2007. For each power plant and for each day, 4 trajectories are simulated at 6 hour intervals throughout the day (midnight, 6am, 12pm, 6pm) to accommodate diurnal variation in wind patterns.     The output of each trajectory provides the exact coordinates (latitude/longitude) and altitude at hourly intervals from a starting location. Continuous trajectory paths are formed by interpolating hourly air mass locations linearly.  Note that the conversion of \SOTwo\, to \PMTwo\, is not a simple diffusion process, as it takes time for \SOTwo\, emitted from a power plant to undergo chemical transformation into \PMTwo.  Thus, each simulated trajectory continues for 168 hours, with a starting vertical height of 500m.
We also truncate trajectories when their altitudes go above 1000m in order to approximate the consideration of pollution below the boundary layer.

All told, the above process results in 2,963,800 trajectories simulated over 4-5 hours on a high-performance computing cluster (the research computing environment supported by the IQSS at Harvard university).  The next step in linking these trajectories to the zip-code level data base is to construct a 30km buffer around each trajectory and enumerate all zip codes that have centroids lying within the buffer.  A zip code is then regarded as ``linked'' to the pollution from a given power plant if trajectories/buffers from that power plant frequently contain that zip code.  Thus, for each possible zip code/power plant pair in the database, we construct a continuous measure of linkage by calculating the percentage, out of a total possible connections ($4 \text{ times per day }\times 365 \text{ days } \times 1 \text{ year}$), that the zip code falls within the trajectory band of the power plant.  Figure \ref{fig1} provides a graphical illustration of linking a power plant to zip codes with HYSPLIT trajectories and their buffers via a PostGIS tool, an open source software program that adds support for geographic objects to the PostgresSQL object database. Section \ref{sec:exposure} discusses how this continuous measure of linkage is used to define a zip-code level emission exposure for the analysis. Note that zip codes that are not linked to any of the power plants are discarded and so are any power plants that are not connected to any of the zip codes.  We have developed an R code (with implementation of a PostGIS tool) for accurate and fast linkage.

\subsection*{Calculating Zip-Code Level Power Plant \SOTwo\, Exposure}\label{sec:exposure}
Once the linkage process is done, we need to determine whether each zip code location is assigned to the treated (exposed to high \SOTwo\, emissions) or the control (exposed to low \SOTwo\, emissions). Towards this end, we extract the monthly total \SOTwo\, emissions from each power plant $j$ denoted by $E_{j,h}$, for $h=1,\cdots, 60$ months.  For zip code $i$, let $W_{j,h,i}$ denote the continuous linkage measure described in Section \ref{sec:linkage}; i.e., the percentage of times trajectories from power plant $j$ pass over the zip code $i$ in month $h$, for every $h=6,18,30,42,54$ and $j=1,2,\ldots P$.  Then, the \SOTwo\ exposure level at zip code $i$ is defined as:
\[\text{\SOTwo\, emission exposure at } i \text{ in month} h \equiv Z_i = \sum_{j \in \mathcal{J}}\sum_{h} \log(E_{j,h}) \times W_{j,h,i},\]
where $\mathcal{J}$ denotes the entire set of the coal-fired power plants.  This provides a continuous measure of the zip-code level \SOTwo, exposure, which we dichotomize it (at the median of all 60 months) to classify each zip code as either exposed to ``high'' or ``low'' \SOTwo\, emissions originated from the coal-fired power plants.

\section*{Appendix B}
$E[Y_{\overline{z}; \overline{M}_{\overline{z}^\prime}(t)}(t)|\overline{\boldsymbol{\theta}}(t)]$ (a priori counterfactuals) as follows 
{\footnotesize
\begin{eqnarray*}
\lefteqn{E[Y_{\overline{z}, \overline{M}_{\overline{z}^\prime}(t)}(t) | \overline{\boldsymbol{\theta}}(t)] }\\
& = & \int E[Y_{\overline{z}, \overline{M}_{\overline{z}^\prime}(t)}(t) \,|\, M_{\overline{z}^\prime}(t)=m, \overline{M}(t-1)=\overline{m}_{t-1}, \overline{Y}(t-1)=\overline{y}_{t-1}, \overline{W}(t-1)=\overline{w}_{t-1}, \overline{Z}(t-1)=\overline{z}_{t-1}, \overline{\boldsymbol{\theta}}(t)] \\
& & \times f(M_{\overline{z}^\prime}(t)=m \,|\, \overline{M}(t-1)=\overline{m}_{t-1}, \overline{Y}(t-1)=\overline{y}_{t-1}, \overline{W}(t-1)=\overline{w}_{t-1}, \overline{Z}(t-1)=\overline{z}_{t-1}, \overline{\boldsymbol{\theta}}(t))\\
& & \underline{\times f(\overline{M}(t-1)=\overline{m}_{t-1}, \overline{Y}(t-1)=\overline{y}_{t-1}, \overline{W}(t-1)=\overline{w}_{t-1}, \overline{Z}(t-1)=\overline{z}_{t-1}, \overline{\boldsymbol{\theta}}(t)) \, d \overline{m}_{t} \, d \overline{y}_{t-1} \, d\overline{w}_{t-1}} \quad \text{(a)}\\
& & \text{(by Assumption 2)}\\
& = & \int E[Y_{\overline{z}, \overline{M}_{\overline{z}}(t)}(t) \,|\,  M_{\overline{z}}(t)=m,\overline{M}(t-1)=\overline{m}_{t-1}, \overline{Y}(t-1)=\overline{y}_{t-1}, \overline{W}(t-1)=\overline{w}_{t-1}, \overline{Z}(t-1)=\overline{z}_{t-1}, \overline{\boldsymbol{\theta}}(t)] \\
& & \times f(M_{\overline{z}^\prime}(t)=m \,|\, \overline{M}(t-1)=\overline{m}_{t-1}, \overline{Y}(t-1)=\overline{y}_{t-1}, \overline{W}(t-1)=\overline{w}_{t-1}, \overline{Z}(t-1)=\overline{z}_{t-1}, \overline{\boldsymbol{\theta}}(t))\\
& & \times f(\overline{M}(t-1)=\overline{m}_{t-1}, \overline{Y}(t-1)=\overline{y}_{t-1}, \overline{W}(t-1)=\overline{w}_{t-1}, \overline{Z}(t-1)=\overline{z}_{t-1}, \overline{\boldsymbol{\theta}}(t)) \, d \overline{m}_{t}\, d \overline{y}_{t-1} \, d\overline{w}_{t-1}) \\
& & \text{(by Assumption 1)}\\
& = & \int E[Y_{\overline{z}, \overline{M}_{\overline{z}}(t)}(t) \,|\,  M(t)(\overline{z}_t)=m,\overline{M}(t-1)=\overline{m}_{t-1}, \overline{Y}(t-1)=\overline{y}_{t-1}, \overline{W}(t-1)=\overline{w}_{t-1}, {\color{red}\overline{Z}(t)}={\color{red}\overline{z}_{t}}, \overline{\boldsymbol{\theta}}(t)] \\
& & \times f(M_{\overline{z}^\prime}(t)=m \,|\, \overline{M}(t-1)=\overline{m}_{t-1}, \overline{Y}(t-1)=\overline{y}_{t-1}, \overline{W}(t-1)=\overline{w}_{t-1}, {\color{red}\overline{Z}(t)}={\color{red}\overline{z}_{t}^\prime}, \overline{\boldsymbol{\theta}}(t))\\
& & \times f(\overline{M}(t-1)=\overline{m}_{t-1}, \overline{Y}(t-1)=\overline{y}_{t-1}, \overline{W}(t-1)=\overline{w}_{t-1}, \overline{Z}(t-1)=\overline{z}_{t-1}, \overline{\boldsymbol{\theta}}(t)) \, d \overline{m}_{t}\, d \overline{y}_{t-1} \, d\overline{w}_{t-1} \\
& & \text{(by SUTVA or consistency assumption)}\\
& = & \int E[Y(t) \,|\,  M(t)=m,\overline{M}(t-1)=\overline{m}_{t-1}, \overline{Y}(t-1)=\overline{y}_{t-1}, \overline{W}(t-1)=\overline{w}_{t-1}, \overline{Z}(t)=\overline{z}_{t-1}, \overline{\boldsymbol{\theta}}(t)] \\
& & \times f(M(t)=m \,|\, \overline{M}(t-1)=\overline{m}_{t-1}, \overline{Y}(t-1)=\overline{y}_{t-1}, \overline{W}(t-1)=\overline{w}_{t-1}, \overline{Z}(t)=\overline{z}_{t}^\prime, \overline{\boldsymbol{\theta}}(t))\\
& & \times f(\overline{M}(t-1)=\overline{m}_{t-1}, \overline{Y}(t-1)=\overline{y}_{t-1}, \overline{W}(t-1)=\overline{w}_{t-1}, \overline{Z}(t-1)=\overline{z}_{t-1}, \overline{\boldsymbol{\theta}}(t)) \, d \overline{m}_{t}\, d \overline{y}_{t-1} \, d\overline{w}_{t-1} \\
& & \text{(by Bayesian dynamic model assumption)}\\
& = & \int E[Y(t) \,|\,  M(t)=m, {Y}(t-1)={y}_{t-1}, {W}(t-1)={w}_{t-1}, {Z}(t)={z}_{t}, {\boldsymbol{\theta}}(t)] \\
& & \times f(M(t)=m \,|\, {M}(t-1)= {m}_{t-1}, {Y}(t-1)={y}_{t-1}, {W}(t-1)={w}_{t-1}, {Z}(t)={z}_{t}^\prime, {\boldsymbol{\theta}}(t))\\
& & \times f(\overline{M}(t-1)=\overline{m}_{t-1}, \overline{Y}(t-1)=\overline{y}_{t-1}, \overline{W}(t-1)=\overline{w}_{t-1}, \overline{Z}(t-1)=\overline{z}_{t-1}, \overline{\boldsymbol{\theta}}(t)) \, d \overline{m}_{t}\, d \overline{y}_{t-1} \, d\overline{w}_{t-1} \\
& = & \int E[Y(t) \,|\,  M(t)=m, {Y}(t-1)={y}_{t-1}, {W}(t-1)={w}_{t-1}, {Z}(t)={z}_{t}, {\boldsymbol{\theta}}(t)] \\
& & \times f(M(t)=m \,|\, {M}(t-1)= {m}_{t-1}, {Y}(t-1)={y}_{t-1}, {W}(t-1)={w}_{t-1}, {Z}(t)={z}_{t}^\prime, {\boldsymbol{\theta}}(t))\\
& & \times \color{red}f(Y(t-1)=y_{t-1})\,|\,\overline{M}(t-1)=\overline{m}_{t-1}, \overline{Y}(t-2)=\overline{y}_{t-2}, \overline{W}(t-1)=\overline{w}_{t-1}, \overline{Z}(t-1)=\overline{z}_{t-1}, \overline{\boldsymbol{\theta}}(t)) \\
& & \times\color{red} f(M(t-1)=m_{t-1}\,|\,\overline{M}(t-2)=\overline{m}_{t-2}, \overline{Y}(t-2)=\overline{y}_{t-2}, \overline{W}(t-1)=\overline{w}_{t-1}, \overline{Z}(t-1)=\overline{z}_{t-1}, \overline{\boldsymbol{\theta}}(t))  \\
& & \times\color{red} f(W(t-1)=w_{t-1}\,|\,\overline{M}(t-2)=\overline{m}_{t-2}, \overline{Y}(t-2)=\overline{y}_{t-2}, \overline{W}(t-2)=\overline{w}_{t-2}, \overline{Z}(t-1)=\overline{z}_{t-1}, \overline{\boldsymbol{\theta}}(t)) \\
& & \times f(\overline{M}(t-2)=\overline{m}_{t-2}, \overline{Y}(t-2)=\overline{y}_{t-2}, \overline{W}(t-2)=\overline{w}_{t-2}, \overline{Z}(t-1)=\overline{z}_{t-1}, \overline{\boldsymbol{\theta}}(t)) \, d \overline{m}_{t}\, d \overline{y}_{t-1} \, d\overline{w}_{t-1} \\
\end{eqnarray*}}where the conditional distributions for $Y(t-1), M(t-1)$, and $W(t-1)$ can be simplified using the Bayesian dynamic model assumption (again),
{\footnotesize
\begin{eqnarray*}
& = & \int E[Y(t) \,|\,  M(t)=m, {Y}(t-1)={y}_{t-1}, {W}(t-1)={w}_{t-1}, {Z}(t)={z}_{t}, {\boldsymbol{\theta}}(t)] \\
& & \times f(M(t)=m \,|\, {M}(t-1)= {m}_{t-1}, {Y}(t-1)={y}_{t-1}, {W}(t-1)={w}_{t-1}, {Z}(t)={z}_{t}^\prime, {\boldsymbol{\theta}}(t))\\
& & \times f(Y(t-1)=y_{t-1}\,|\, {M}(t-1)={m}_{t-1}, {Y}(t-2)={y}_{t-2}, {W}(t-2)={w}_{t-2}, {Z}(t-1)={z}_{t-1}, {\boldsymbol{\theta}}(t-1)) \\
& & \times f(M(t-1)=m_{t-1}\,|\,{M}(t-2)={m}_{t-2}, {Y}(t-2)={y}_{t-2}, {W}(t-2)={w}_{t-2}, {Z}(t-1)={z}_{t-1}, {\boldsymbol{\theta}}(t-1))  \\
& & \times f(W(t-1)=w_{t-1}\,|\,{M}(t-2)={m}_{t-2}, {Y}(t-2)={y}_{t-2}, {W}(t-2)={w}_{t-2}, {Z}(t-1)={z}_{t-1}, {\boldsymbol{\theta}}(t-1)) \\
& & \times{\color{red} f(\overline{M}(t-2)=\overline{m}_{t-2}, \overline{Y}(t-2)=\overline{y}_{t-2}, \overline{W}(t-2)=\overline{w}_{t-2}, \overline{Z}(t-1)=\overline{z}_{t-1}, \overline{\boldsymbol{\theta}}(t))} \, d \overline{m}_{t}\, d \overline{y}_{t-1} \, d\overline{w}_{t-1}. 
\end{eqnarray*}}For the joint distribution of $(\overline{M}(t-2), \overline{Y}(t-2), \overline{W}(t-2), \overline{Z}(t-1), \overline{\boldsymbol{\theta}}(t))$, we follow the same steps we did for the joint distribution in (a).

\section*{Plots and Tables}

.
\begin{table}[p]
\centering
{\footnotesize
\begin{tabular}{lcccccc}
\hline
\hline
 && \multicolumn{2}{c}{\bf Low \SOTwo\, emissions } & &
 \multicolumn{2}{c}{\bf High \SOTwo\, emissions}\\
& & Median & IQR & & Median & IQR\\
\hline
\bf Exposure & & & & & & \\
\quad Year 2003 & & \multicolumn{2}{c}{n=9432} & & \multicolumn{2}{c}{n=6040} \\
\quad Year 2004 & & \multicolumn{2}{c}{n=7866} & & \multicolumn{2}{c}{n=7606} \\
\quad Year 2005 & & \multicolumn{2}{c}{n=7267} & & \multicolumn{2}{c}{n=8205} \\
\quad Year 2006 & & \multicolumn{2}{c}{n=9089} & & \multicolumn{2}{c}{n=6383} \\
\quad Year 2007 & & \multicolumn{2}{c}{n=9783} & & \multicolumn{2}{c}{n=5689} \\
& & & & & &\\
\bf Medicare Data & & & & & & \\
\quad Person-year 2003  &&     563.5  & (248.4, 1560.6) & & 662.4  & (274.2, 1840.7) \\
\quad Person-year 2004   &&     530.8  & (241.3, 1482.2)& & 694.3  & (285.4, 1886.2) \\
\quad Person-year 2005   &&     489.4  & (232.8, 1356.8)& & 750.0  & (300.2, 1946.6) \\
\quad Person-year 2006   &&     524.4  & (236.0, 1464.4)& & 725.4  & (283.6, 1865.3) \\
\quad Person-year 2007   &&     505.2  & (230.8, 1401.0)& & 742.2  & (286.2, 1892.8) \\
\quad Respiratory Hospitalization 2003  & & 10 &  (4, 27)    &&    13   &  (5, 33) \\
\quad Respiratory Hospitalization 2004  & & 8.5 &  (3, 24)    &&    12  &  (5, 30) \\
\quad Respiratory Hospitalization 2005  & & 8 &  (3, 22)    &&    12   &  (5,31) \\
\quad Respiratory Hospitalization 2006  & & 8 &  (3, 22)    &&    12   &  (5, 30) \\
\quad Respiratory Hospitalization 2007  & & 8 &  (3, 22)    &&    12   &  (5, 30) \\
& & & & & &\\
\bf Ambient Air Quality and Other Data & & & & & & \\
\quad Ambient $\text{PM}_{2.5}$ 2003 & &13.7    &  (11.3, 15.3)    &  &    16.8     & (15.8, 17.9)\\
\quad Ambient $\text{PM}_{2.5}$ 2004 & &12.2    &  (10.6, 13.8)    &  &    16.2     & (14.9, 17.3)\\
\quad Ambient $\text{PM}_{2.5}$ 2005 & &15.6    &  (12.4, 17.8)    &  &    17.7     & (16.3, 18.9)\\
\quad Ambient $\text{PM}_{2.5}$ 2006 & &14.9    &  (13.1, 16.6)    &  &    18.5     & (17.2, 19.9)\\
\quad Ambient $\text{PM}_{2.5}$ 2007 & &13.6    &  (12.1, 15.7)    &  &    16.6     & (15.8, 17.5)\\
\quad Temperature 2003& & 27.3  &  (26.1, 28.9)     & &   25.9  &    (24.7, 26.9)\\
\quad Temperature 2004& & 28.2  &  (25.5, 29.8)     & &   25.9  &    (24.4, 27.6)\\
\quad Temperature 2005& & 29.2  &  (27.5, 30.4)     & &   27.1  &    (26.1, 28.2)\\
\quad Temperature 2006& & 26.9  &  (24.8, 30.5)     & &   27.0  &    (25.4, 28.4)\\
\quad Temperature 2007& & 27.6  &  (26.0, 30.1)     & &   27.3  &    (26.5, 28.9)\\
\quad Ground Elevation$^\bigstar$ & & 175  &  (0, 282)     & &   232  &    (141, 317)\\
& & & & & &\\
\bf Census Data & & & & & & \\
\quad Total Population (1,000)$^\bigstar$ &&   6.2 & (2.3, 17.9)  & &   6.7  & (2.7, 19.2)\\
\quad \% Urban$^\bigstar$ && 43.9   &   (0.0, 93.5)  &&      58.4    &  (0.0, 99.2)\\
\quad \% White$^\bigstar$ && 92.9 & (76.5, 97.4)  &&  94.4     & (81.9, 98.0)\\
\quad \% High School$^\bigstar$ && 34.4    & (28.5, 39.4)  &  &     36.8   &  (29.9, 44.0)\\
\quad \% Female$^\bigstar$ && 51.0   &   (49.8, 52.3)  &&       51.2    &  (49.9, 52.3)\\
\quad \% Poor$^\bigstar$ & & 10.5  &   (6.1, 16.5)  &&       9.1  &  (5.3, 14.4)\\
\quad Median House Income (\$1000)$^\bigstar$ && 37.6  & (30.8, 47.6)  & &    39.7 & (32.3, 50.4)\\
& & & & & &\\
\bf CDC-BRFSS & & & & & & \\
\quad \% Smoke$^\bigstar$ &&  26.6   &   (24.4, 28.4)   &&        27.7   &   (25.0, 29.6)\\
\end{tabular}
\caption{Summary statistics for covariates and outcomes available. $^\bigstar$ summarizes the confounders based on the 2003 exposure level. }}\label{Data}
\end{table}
.
\begin{figure}[pt]
    \centering
\includegraphics[width=12cm]{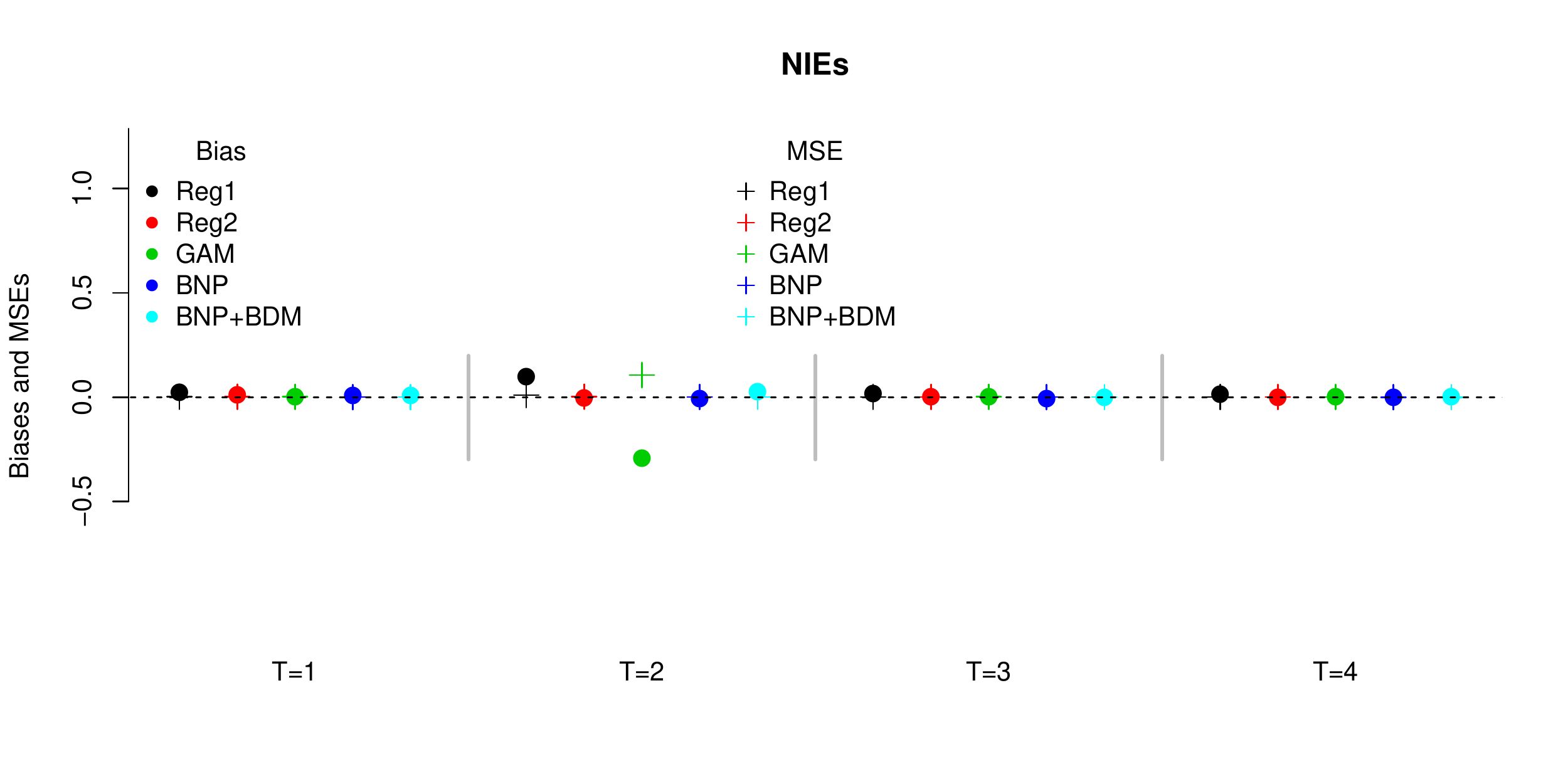}
\includegraphics[width=12cm]{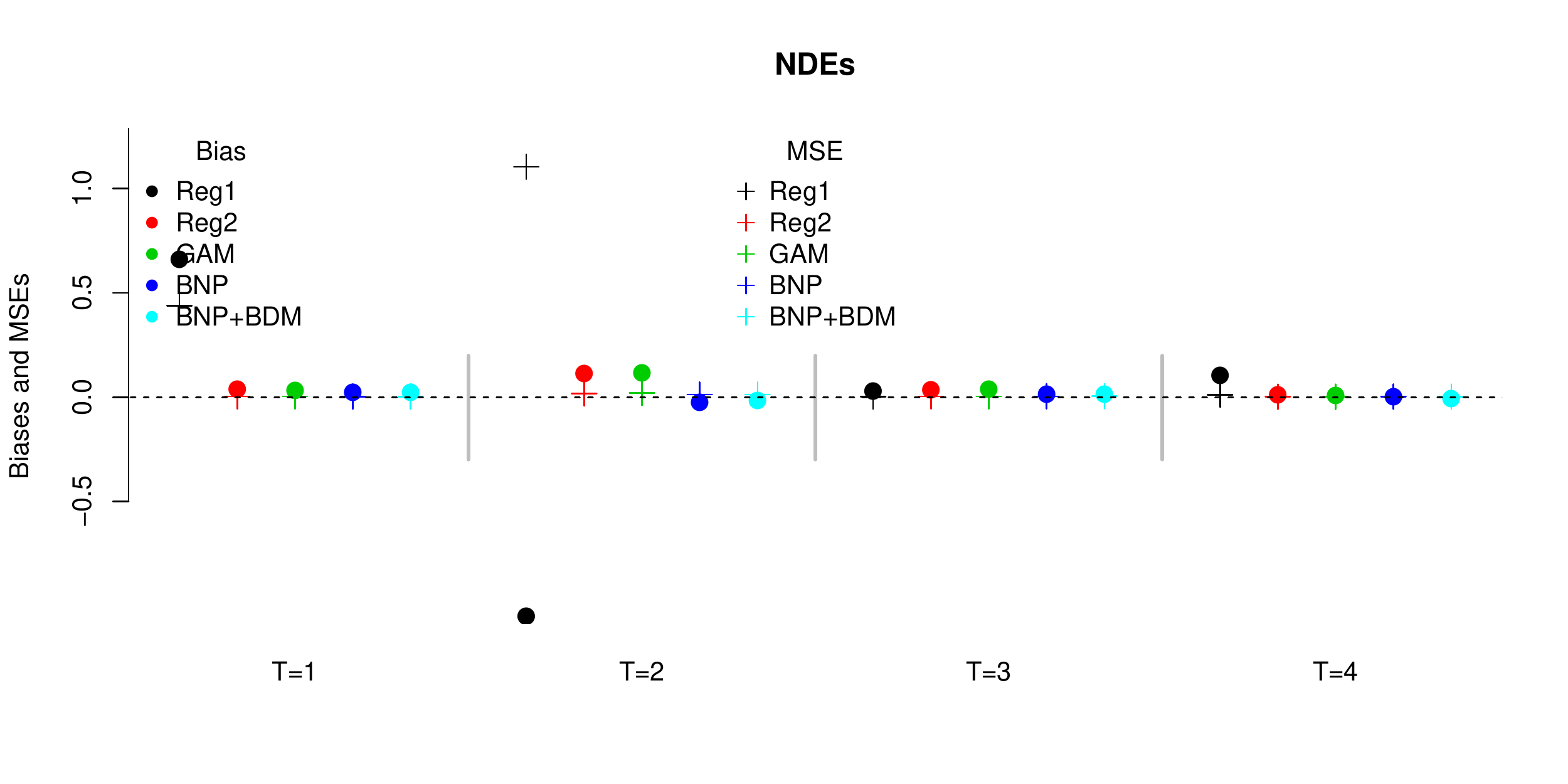}
\includegraphics[width=12cm]{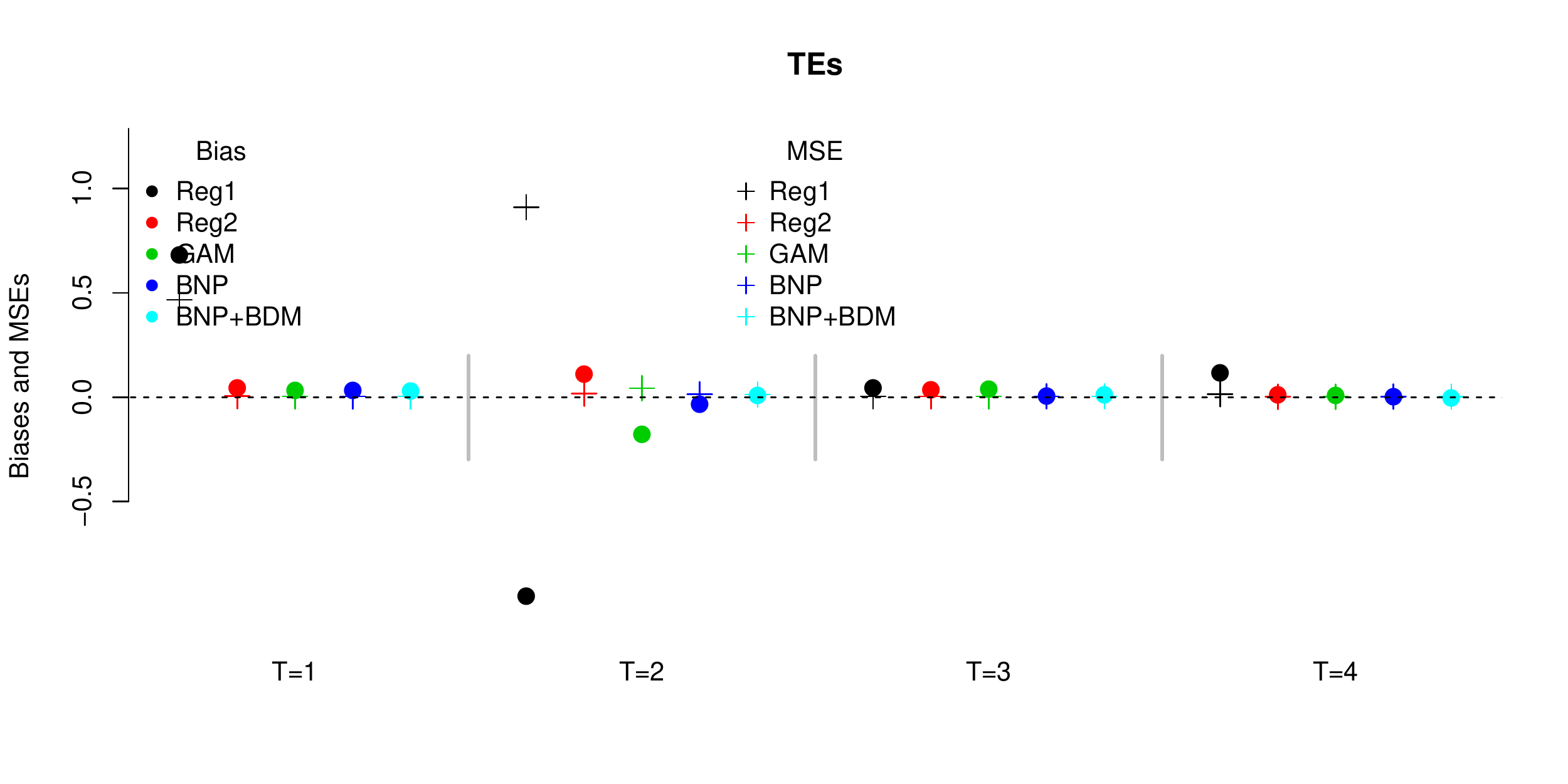}
\caption{Case 2. The coefficients of the one-time preceding variables decrease by \% 30 at each time. Simulation results for natural indirect, direct and total effects over 400 replications. Solid circles are biases and crosses are mean squared errors (MSEs) from 5 different models: Reg 1 (regression models with all previous predictors); Reg 2 (regression models with one-time preceding predictors); GAM (Generalized Additive Models with one-time preceding predictors); BNP (Bayesian Nonparametric models w/o Bayesian dynamic model); BNP+BDM (Bayesian Nonparametric models w/ Bayesian dynamic model)}\label{simulation2}
\end{figure}

\begin{figure}[pt]
    \centering
\includegraphics[width=12cm]{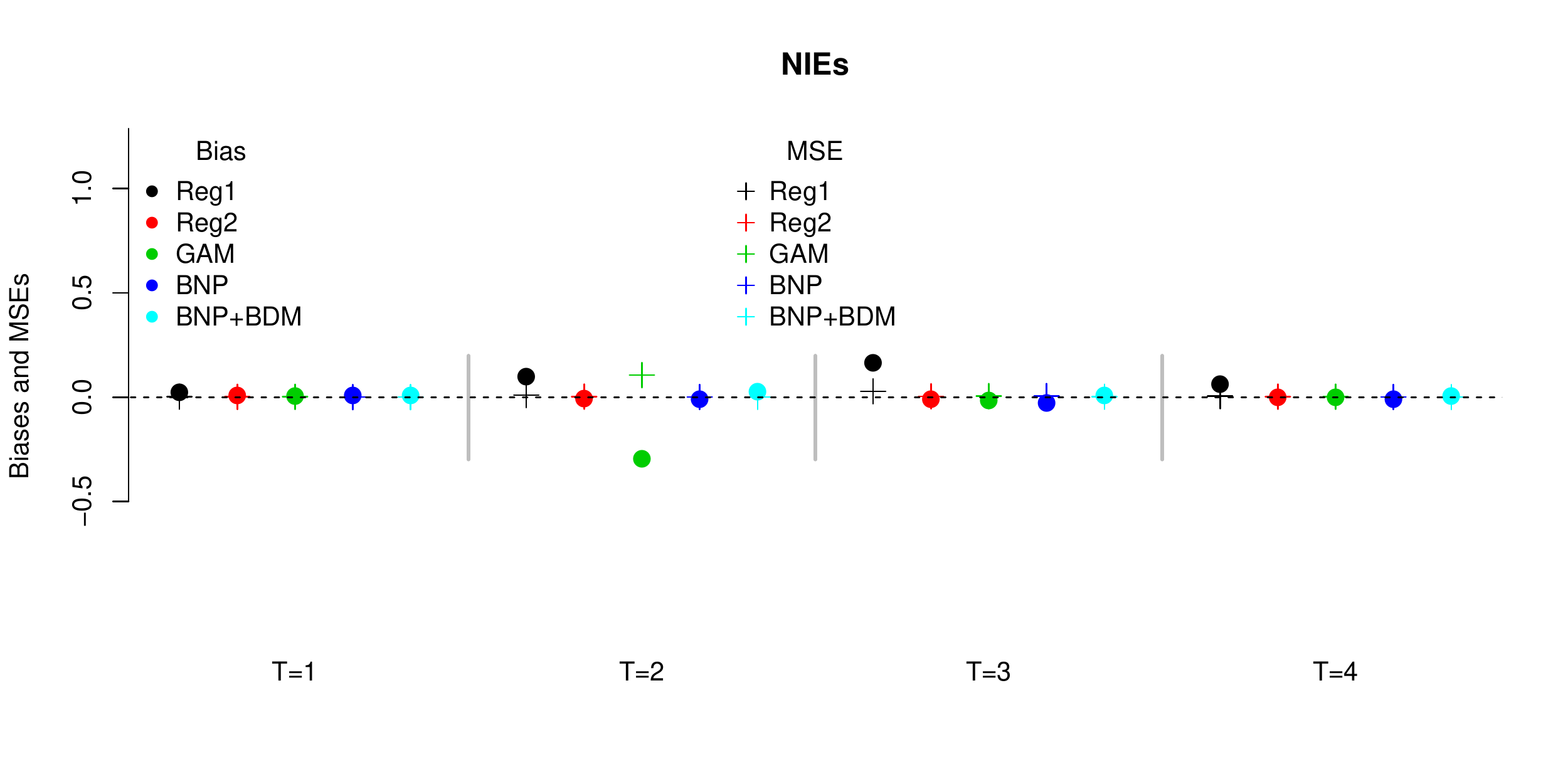}
\includegraphics[width=12cm]{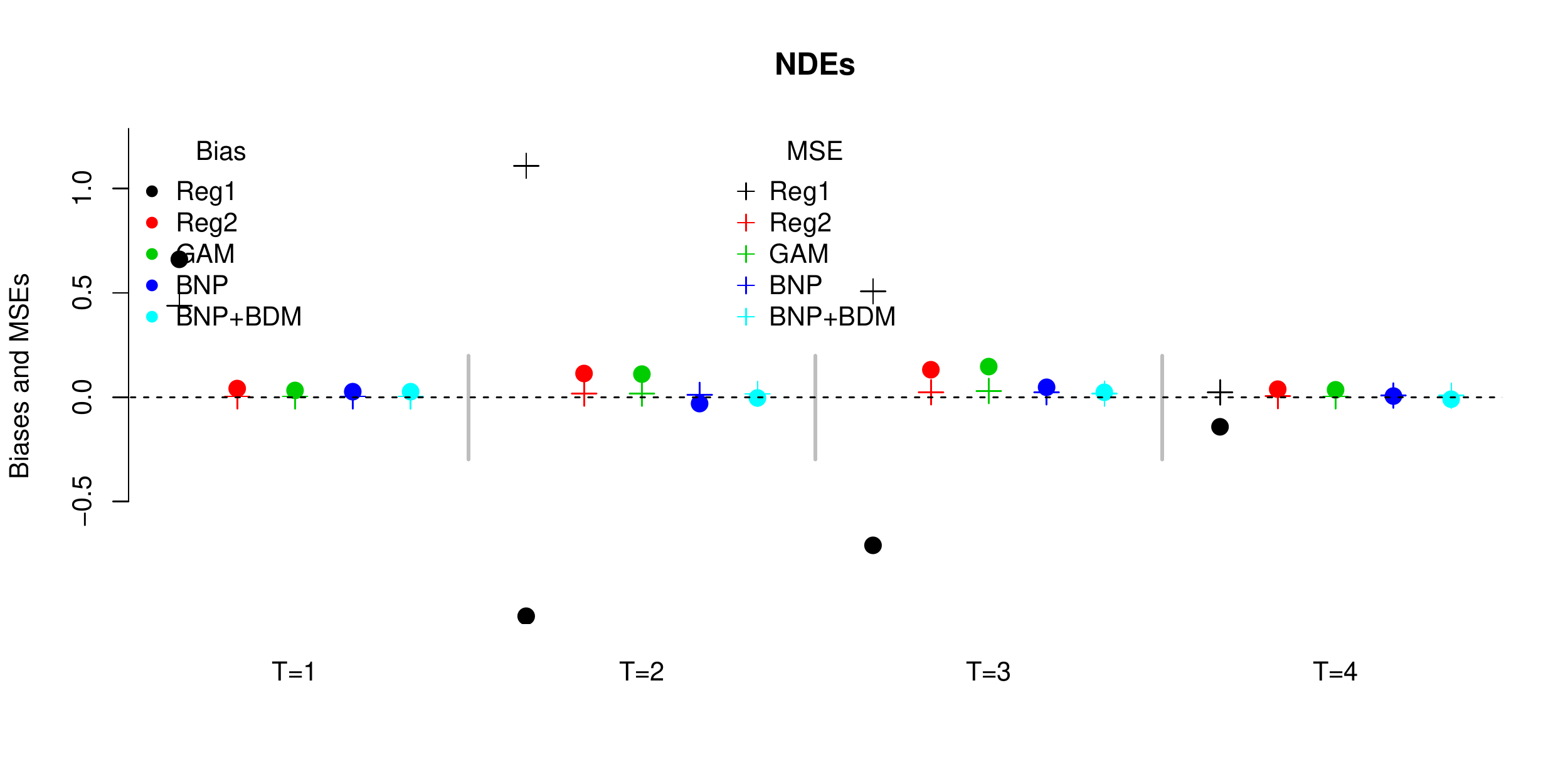}
\includegraphics[width=12cm]{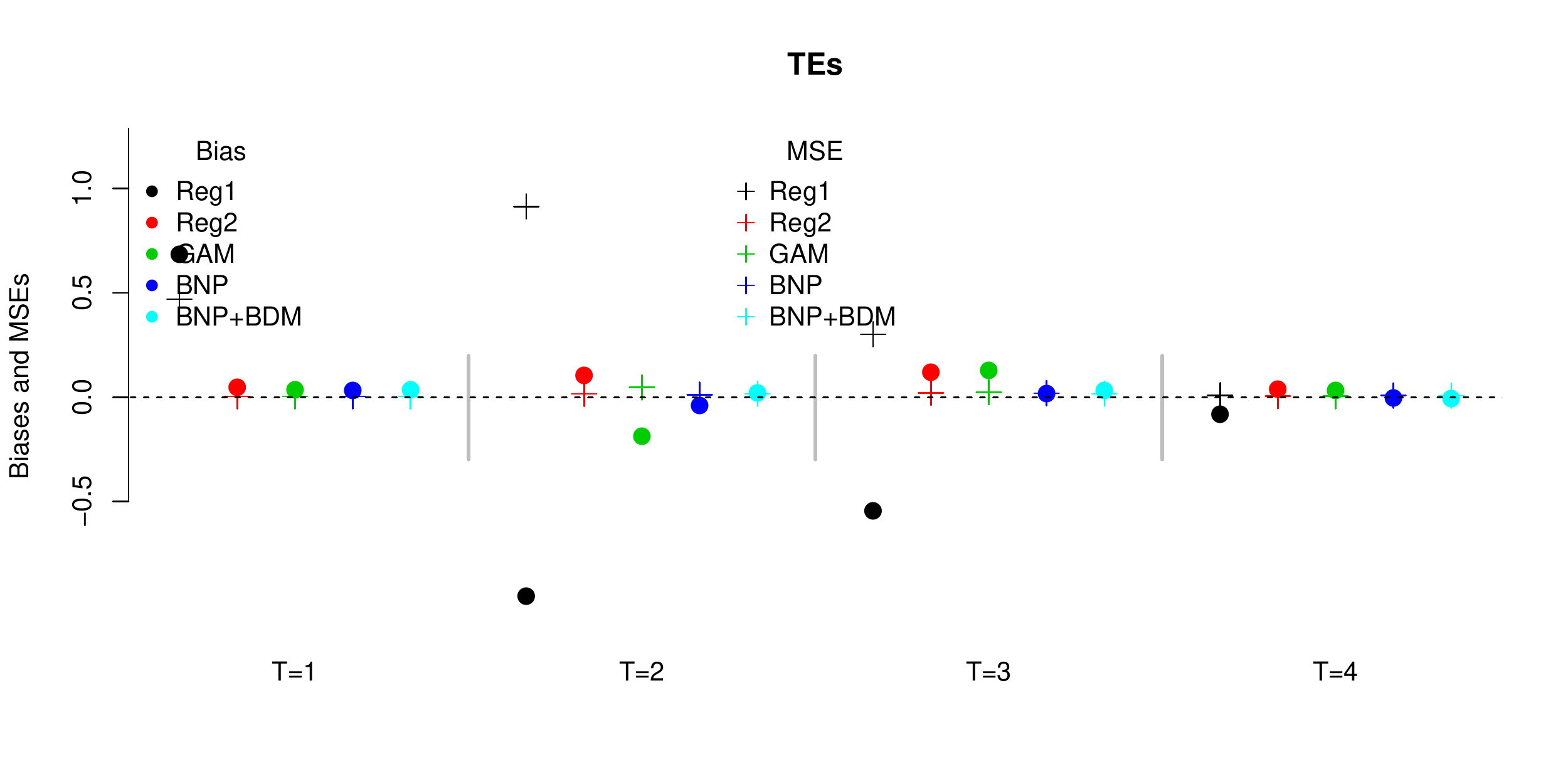}
\caption{Case 3. The coefficients of the one-time preceding variables change irregularly at each time. Simulation results for natural indirect, direct and total effects over 400 replications. Solid circles are biases and crosses are mean squared errors (MSEs) from 5 different models: Reg 1 (regression models with all previous predictors); Reg 2 (regression models with one-time preceding predictors); GAM (Generalized Additive Models with one-time preceding predictors); BNP (Bayesian Nonparametric models w/o Bayesian dynamic model); BNP+BDM (Bayesian Nonparametric models w/ Bayesian dynamic model)}\label{simulation3}
\end{figure}

\end{document}